\documentclass[11pt]{article}
\usepackage{amssymb}
\usepackage{latexsym}
\usepackage[mathscr]{eucal}
\usepackage{citesort}
\usepackage{url}
\usepackage{deleq}
\usepackage{a4}
\usepackage{cite}
\usepackage{ntheorem}

\newcommand{\verbatimy}[1]{}

\newcommand{\mcW}{{\mycal W}}
\newcommand{\ApSwSw}{Appendix~\ref{SwSs}}
\newcommand{\abs}[1]{\left\vert#1\right\vert}

\newcommand{\Lie}{\EuScript L}

\newcommand{\nablash}{\nabla{\kern -.75 em
     \raise 1.5 true pt\hbox{{\bf/}}}\kern +.1 em}
\newcommand{\Deltash}{\Delta{\kern -.69 em
     \raise .2 true pt\hbox{{\bf/}}}\kern +.1 em}
\newcommand{\Rslash}{R{\kern -.60 em
     \raise 1.5 true pt\hbox{{\bf/}}}\kern +.1 em}
\newcommand{\Div}{\operatorname{div}}

\newcommand{\gammab}{\bar\gamma}
\newcommand{\chib}{\bar\chi}

\newcommand{\Hb}{\bar H}
\newcommand{\Ab}{\bar A}
\newcommand{\Bb}{\bar B}
\newcommand{\betah}{{\hat\beta}}

\newcommand{\Ric}{\operatorname{Ric}}

\newcommand{\hbound}{\mathring{\mathsf{H}}}
\newcommand{\cbord}{{\mathsf{C}}}
\newcommand{\pM}{\partial M}

\newcommand{\hhyp}{\mathring{\mathcal{H}}}

\newcommand{\maclK}{{\mathcal K}}
\newcommand{\maclKzo}{{\mathcal K}^{\bot_g}_0}
\newcommand{\maclKz}{{\mathcal K}_0}
\newcommand{\hbord}{\hbound}%
\newcommand{\cKi}{{\mycal K}_{i^0}}
\newcommand{\fg}{{}^4g}
\newcommand{\mcO}{{\mycal O}}
\newcommand{\mcT}{{\mycal T}}
\newcommand{\mcU}{{\mycal U}}
\newcommand{\mcV}{{\mycal V}}

\newcommand{\rd}{\,{ d}} 

\newcommand{\ourU}{\mathbb U}

\newcommand{\hyp}{{\mycal S}}

\newcommand{\bg}{\overline{g}}

\newcommand{\bmetric}{{b}} 

\newcommand{\znabla} {\mathring{\nabla}} 
\newcommand{\mn} {M} 

\newcommand{\mcM}{{\mycal M}}
\newcommand{\mcP}{{\mycal P}}
\newcommand{\mcK}{{\mycal K}}
\newcommand{\notreP}{{\widehat P}}

\newcommand{\bea}{\begin{eqnarray}}

\newcommand{\Mext}{M_{\mbox{\scriptsize \rm ext}}}
\newcommand{\bel}[1]{\begin{equation}\label{#1}}
\newcommand{\beal}[1]{\begin{eqnarray}\label{#1}}
\newcommand{\beadl}[1]{\begin{deqarr}\label{#1}}
\newcommand{\eeadl}[1]{\arrlabel{#1}\end{deqarr}}
\newcommand{\eeal}[1]{\label{#1}\end{eqnarray}}
\newcommand{\eead}[1]{\end{deqarr}}
\newcommand{\eea}{\end{eqnarray}}
\newcommand{\nn}{\nonumber}
\newcommand{\Hess}{\mathrm{Hess}\,}
\newcommand{\Ricc}{\mathrm{Ric}\,}
\newcommand{\Riccg}{\Ricc(g)}
\newcommand{\Lpsi}{L^2_{\psi}}
\newcommand{\Lpsione}{\zH^1_{\phi,\psi}}
\newcommand{\Lpsitwo}{\zH^2_{\phi,\psi}}

\newcommand{\Lpsig}{L^2_{\psi}(g)}
\newcommand{\Lpsioneg}{\zH^1_{\phi,\psi}(g)}
\newcommand{\Lpsitwog}{\zH^2_{\phi,\psi}(g)}
\newcommand{\Lpsikg}[2]{\zH^{#1}_{\phi,\psi}(#2)}

\newcommand{\be}{\begin{equation}}
\newcommand{\ee}{\end{equation}}
\newcommand{\divr }{\mbox{\rm div}\,}
\newcommand{\tr}{\mbox{\rm tr}\,}
\newcommand{\trg}{\tr}
\newcommand{\ext}{\mbox{\rm \scriptsize ext}\,}
\newcommand{\J}{\delta J}
\newcommand{\source}{\delta \rho}
\newcommand{\eq}[1]{\eqref{#1}}
\newcommand{\Eq}[1]{Equation~(\ref{#1})}

\DeclareFontFamily{OT1}{rsfs}{}
\DeclareFontShape{OT1}{rsfs}{m}{n}{ <-7> rsfs5 <7-10> rsfs7 <10->
rsfs10}{} \DeclareMathAlphabet{\mycal}{OT1}{rsfs}{m}{n}
\def\scri{{\mycal I}}%
\def\scrip{\scri^{+}}%

\newcommand{\zmcH }{\,\,\,\,\mathring{\!\!\!\!\mycal H}{}}

\usepackage{amsmath} 

\renewcommand{\theequation}{\thesection.\arabic{equation}}

\let\ssection=\section

\renewcommand{\section}{\setcounter{equation}{0}\ssection}

\newtheorem{defi}{\sc Definition\rm}[section]

\newtheorem{theorem}[defi]{\sc Theorem\rm}

\newtheorem{prop}[defi]{\sc Proposition\rm}

\newtheorem{Definition}[defi]{\sc Definition\rm}
\newtheorem{Def}[defi]{\sc Definition\rm}

\newtheorem{theor}[defi]{\sc Theorem\rm}

\newtheorem{Proposition}[defi]{\sc Proposition\rm}
\newtheorem{lem}[defi]{\sc Lemma\rm}
\newtheorem{Lemma}[defi]{\sc Lemma\rm}

\newtheorem{cor}[defi]{\sc Corollary\rm}
\theorembodyfont{\upshape}
\newtheorem{Remark}[defi]{\sc Remark\rm}

\newtheorem{remark}[defi]{\sc Remark\rm}

\newcommand{\qed}{\hfill $\Box$\bigskip}

\newcommand{\proof}{\noindent {\sc Proof:\ }}

\def \Reel{\mathbb{R}}
\def \R {\Reel}

\def \Nat{\mathbb{N}}
\def \Z{\mathbb{Z}}
\def \N {\Nat}

\newcommand{\bM}{\,\overline{\!M}}

\newcommand{\zHkpp}{\zHk_{\phi,\psi}}
\newcommand{\Hkpp}{H^k_{\phi,\psi}}
\newcommand{\zHk}{\zH^k}
\newcommand{\zH}{\mathring{H}}

\newcounter{mnotecount}[section]

\renewcommand{\themnotecount}{\thesection.\arabic{mnotecount}}

\newcommand{\mnote}[1]
{\protect{\stepcounter{mnotecount}}$^{\mbox{\footnotesize $
\bullet$\themnotecount}}$ \marginpar{
\raggedright\tiny\em $\!\!\!\!\!\!\,\bullet$\themnotecount: #1} }

\newcommand{\rmnote}[1]{}

\newcommand{\loc}{{\textrm{loc}}}

\begin{document}
\title{On mapping properties of the general relativistic
constraints operator in weighted function spaces, with
applications}
\author{
Piotr T. Chru\'sciel\thanks{Partially supported by a Polish
Research Committee grant 2 P03B 073 24; email \protect\url{
piotr@gargan.math.univ-tours.fr}} \\
Albert Einstein Institute\thanks{Visiting Scientist. Permanent
address: D\'epartement de Math\'ematiques, Facult\'e des
Sciences, Parc de Grandmont, F37200 Tours, France.}\\
 Golm, Germany \\
 \\ Erwann Delay\thanks{Partially supported by the ACI program of the French Ministry of Research; email
  \protect\url{delay@gargan.math.univ-tours.fr}}\\ D\'epartement de Math\'ematiques\\
Facult\'e des Sciences\\ Parc de Grandmont\\ F37200 Tours, France
}
\date{\today}
\maketitle


\begin{abstract}
Generalising an analysis of Corvino and Schoen, we study
surjectivity properties of the constraint map in general
relativity in a large class of weighted Sobolev spaces. As a
corollary we prove several perturbation, gluing, and extension
results: we show existence of non-trivial, singularity-free,
vacuum space-times which are \emph{stationary} in a neighborhood
of $i^0$; for small perturbations of \emph{parity-covariant}
initial data sufficiently close to those for Minkowski space-time
this leads to space-times with a smooth global $\scri$; we prove
existence of initial data for many black holes which are exactly
Kerr -- or exactly Schwarzschild -- both near infinity and near
each of the connected components of the apparent horizon; under
appropriate conditions we obtain existence of vacuum extensions of
vacuum initial data across compact boundaries; we show that for
generic metrics the deformations in the  Isenberg-Mazzeo-Pollack
gluings can be localised, so that the initial data on the
connected sum manifold coincide with the original ones except for
a small neighborhood of the gluing region;  we prove existence of
asymptotically flat solutions which are static or stationary up to
$r^{-m}$ terms, for any fixed $m$, and with multipole moments
freely prescribable within certain ranges.
\end{abstract}

\newpage
\tableofcontents





\section{Introduction}\label{Sintro}

 In a recent
significant paper~\cite{Corvino} Corvino has presented a new
gluing construction of scalar flat metrics, leading to  the
striking result of existence of non-trivial scalar flat metrics
which are exactly Schwarzschildian at large distances; see
also~\cite{CorvinoSchoen}. Extensions of the results
in~\cite{Corvino} have been announced
in~\cite{CorvinoOberwolfach}, and those results should be
available\footnote{This paper has been written
after~\cite{Corvino,CorvinoOberwolfach}, but independently
of~\cite{CorvinoSchoenprep}.} in a near
future~\cite{CorvinoSchoenprep}. A reading of the proofs
in~\cite{Corvino} reveals that the arguments there can be
simplified or streamlined using known techniques for PDE's in
weighted Sobolev spaces ({\em cf.,
e.g.}\/~\cite{choquet-bruhat:christodoulou:elliptic,Bartnik:mass,AndElli,AndChDiss,GL,Lee:fredholm}).
Further, the methods introduced by Corvino and Schoen can be
applied in other contexts to obtain new classes of solutions of
the general relativistic constraint equations. The object of this
paper is to present an abstract version, in a large class of
weighted Sobolev spaces, of the arguments of Corvino and Schoen.
Specific results on compact manifolds with boundary (as considered
by Corvino), or on asymptotically flat manifolds, or on
asymptotically hyperboloidal manifolds, can then be obtained  by
an appropriate choice of the weight functions.
 More precisely, we develop a general theory of mapping properties
 of the solutions of the linearised constraint operator in a class
 of weighted Sobolev spaces, assuming certain inequalities. The
 class of weighted Sobolev spaces includes those of Christodoulou ---
 Choquet-Bruhat~\cite{choquet-bruhat:christodoulou:elliptic}, appropriate in the asymptotically
 Euclidean context, as well as an exponentially weighted version
 thereof, and distance--weighted spaces near a boundary, or an
 exponentially weighted version thereof; the latter two classes are relevant near a compact boundary,
 or in an asymptotically hyperboloidal context. We establish the required
 inequalities in all the spaces just mentioned. An appropriate version of the
 inverse function  theorem allows one to produce new classes of
 solutions of interest. One application is that of existence of
 space-times which are Kerrian near spatial infinity; this has
 already been observed in~\cite{CorvinoOberwolfach}. We apply our
 techniques
 to produce
 two further large classes of initial data sets with controlled asymptotic behavior at spatial infinity.
The first class is obtained by gluing any asymptotically flat
initial data with data in the exterior region which are exactly
stationary there. This leads to a large class of space-times which
are exactly stationary away from the domain of influence of a
compact set. The second class consists of initial data which are
approximately stationary in the asymptotic region, with the
non-stationary part decaying at a prescribed (as high as desired)
order in terms of powers of $r$. On the other hand the stationary
part is controlled by a set of multipole moments which are freely
prescribable within certain ranges. Such initial data are relevant
to the program of~\cite{Friedrich:Pune,friedrich:static:1988}.
 Yet another
 application is an extension result for initial data near the
 Minkowskian ones, which leads to asymptotically simple
 space-times, or to new ``many black hole" space-times. Our final
 application here
 is a gluing construction for generic CMC initial data sets, in which the perturbation
 of the metric is localised in a small neighborhood of the points where the gluing
 is performed. This makes use of, and refines, the recent gluing construction
 of Isenberg, Mazzeo and Pollack~\cite{IMP,IMP2}.
 Some
 further applications, involving local extensions near positively
 or negatively curved space forms, or concerning the construction
 of initial data with controlled Bondi functions, will be
 discussed elsewhere.

 We note that all the results in Section~\ref{Sisoth} are  valid
 when $M$ is a compact manifold without boundary by setting all the
 weight functions  to one, $\varphi=\phi=\psi=1$, and by taking the compact
 set $\mcK$ appearing in Proposition~\ref{P:estproj} and elsewhere
 equal to $M$.

\section{The constraints map}

The aim of this section is to establish some
algebraic-differential properties of the constraints map, and some
elementary properties of the associated differential operators in
a class of weighted Sobolev spaces. The reader is referred to
\ApSwSw{} for the definition of the latter.

 Initial data $(g,K)$
for the vacuum Einstein equations belong to the zero level set of
the {\em constraints map}: \be \label{1} \left(
\begin{array}{c}
J\\
  \\
\rho
\end{array}
\right)
(K,g):=
\left(
\begin{array}{l}
2(-\nabla^jK_{ij}+\nabla_i\;\tr  K)\\
  \\
R(g)-|K|^2 + (\tr  K)^2
\end{array}
\right)
=
\left(
\begin{array}{l}
0\\
  \\
0
\end{array}
\right)\;. \ee  These are the general relativistic constraint
equations whatever the space-dimension $n$. As Equations~\eq{1}
are trivial in space-dimension zero and one, in the remainder of
this paper we shall assume that $n\ge 2$.

Let $h=\delta g$ and $Q=\delta K$, the linearisation of the
constraints map at $(K,g)$ reads \be \label{2} P(Q,h)= \left(
\begin{array}{l}
-K^{pq}\nabla_i h_{pq}+K^q{}_i(2\nabla^j h_{qj}-\nabla_q h^l{}_l)\\
\;\;\;\;\;\;\;-2\nabla^jQ_{ij}+2\nabla_i\;\tr  Q
-2(\nabla_iK^{pq}-\nabla^qK^p{}_i)h_{pq}\\
  \\
-\Delta(\tr  h)+\divr    \divr    h-\langle h,\Ricc(g)\rangle +2K^{pl}K^q{}_l h_{pq}\\
\;\;\;\;\;\;\;-2\langle K,Q\rangle +2\tr  K(-\langle h,K\rangle
+\tr Q)
\end{array}
\right)\;.
\ee
\begin{remark}
We note that for any real numbers $a$ and  $b$ it holds
\be\label{1.1} P(aK,bg)= \left(
\begin{array}{c}
(a-b)J(K,g)\\
  \\
-bR(g)+2(b-a)[|K|^2-(\tr K)^2]
\end{array}
\right)\;.
\ee
\end{remark}
The order of the differential operators that appear in  $P$ is
$$
\left(
\begin{array}{cc}
1&1\\
0&2\\
\end{array}
\right)
$$
which can be written in the Agmon-Douglis-Nirenberg form
({\em cf., e.g.}\/~\cite[p.~210]{Morrey})
$$
\left(
\begin{array}{cc}
s_1+t_1&s_1+t_2\\
s_2+t_1&s_2+t_2\\
\end{array}
\right)\;,
$$
with $s_1=-1$, $s_2=0$, $t_1=t_2=2$; here it is understood that an operator of order $0$ is
also an operator of order $2$ with vanishing coefficients in front of the first and second derivatives.
It follows that the symbol $P'$ of the principal part of $P$ in the sense of Agmon-Douglis-Nirenberg
reads
$$
P'(x,\xi)(Q,h)=
\left(
\begin{array}{cc}
2(-\xi^s\delta^t_i +\xi_ig^{st})&
-K^{pq}\xi_i+2K^q{}_i\xi^p-K^l{}_i\xi_lg^{pq}\\
0 & -|\xi|^2g^{pq}+\xi^p\xi^q\\
\end{array}
\right)
\left(
\begin{array}{c}
Q_{st}\\
h_{pq}
\end{array}
\right)\;,
$$
while the formal $L^2$-adjoint of $P$ takes the form
\be
\label{4}
P^*(Y,N)=
\left(
\begin{array}{l}
2(\nabla_{(i}Y_{j)}-\nabla^lY_l g_{ij}-K_{ij}N+\tr K\; N g_{ij})\\
 \\
\nabla^lY_l K_{ij}-2K^l{}_{(i}\nabla_{j)}Y_l+
K^q{}_l\nabla_qY^lg_{ij}-\Delta N g_{ij}+\nabla_i\nabla_j N\\
\; +(\nabla^{p}K_{lp}g_{ij}-\nabla_lK_{ij})Y^l-N \Ricc(g)_{ij}
+2NK^l{}_iK_{jl}-2N (\tr \;K) K_{ij}
\end{array}
\right) \;.\ee {}From this we obtain the Agmon-Douglis-Nirenberg
symbol ${P^{*}}'$ of  the principal part of $P^*$, \be\label{5}
{P^{*}}'(x,\xi)(Y,N)= \left(
\begin{array}{cc}
2(\xi_{(i}\delta^l_{j)}-\xi^lg_{ij}) & 0\\
K_{ij}\xi^l-2K^l{}_{(i}\xi_{j)}+K^{pl}\xi_qg_{ij}&
\xi_i\xi_j-|\xi|^2g_{ij}
\end{array}\right)
\left(
\begin{array}{c}
Y_l\\
N
\end{array}
\right)
\;.\ee
\begin{remark}\label{R1intro}
Recall that the formal adjoint $P^*$ is defined by the requirement
that for all smooth $(Q,h)$'s and for all compactly supported
smooth $(Y,N)$'s we have
$$ \langle P^*(Y,N),(Q,h)\rangle_{L^2(g)\oplus L^2(g)} = \langle (Y,N), P(Q,h)\rangle_{L^2(g)\oplus L^2(g)}
\;.
$$
It is easily seen by continuity and density arguments that this
equation still holds for\footnote{\label{fspaces}See
Appendix~\ref{SwSs} for the definitions of the function spaces we
use.} all $(Q,h)\in H^1_\loc\oplus H^2_\loc$ and for all $(Y,N)\in
\zH^1_{\phi,\psi}\times \zH^2_{\phi,\psi}$.
\end{remark}

We wish to check ellipticity of $PP^*$, for this we need the
following:
\begin{lem}\label{L1.1} Suppose that $\dim M\ge 2$, then ${P^*}'(x,\xi)$ is injective
for $\xi\ne 0$.
\end{lem}
\begin{proof}
We define a linear map $\alpha$ from the space $S_2$ of two-covariant symmetric tensors
into itself by the formula
\be\label{alphadef}
\alpha(S)=S-(\tr S)g\;.
\ee
Let $\xi\neq 0$, if $(Y,N)$ is in the kernel of  ${P^*}'(x,\xi)$
then
$$
\alpha(\xi_{(i}Y_{j)})=0\;,
$$
so that $\xi_{(i}Y_{j)}=0$, and $Y=0$.
It follows that
$$
\alpha(\xi_{i}\xi_{j})N=0\;,
$$
which implies $N=0$.
\end{proof}
\qed

The lemma implies:
\begin{cor}\label{C1.1}
The operator $L:=PP^*$ is elliptic in the sense of Agmon-Douglis-Nirenberg
({\em cf., e.g.}\/~\cite[Definition~6.1.1, p.~210]{Morrey}).
\end{cor}
\proof The differential order of the various entries of $L$ is
$$
\left(
\begin{array}{cc}
2&3\\
3&4\\
\end{array}
\right)
=
\left(
\begin{array}{cc}
s_1+t_1&s_1+t_2\\
s_2+t_1&s_2+t_2\\
\end{array}
\right)\;,
$$
with $s_1=-1$, $s_2=0$, $t_1=3$, $t_2=4$.  Now, $P'(x,\xi)$ is of
the form
$$
E:=\left(
\begin{array}{cc}
A&B\\
0&D\\
\end{array}
\right)\;,
$$
while ${P^*}'(x,\xi)$ can be written as
$$
\left(
\begin{array}{cc}
-^tA&0\\
-^tB& ^tD\\
\end{array}
\right)\;,
$$
where $^t X$ denotes the transpose of $X$. Let $\xi\ne 0$; by Lemma \ref{L1.1}
$^tA$ and $^tD$ are injective (hence $A$ and $D$ are surjective), which implies that
 $^tE$ is injective (hence $E$ is surjective). This shows that
$E\;^tE$ is bijective: indeed, $E\;^tEX=0$ implies $^tXE\;^tEX=0$,
which is the same as $|^tEX|^2=0$, hence $X=0$. It is
straightforward to check that the  Agmon-Douglis-Nirenberg symbol
of $PP^*$, defined as the symbol built from those terms which are
precisely of order $s_i+t_j$, equals
$$P'(x,\xi){P^*}'(x,\xi)=E\;^tE
\left(
\begin{array}{cc}
-I&0\\
0&1\\
\end{array}
\right)\;,
$$
and its bijectivity for $\xi\ne 0$ follows. This is precisely the
ellipticity condition  of Agmon, Douglis, and Nirenberg, whence
the result. \qed

\medskip

We note the following simple fact:$^{\mbox{\scriptsize
\ref{fspaces}}}$

\begin{lem} \label{L4.3} Let $k\in\Z$, $k\ge -2$. Suppose that\footnote{The local
differentiability conditions follow from the requirement that the
$k+$ fourth covariant derivatives of $N$ and the $k+$ third ones
of $Y$ can be defined in a distributional sense; both of those
conditions are fulfilled by a metric $g\in W^{k+3,\infty}_\loc$
--- the reader should note that the first covariant derivatives of
$N$ do not involve the Christoffel symbols of $g$ since $N$ is a
function.}  $g\in W^{k+3,\infty}_\loc$ and that  \be
\label{Richyp} \Ricc(g)\in\phi^{-2}W^{k+2,\infty}_{\phi}, \ee \be
\label{Khyp} K\in W^{k+3,\infty}_{\phi}\cap
\phi^{-2}W^{k+2,\infty}_{\phi}\;. \ee If \eq{lcond} holds with
$0\le i\le k+2$, then the linear operators
$$
P^*: \phi \zH^{k+3}_{\phi,\psi}\times \phi^2
\zH^{k+4}_{\phi,\psi}\longrightarrow
 \zH^{k+2}_{\phi,\psi}\times  \zH^{k+2}_{\phi,\psi},\quad k\ge -2\;,
$$
$$
P:\psi^2(\zH^{k+2}_{\phi,\psi}\times  \zH^{k+2}_{\phi,\psi})
\longrightarrow
\psi^2(\phi^{-1}\zH^{k+1}_{\phi,\psi}\times\phi^{-2}
\zH^{k}_{\phi,\psi}), \quad k\ge 0\;,
$$
are well defined, and bounded.
\end{lem}

\proof The result follows immediately from \eq{mpp}; we simply mention the
inequality
$$ |\phi K| \le \frac 12 \left( \phi^2 |K| + |K|\right)\;,
$$
which shows that under \eq{Khyp} we have $K\in
\phi^{-1}W^{k+2,\infty}_{\phi}$;
this is used to control the $K^2$ terms in $P^*$ and in $P$.
\qed

We define  a map $\Phi$  by \be\label{DefPhi} \Phi(x,y):=(\phi
x,\phi^2 y)\;.\ee As before, we have the
\begin{lem} \label{L4.3bis} Let $k\in\Z$.
Suppose that $g\in W^{k+3,\infty}_\loc$ and that  \be
\label{Richypbis} \Ricc(g)\in\phi^{-2}W^{k+2,\infty}_{\phi}, \ee
\be \label{Khypbis} K\in \phi^{-1}W^{k+3,\infty}_{\phi}\;. \ee If
\eq{lcond} holds with $0\le i\le k+2$, then the linear operators
$$
\Phi P^*: \zH^{k+3}_{\phi,\psi}\times
\zH^{k+4}_{\phi,\psi}\longrightarrow
 \zH^{k+2}_{\phi,\psi}\times  \zH^{k+2}_{\phi,\psi},\quad k\ge -2\;,
$$
$$
\psi^{-2}P\Phi\psi^2:\zH^{k+2}_{\phi,\psi}\times
\zH^{k+2}_{\phi,\psi} \longrightarrow \zH^{k+1}_{\phi,\psi}\times
\zH^{k}_{\phi,\psi}, \quad k\ge 0\;,
$$
are well defined, and bounded.
\end{lem}

\medskip

Let us establish now some estimates satisfied by $P^*$:

\begin{lem}\label{L5}
Suppose that $g\in W^{1,\infty}_\loc$, that  \eq{lcond} holds with
$0\le i\le 2$, and that
 \be\label{Riccond} \Ricc(g) \in \phi^{-2}L^\infty 
 \;, \ee
\be\label{Kcond} K\in W^{1,\infty}_{\phi}\cap\phi^{-2} L^\infty
\;. \ee Then for any $C^1$ vector field $Y$ and $C^2$ function
$N$, both compactly supported on $\bM$, we have \be\label{eL5}
C\left(\|P^*(\phi Y,\phi^2 N)\|_{\Lpsi}+\|Y\|_{\Lpsi}
+\|N\|_{\Lpsione}+|b(\phi\psi Y)|^{1/2}\right) \geq
\|Y\|_{\Lpsione}+\|N\|_{\Lpsitwo}, \ee where
$$
b(Y)=\int_{\partial
  M}(\nabla_iY_jY^i-\nabla_iY^iY_j)\nu^j\;.
$$
\end{lem}
\begin{proof} Throughout this work the letter $C$ denotes a constant which
might change from term to term and line to line. The leading order
terms in $P^*$ are of the form
\begin{eqnarray}\nn
P^*(\phi Y, \phi^2 N)-\mbox{sub-leading terms} &=:& \left(
\begin{array}{l}
\notreP^*_1(\phi Y,\phi^2 N)\\
\notreP^*_2(\phi Y,\phi^2 N)\\
\end{array}
\right)\\
&=:& \left(
\begin{array}{l}
2\alpha(\nabla_{(i}(\phi Y_{j)})\\
\beta(\nabla_{i}(\phi Y_j))+\alpha(\nabla_{i}\nabla_{j}(\phi ^2N))\\
\end{array}
\right)\;, \nn \\ &&\label{P12}
\end{eqnarray}
and this defines the $\notreP^*_1$, $\notreP^*_2$ and $\beta$
operations (recall that $\alpha$ has been defined in
\eq{alphadef}). Invertibility of $\alpha$ shows that
$$
\|2\alpha(\nabla_{(i} Y_{j)})\|_{L^2}\geq C \|\nabla_{(i}
Y_{j)}\|_{L^2}.
$$
We have
$$ \int_M\nabla_{(i}Y_{j)}\nabla^{(i}Y^{j)}
=\frac{1}{2}(\int_M\nabla_{i}Y_{j}\nabla^{i}Y^{j}+
\int_M\nabla_{i}Y_{j}\nabla^{j}Y^{i})\;,
$$
and Stokes' theorem gives
$$ \int_M\nabla_{i}Y_{j}\nabla^{j}Y^{i}=
-\int_M(\nabla^{j}\nabla_{i}Y_{j})Y^{i}+\int_{\partial M}
(\nabla_{i}Y_{j})Y^{i}\nu^j\;,
$$
supposing for the moment that $Y$ is $C^2$. Using
$\nabla^{j}\nabla_{i}Y_{j}=\nabla_{i}\nabla^{j}Y_{j}+(\Ricc
(Y,\cdot))_i$, and integrating again by parts,
$$
-\int_M\nabla_{i}\nabla^{j}Y_{j}Y^{i}
=\int_M\nabla^{j}Y_{j}\nabla_{i}Y^{i}-\int_{\partial M}
\nabla^{j}Y_{j}Y^{i}\nu_i\;,
$$
one is led to
\begin{eqnarray*}
\int_M\nabla_{(i}Y_{j)}\nabla^{(i}Y^{j)} &=&\frac{1}{2}( \int_M
|\nabla Y|^2+ (\divr   Y)^2 -\Ricc (Y,Y) \\
& &\;\;
+\int_{\partial M}[(\nabla_iY_j)Y^i-(\nabla_iY^i)Y_j]\nu^j).
\end{eqnarray*}
We have thus showed that for $C^2$ compactly supported vector
fields we have \be\label{fi} |b(Y)|+\|2\alpha(\nabla_{(i}
Y_{j)})\|_{L^2}+\|\Ricc (Y,Y)\|_{L^1}\geq C \|\nabla Y\|_{L^2}\;,
\ee
 and it should be clear that this remains true for
vector fields which are only differentiable once. To continue, we
use \eq{fi} with $Y$ replaced with $\phi \psi Y$; the hypothesis
that $\Riccg\in \phi^{-2}L^{\infty}_\phi$ allows us to write
$$
|b(\phi\psi Y)|+\|2\alpha(\nabla_{(i}(\phi\psi Y_{j)}))\|_{L^2}
+\|\psi Y\|_{L^2}\geq c \|\nabla(\phi\psi Y)\|_{L^2}.
$$
We have
\begin{eqnarray}\nn \|2\alpha(\nabla_{(i}(\phi
Y_{j)}))\|_{\Lpsi}
&= & \|2\alpha(\psi\nabla_{(i}(\phi Y_{j)}))\|_{L^2}\\
\nn &= & \|2\alpha(\nabla_{(i}(\psi\phi Y_{j)}))
-2\alpha((\nabla_{(i}\psi)\phi Y_{j)}))\|_{L^2}\\
\nn&\geq &\|2\alpha(\nabla_{(i}(\phi\psi Y_{j)}))\|_{L^2}
-C\|(\nabla_{(i}\psi)\phi Y_{j)}\|_{L^2}\\
\nn&\geq&C\|(\nabla(\phi\psi Y)\|_{L^2}-C |b(\phi\psi Y)|-C\|\psi Y\|_{L^2}\\
\nn& &\;-C\|(\nabla_{(i}\psi)\phi Y_{j)}\|_{L^2}\\
\nn&\geq&
C\|(\nabla(\phi\psi) Y+\phi\psi\nabla Y\|_{L^2}-C |b(\phi\psi Y)|\\
\nn& &\;-C\|\psi Y\|_{L^2}
-C\|(\nabla_{(i}\psi)\phi Y_{j)}\|_{L^2}\\
\nn&\geq&
C\|\phi\psi\nabla Y\|_{L^2}-C\|\nabla(\phi\psi) Y\|_{L^2}-C |b(\phi\psi Y)|\\
& &\;-C\|\psi Y\|_{L^2} -C\|(\nabla_{(i}\psi)\phi
Y_{j)}\|_{L^2}\;,\label{Ymin}
\end{eqnarray}
which finally gives \bel{fi.1} |b(\phi\psi Y)|+\|Y\|_{\Lpsi}+
\|2\alpha(\nabla_{(i}(\phi Y_{j)}))\|_{\Lpsi} \geq
C\|\phi\psi\nabla Y\|_{L^2}\;. \ee Invertibility of $\alpha$ leads
us to
$$
\begin{array}{ll}
\|\alpha(\nabla\nabla(\phi ^2N))\|_{\Lpsi}
&\geq C\|\nabla\nabla(\phi ^2N)\|_{\Lpsi}\\
&\geq C\|\phi ^2 \nabla\nabla N\|_{\Lpsi}
-2C \|\nabla(\phi^2) \nabla N)\|_{\Lpsi}\\
&\;-C \| \nabla\nabla (\phi ^2)N\|_{\Lpsi}\;,
\end{array}
$$
so that \be\label{Nmin} \|\alpha(\nabla\nabla(\phi
^2N))\|_{\Lpsi}+ \|N\|_{\Lpsione}\geq C\|\phi ^2 \nabla\nabla
N\|_{\Lpsi}\;.\ee

Using the hypothesis that $K\in W^{0,\infty}_\phi$ we obtain
\begin{eqnarray*}
\|\alpha(\nabla_{i}\nabla_{j}(\phi ^2N))\|_{\Lpsi}
&=&\|\notreP^*_2(\phi Y,\phi^2 N)-\beta(\nabla_{i}(\phi Y_j))\|_{\Lpsi}\\
&\leq& \|\notreP^*_2(\phi
Y,\phi^2N)\|_{\Lpsi}+\|\beta(\nabla_{i}(\phi Y_j))
\|_{\Lpsi}\\
&\leq& \|\notreP^*_2(\phi Y,\phi^2N)\|_{\Lpsi}+C\|\nabla_{i}(\phi Y_j)\|_{\Lpsi}\\
&\leq& \|\notreP^*_2(\phi Y,\phi^2N)\|_{\Lpsi}+C\|\notreP^*_1(\phi
Y,\phi^2
N)\|_{\Lpsi}\\
& &\;+ C\|Y\|_{\Lpsi} +C|b(\phi\psi Y)|\;,
\end{eqnarray*}
and in the last step we have used \eq{fi.1}. The lower order terms
are controlled using the hypotheses
 $K\in W^{1,\infty}_\phi\cap \phi^{-2} W^{0,\infty}_\phi$
and $\Ricc(g)\in\phi^{-2} W^{0,\infty}_\phi$ (compare the proof of
Lemma~\ref{L4.3}), leading to \eq{eL5}.
$$
C(\|P^*(\phi Y,\phi^2 N)\|_{\Lpsi}+\|Y\|_{\Lpsi}
+\|N\|_{\Lpsione}+|b(\phi\psi Y))| \geq \|\phi\nabla Y\|_{\Lpsi}+
\|\phi ^2 \nabla\nabla N\|_{\Lpsi}\;.
$$
\end{proof}
\qed

We have  the following equivalent of Lemma~\ref{L5} for the map
considered in Lemma~\ref{L4.3bis}:
\begin{lem}\label{L5bis}
Suppose that $g\in W^{1,\infty}_\loc$, that  \eq{lcond} holds with
$0\le i\le 2$, and that
 \be\label{Riccondbis} \Ricc(g) \in \phi^{-2}L^\infty
 \;, \ee
\be\label{Kcondbis} K\in\phi^{-1} W^{1,\infty}_{\phi} \;. \ee Then
for any $C^1$ vector field $Y$ and $C^2$ function $N$, both
compactly supported on $\bM$, we have \be\label{eL5bis}
C\left(\|\Phi P^*( Y, N)\|_{\Lpsi}+\|Y\|_{\Lpsi}
+\|N\|_{\Lpsione}+|b(\phi\psi Y)|^{1/2}\right) \geq
\|Y\|_{\Lpsione}+\|N\|_{\Lpsitwo}, \ee where
$$
b(Y)=\int_{\partial
  M}(\nabla_iY_jY^i-\nabla_iY^iY_j)\nu^j\;.
$$
\end{lem}
\begin{proof}
The proof is essentially identical with that of Lemma~\ref{L5},
with the inequality \eq{Ymin} replaced by
\begin{eqnarray*}
 \|2\phi\alpha(\nabla_{(i}
Y_{j)})\|_{\Lpsi} &\geq&
C\|\phi\psi\nabla Y\|_{L^2}-C\|\nabla(\phi\psi) Y\|_{L^2}-C |b(\phi\psi Y)|\\
& &\;-C\|\psi Y\|_{L^2} -C\|Y_{(i}\nabla_{j)}(\psi\phi)
\|_{L^2}\;,
\end{eqnarray*}
and inequality \eq{Nmin} replaced by
$$\|\phi^2\alpha(\nabla\nabla  N)\|_{\Lpsi}\geq
C\|\phi ^2 \nabla\nabla N\|_{\Lpsi}\;.$$
\end{proof}
\qed

\section{Isomorphism theorems}\label{Sisoth}
In this section, we assume that we have a solution $(K_0,g_0)$ to
the constraint map, with possibly a non-trivial kernel for the
associated operator $P^*_0$, defined as $P^*$ with $(K,g)$
replaced by $(K_0,g_0)$. We present here a general abstract method
to construct ``solutions-up-to-kernel" to the constraint equations
which are close to $(K_0,g_0)$; our argument is a straightforward
generalisation of \cite{Corvino}. (In particular if the kernel is
trivial we obtain solutions.)

\begin{Proposition}\label{P:estproj}
Under the hypotheses of Lemma~\ref{L5} with $(K,g)=(K_0,g_0)$, let
${\maclKz   }$ be the kernel of
$$P_0^*\Phi:
\Lpsikg{1}{g_0}\times \Lpsikg{2}{g_0}\longrightarrow
{\Lpsi}(g_0)\times {\Lpsi}(g_0),$$ and let ${\maclKz
}^{\bot_{g_0}}$ be its $\Lpsi(g_0)\oplus\Lpsi(g_0)$-orthogonal.
Assume there exists a compact set $\mcK\subset M$ such that for
all $\Lpsione(g_0)$ vector fields $Y$ and $\Lpsitwo(g_0)$
functions $N$, both supported in $M\setminus \mcK$ we have \be
\label{it1ker} C\|P_0^*\Phi(Y, N)\|_{\Lpsi(g_0)} \geq
\|Y\|_{\Lpsi(g_0)}+\|N\|_{\Lpsione(g_0)}\;. \ee Then there exists
a constant $C'$ such that for all $(K,g)$ close to $(K_0,g_0)$ in
$(W^{1,\infty}_\phi(g_0)\cap \phi^{-2}L^\infty(g_0))\times
W^{2,\infty}_\phi(g_0)$ norm, and for all $(Y,N)\in \maclKzo \cap
(\Lpsikg{1}{g}\times  \Lpsikg{2}{g})$ it holds that \bel{broker}
C'\|P^*\Phi( Y, N)\|_{\Lpsig} \geq
\|Y\|_{\Lpsioneg}+\|N\|_{\Lpsitwog }\;. \ee
\end{Proposition}
\begin{remark}\label{Rem1}
The conclusion still
holds if \eq{it1ker} is replaced by \bel{it1alt} C\Big(\|P^*(\phi
Y,\phi^2 N)\|_{\Lpsi}+ \|(\phi Y,\phi^2 N)\|_{X}\Big)\geq
\|Y\|_{\Lpsi}+\|N\|_{\Lpsione}\;, \ee where $X$ is a normed space
such that we have a compact inclusion $\phi\Lpsione\times
\phi^2\Lpsitwo\subset X$; however, \eq{it1ker} is sufficient for
our purposes.
\end{remark}
\begin{proof} For $(K,g)=(K_0,g_0)$, this is proved by a standard argument,
compare~\cite{choquet-bruhat:christodoulou:elliptic,AndElli}:
assuming that the inequality fails,  there is a sequence
$(Y_n,N_n)\in (\zH^1_{\phi,\psi}(g_0)\times
\zH^2_{\phi,\psi}(g_0))\cap {\maclKz   }^{\bot_{g_0}} $ with norm
$1$ such that $\|P_0^*\Phi( Y_n,N_n)\|_{\Lpsi(g_0)}$ approaches
zero as $n$ tends to infinity. One obtains a contradiction with
injectivity on $(\zH^1_{\phi,\psi}(g_0)\times
\zH^2_{\phi,\psi}(g_0))\cap {\maclKz   }^{\bot_{g_0}} $ by using
the Rellich-Kondrakov compactness on a conditionally compact open
set ${\mycal O}\supset \mcK$, applying \eq{eL5} with $b(\phi\psi
Y)=0$, and \eq{it1ker}, to $Y$ and $N$ multiplied by suitable
cut-off functions; we simply note that \eq{eL5} holds without the
boundary term for smooth compactly supported
fields\footnote{\label{footmanif}We use the analysts' convention
that a manifold $M$ is always open; thus a manifold $M$ with
non-empty boundary $\partial M$ does not contain its boundary;
instead, $\bM:= M \cup
\partial M$ is a manifold with boundary in the differential
geometric sense. Unless explicitly specified otherwise \emph{no}
conditions on $M$ are made  --- \emph{e.g.}\/ that $\partial M$,
if non-empty, is compact
--- except that $M$ is a smooth manifold; similarly no conditions
\emph{e.g.}\/ on completeness of $(M,g)$, or on its radius of
injectivity, are made.} , hence on ${\maclKz }^{\bot_{g_0}}\cap(
\zH^1_{\phi,\psi}(g_0)\times \zH^2_{\phi,\psi}(g_0))$ by density.
Increasing $C'$ if necessary, the inequality at $(K_0,g_0)$
together with straightforward algebra shows that the inequality
remains true for $(K,g)$ close to $(K_0,g_0)$.
\end{proof}
\qed

Similarly one obtains:
\begin{Proposition}\label{P:estprojbis}
Under the hypotheses of Lemma~\ref{L5bis} with $(K,g)=(K_0,g_0)$,
let ${\maclKz   }$ be kernel of
$$\Phi P_0^*:
\Lpsikg{1}{g_0}\times \Lpsikg{2}{g_0}\longrightarrow
{\Lpsi}(g_0)\times {\Lpsi}(g_0),$$ and let ${\maclKz
}^{\bot_{g_0}}$ be its $\Lpsi(g_0)\oplus\Lpsi(g_0)$-orthogonal.
Assume there exists a compact set $\mcK\subset M$ such that for
all $\Lpsione(g_0)$ vector fields $Y$ and $\Lpsitwo(g_0)$
functions $N$, both supported in $M\setminus \mcK$ we have \be
\label{it1kerbis} C\|\Phi P_0^*(Y, N)\|_{\Lpsi(g_0)} \geq
\|Y\|_{\Lpsi(g_0)}+\|N\|_{\Lpsione(g_0)}\;. \ee Then there exists
a constant $C'$ such that for all $(K,g)$ close to $(K_0,g_0)$ in
$\phi^{-1}W^{1,\infty}_\phi(g_0)\times W^{2,\infty}_\phi(g_0)$
norm, and for all $(Y,N)\in \maclKzo \cap (\Lpsikg{1}{g}\times
\Lpsikg{2}{g})$ it holds that \bel{brokerbis} C'\|\Phi P^*( Y,
N)\|_{\Lpsig} \geq \|Y\|_{\Lpsioneg}+\|N\|_{\Lpsitwog }\;. \ee
\end{Proposition}
Set
$$ {\mathcal L}_{\phi,\psi}:=
\Phi \psi^{-2} P\psi^2 P^* \Phi
\;.$$ We denote by $\pi_{\maclKzo }$ the $L^2_\psi(g)$ projection
onto $\maclKzo $. We are ready now to prove:

\begin{theor}\label{Til1proj} Let $k\ge 0$, $g_0\in
W^{k+4,\infty}_\loc$,
 suppose that \eq{lcond} holds with $0\le i\le 4+k$, and
that
$$ \Ricc(g_0)\in\phi^{-2}W^{k+2,\infty}_{\phi}(g_0)\;,
$$
$$
K_0\in W^{k+3,\infty}_{\phi}(g_0)\cap
\phi^{-2}W^{k+2,\infty}_{\phi}(g_0)\;.
$$
We further assume that the weights $\phi$ and $\psi$ have the
\emph{scaling property}, \emph{cf.\/} the end of \ApSwSw{} and
Appendix~\ref{Sscaling}. If there exists a compact set
$\mcK\subset M$ such that for all $\Lpsione(g_0)$ vector fields
$Y$ and $\Lpsitwo(g_0)$ functions $N$, both supported in
$M\setminus \mcK$, the inequality \eq{it1ker} holds, then for all
$(K,g)$ close to $(K_0,g_0)$ in $(W^{k+3,\infty}_\phi(g_0)\cap
\phi^{-2}W^{k+2,\infty}_\phi(g_0))\times W^{k+4,\infty}_\phi(g_0)$
norm, the map \bel{iso1} \pi_{\maclKzo } {\mathcal L}_{\phi,\psi}
:{\maclKzo }\cap (\Lpsikg{k+3}{g}\times \Lpsikg{k+4}{g})
\longrightarrow {\maclKzo }\cap (\Lpsikg{k+1}{g}\times
\Lpsikg{k}{g}) \ee is an  isomorphism such that the norm of its
inverse is bounded independently\footnote{The bound on the norm
might depend upon $(K_0,g_0)$.} of $(K,g)$.
\end{theor}
\begin{remark}
It is easily seen (see \Eq{orjus} below and Remark~\ref{R1intro})
that, in our context,  the image of ${\mathcal L}_{\phi,\psi}$ is
orthogonal to the kernel of $P^*\Phi$. We emphasise, however, that
the projection $\pi_{\maclKzo }$ in \eq{iso1} is on the orthogonal
to the kernel of $P_0^*\Phi$, and {\em not} on that of $P^*\Phi$.
\end{remark}

 \proof
For $(\J,\source )\in {\maclKzo }\cap({\Lpsi}(g)\times \Lpsig)$
let $\mathcal F$ be the following (continuous) functional defined
on ${\maclKzo }\cap (\Lpsikg{1}{g}\times \Lpsikg{2}{g})$:
$$
{\mathcal F}(Y,N):=\int_M (\frac{1}{2}|P^*\Phi( Y,N)|^2_g -\langle
(Y, N),(\J, \source )\rangle _g)\psi^2 d\mu_g\;;
$$
we set
$$
\mu_F=\inf_{(Y,N)\in {\maclKzo }\cap (\Lpsikg{1}{g}\times
\Lpsikg{2}{g})}{\mathcal F}(Y,N)\;.
$$
We claim that ${\mathcal F}$ is coercive: indeed,
Proposition~\ref{P:estproj} and the Schwarz inequality give
$$
\begin{array}{lll}
{\mathcal F}(Y,N)&\geq& C(\|Y\|_{\Lpsioneg}+\|N\|_{\Lpsitwog })^2-
\|(Y,N)\|_{\Lpsig}\|(\J,\source )\|_{\Lpsig}\\
&\geq & C(\|Y\|_{\Lpsioneg}+\|N\|_{\Lpsitwog })^2-
(\|Y\|_{\Lpsioneg}+\|N\|_{\Lpsitwog })\|(\J,\source
)\|_{\Lpsig}\;.
\end{array}
$$
Standard results on convex, proper, coercive, l.s.c.\ ({\em cf.,
e.g.,}~\cite[Proposition~1.2, p.~35]{EkelandTemam}) functionals
 show that $\mu_F$ is
achieved by some $(Y,N)\in {\maclKzo }\cap
(\zH^1_{\phi,\psi}(g)\times \zH^2_{\phi,\psi}(g))$ satisfying
\begin{eqnarray}
\nonumber \lefteqn {\forall \ (\delta Y,\delta N)\in
\zH^1_{\phi,\psi}(g)\times \zH^2_{\phi,\psi}(g)}
&&%
\\&& \nonumber
\int_M \left\langle P^*\Phi(Y,N),P^*\Phi(\delta Y,\delta N)\rangle
_g -\langle (\delta Y, \delta N),( \J, \source )\rangle
_g\right)\psi^2 d\mu_g=0.
\\ &&\label{p16}
\end{eqnarray}
It follows that  $(Y,N)\in {\maclKzo }\cap
(\zH^1_{\phi,\psi}(g)\times \zH^2_{\phi,\psi}(g))$ is a weak
solution of the equation
$$\Phi \psi^{-2} P\psi^2 P^*\Phi(Y,
N)=( \J,\source ).
$$
The variational equation \eq{p16} satisfies the hypotheses of
\cite[Section~6.4, pp.~242-243]{Morrey} with $s_j$, $t_k$ as in
Corollary~\ref{C1.1}, and with $m_1=1$, $m_2=2$, $h_0=-2$. By
elliptic regularity \cite[Theorem~6.4.3, p.~246]{Morrey} and by
standard scaling arguments ({\em cf.\/} the discussion at the end
of \ApSwSw{}) for $(\J,\source )\in \Lpsikg{k+1}{g}\times
\Lpsikg{k}{g}$, we have $(Y,N)\in \Lpsikg{k+3}{g}\times
\Lpsikg{k+4}{g}$, and surjectivity follows. To prove bijectivity,
we note that  the operator $\pi_{\maclKzo }{\mathcal
L}_{\phi,\psi}$ is injective: indeed, if $(Y,N)\in{\maclKzo }$ is
in the kernel of $\pi_{\maclKzo } {\mathcal L}_{\phi,\psi}$, then
(see Remark~\ref{R1intro}) \bel{orjus}0=\langle {\mathcal
L}_{\phi,\psi}(Y,N),(Y,N)\rangle _{\Lpsig\oplus\Lpsi(g)} =\langle
P^*\Phi(Y, N),P^*\Phi(Y,N)\rangle _{\Lpsig\oplus\Lpsi(g)}\;, \ee
so $(Y,N)=0$ from inequality \eq{broker}. \qed

There is yet another operator which is of interest in our context,
\bel{uprightL} {L}_{\phi,\psi}:=  \psi^{-2} P\Phi\psi^2 \Phi P^*
\;.\ee Similarly to Theorem~\ref{Til1proj}, using
Proposition~\ref{P:estprojbis} instead of \ref{P:estproj}, we
have:

\begin{theor}\label{Til1projbis} Let $k\ge 0$, $g_0\in W^{k+4,\infty}_\loc$, suppose that \eq{lcond}
holds with $0\le i\le 4+k$, that
$$ \Ricc(g_0)\in\phi^{-2}W^{k+2,\infty}_{\phi}(g_0)\;,
$$
$$
K_0\in \phi^{-1}W^{k+3,\infty}_{\phi}(g_0)\;,
$$
and that the weights $\phi$ and $\psi$ have the \emph{scaling
property}, cf.\ end of \ApSwSw{}. If there exists a compact set
$\mcK\subset M$ such that for all $\Lpsione(g_0)$ vector fields
$Y$ and $\Lpsitwo(g_0)$ functions $N$, both supported in
$M\setminus \mcK$, the inequality \eq{it1kerbis} holds, then for
all $(K,g)$ close to $(K_0,g_0)$ in
$\phi^{-1}W^{k+3,\infty}_\phi(g_0)\times W^{k+4,\infty}_\phi(g_0)$
norm, the map$$ \pi_{\maclKzo }{L}_{\phi,\psi} :{{\maclKz
}^{\bot_g}}\cap (\Lpsikg{k+3}{g}\times \Lpsikg{k+4}{g})
\longrightarrow {\maclKzo }\cap (\Lpsikg{k+1}{g}\times
\Lpsikg{k}{g})
$$
is an  isomorphism such that the norm of its inverse is bounded
independently of $(K,g)$.

 \qed
\end{theor}


Whenever the weighted Sobolev spaces are such that the constraints
map is defined and differentiable we obtain:

\begin{theor}\label{theor:proj}
Under the hypotheses of Theorem~\ref{Til1proj}, if the map
\be\label{et1proj}
\begin{array}{c}
 {\maclKzo }\cap (\Lpsikg{k+3}{g}\times  \Lpsikg{k+4}{g})
\longrightarrow
{\maclKzo }\cap (\Lpsikg{k+1}{g}\times \Lpsikg{k}{g})\\
(Y,N) \longmapsto \pi_{\maclKzo }\psi^{-2}\Phi\left\{\left(
\begin{array}{c}
J\\
\rho
\end{array}
\right)[(K,g)+\psi^2 P^*\Phi(Y,N)] -\left(
\begin{array}{c} J\\
\rho
\end{array}
\right)(K,g)\right\}
\end{array}
\ee is differentiable in a neighborhood  $\mcU_k$ of zero, then it
is bijective in a (perhaps smaller) neighborhood $\mcV_k$ of zero.
In particular there exists $\epsilon>0$ such that for all $(K,g)$
close to $(K_0,g_0)$ in $(W^{k+3,\infty}_\phi(g_0)\cap
\phi^{-2}W^{k+2,\infty}_\phi(g_0))\times
W^{k+4,\infty}_\phi(g_0)$, and for all pairs $(\delta J,
\delta\rho)\in  \psi^2\Phi^{-1}\Big(\Lpsikg{k+1}{g}\times
\Lpsikg{k}{g}\Big)$ with norm less than $\epsilon$,
 there exists a solution $(\delta
K,\delta g)=\psi^2 P^*\Phi(Y,N)
\in\psi^2(\zH^{k+2}_{\phi,\psi}(g)\times
\zH^{k+2}_{\phi,\psi}(g))$, close to zero, of the equation
\bel{et1proj1}\pi_{\maclKzo }\psi^{-2}\Phi \left\{\left(
\begin{array}{c}
J\\
\rho
\end{array}
\right) (K+\delta K,g+\delta g) - \left(
\begin{array}{c}
J\\
\rho
\end{array}
\right) (K,g) \right\}= \pi_{\maclKzo }\psi^{-2}\Phi\left(
\begin{array}{c}
\delta J\\
\delta \rho
\end{array}
\right) \ee
\end{theor}

\begin{remark}
The question of  differentiability of the map \eq{et1proj}, or
even of its existence, will depend upon the weight functions
$\phi$ and $\psi$, and requires a case-by-case treatment.
\end{remark}

\proof We apply  Proposition~\ref{prop:invloc} with $A$  a
neighborhood of $(K_0,g_0)$ in $(W^{k+3,\infty}_\phi(g_0)\cap
\phi^{-2} W^{k+2,\infty}_\phi(g_0)) \times
W^{k+4,\infty}_\phi(g_0)$, $x=(K,g)$, $\delta x=(\delta K, \delta
g)$, $V_x= \psi^2(\zH^{k+2}_{\phi,\psi}(g)\times
\zH^{k+2}_{\phi,\psi}(g))$, $W_x={\maclKzo }\cap
(\Lpsikg{k+1}{g}\times \Lpsikg{k}{g})$ and \bel{et1proja}
f_x(\delta x)= \pi_{\maclKzo }\psi^{-2}\Phi \left\{\left(
\begin{array}{c}
J\\
\rho
\end{array}
\right) (K+\delta K,g+\delta g) - \left(
\begin{array}{c}
J\\
\rho
\end{array}
\right) (K,g) \right\}\;. \ee \qed

We also have the following analogue of Theorem~\ref{theor:proj},
with an identical proof, based on Theorem~\ref{Til1projbis}:
\begin{theor}\label{theor:projbis}
Under the hypotheses of Theorem~\ref{Til1projbis}, if the map
\be\label{et1projbis}
\begin{array}{c}
 {\maclKzo }\cap (\Lpsikg{k+3}{g}\times  \Lpsikg{k+4}{g})
\longrightarrow
{\maclKzo }\cap (\Lpsikg{k+1}{g}\times \Lpsikg{k}{g})\\
(Y,N) \longmapsto \pi_{\maclKzo }\psi^{-2}\left\{\left(
\begin{array}{c}
J\\
\rho
\end{array}
\right)[(K,g)+\psi^2\Phi^2 P^*(Y,N)] -\left(
\begin{array}{c} J\\
\rho
\end{array}
\right)(K,g)\right\}
\end{array}
\ee is differentiable in a neighborhood of zero, then it is
bijective in a (perhaps smaller) neighborhood of zero. Thus, there
exists $\epsilon>0$ such that for all $(K,g)$ close to $(K_0,g_0)$
in $\phi^{-1}W^{k+3,\infty}_\phi(g_0)\times
W^{k+4,\infty}_\phi(g_0)$, and for all pairs $(\delta J,
\delta\rho)\in   \psi^2\Big(\Lpsikg{k+1}{g}\times
\Lpsikg{k}{g}\Big)$ with norm less than $\epsilon$,
 there exists a solution $(\delta
K,\delta g)=\Phi\psi^2\Phi P^*(Y,N) \in\psi^2(\phi
H^{k+2}_{\phi,\psi}(g)\times \phi^2 H^{k+2}_{\phi,\psi}(g))$,
close to zero, of the equation \bel{et1proj2} \pi_{\maclKzo
}\psi^{-2} \left\{\left(
\begin{array}{c}
J\\
\rho
\end{array}
\right) (K+\delta K,g+\delta g) - \left(
\begin{array}{c}
J\\
\rho
\end{array}
\right) (K,g) \right\}= \pi_{\maclKzo }\psi^{-2}\left(
\begin{array}{c}
\delta J\\
\delta \rho
\end{array}
\right) \;.\ee

\qed
\end{theor}

The last results allow us to construct solutions of the nonlinear
equation in weighted Sobolev spaces. The drawback of working in
such spaces is that the differentiability of the perturbative
solutions is considerably worse than that of the starting data
$(K_0,g_0)$, even when solutions with zero sources are considered.
In the usual analysis of nonlinear PDE's with implicit-function
techniques the higher regularity is obtained by bootstrap
arguments. In our set-up this does not work, because the
coefficients of the equations do not have enough regularity for
the bootstrap. It has been shown  by Corvino~\cite{Corvino} that
there exists a (non-standard) way of getting  a partial
improvement on the regularity of solutions. This carries over to
the general weighted spaces setting considered here provided some
further properties of the weights are assumed:

\begin{enumerate} \item First,  note that \eq{lcond} can be
rewritten as $\phi\in C^{\ell-1}_{\phi,\phi^{-1}}$, $\psi\in
C^{\ell-1}_{\phi,\psi^{-1}}$, $\varphi\in
C^{\ell-1}_{\phi,\varphi^{-1}}$. When dealing with H\"older spaces
one also needs to assume H\"older continuity of the derivatives of
the weights, so (renaming $\ell-1$ to $\ell$) we will assume:
\bel{threecB}\phi\in
C^{\ell,\alpha}_{\phi,\phi^{-1}}\;, \quad \psi\in
C^{\ell,\alpha}_{\phi,\psi^{-1}}\;, \quad \varphi\in
C^{\ell,\alpha}_{\phi,\varphi^{-1}}\;.\ee
\item As discussed in Appendix~\ref{Sscaling},  the following conditions
are useful for deriving the scaling property: Let us denote by
$B_p $ the open ball of centre $p$ with radius $\phi(p) /2$. We
assume that there exist  constants $C_1,C_2,C_3>0$ such that for
all $p\in M $ and all $y\in B_p $, we have
\beal{nscalprop1}
C_1^{-1}\phi(p) \leq \phi (y)\leq C_1\phi(p) \;,\\
\label{nscalprop2}
C_2^{-1}\varphi(p)\leq \varphi (y)\leq C_2\varphi(p)\;,\\
\label{nscalprop3} C_3^{-1}\psi(p)\leq \psi (y)\leq C_3\psi(p)\;.
\eea

\item Since the tool to handle non-linearities in this paper is the inverse function  theorem,
we need to make sure that the changes in the initial data are
small as compared to the data themselves. A necessary condition
for that is that the new metric be uniformly equivalent to the
original one. For example, in the setting of
Theorem~\ref{Til1projbis}, one way of ensuring this is
\bel{threedB}
\psi^2\phi^2C^{k,\alpha}_{\phi,\varphi}(g_0)\subset
C^{k,\alpha}_{\phi,1}(g_0) \;.\ee This will hold under the
following condition:
\begin{Proposition}
\label{Linclus} The
inequality\bel{threeaB}\psi^2\phi^2\varphi^{-1}\leq C\;.\ee
 implies \eq{threedB}.
\end{Proposition} In order to
check this the reader might wish to prove first that the
conditions imposed so far imply that
\begin{Lemma}\label{Lproduit}
If $u\in C^{k,\alpha}_{\phi,\varphi_1}(g)$ and $v\in
C^{k,\alpha}_{\phi,\varphi_2}(g)$, with one of the $\varphi_a$'s
satisfying
 \eq{nscalprop2} and $\phi$ satisfying \eq{threecB} with $\ell\ge k$, then $uv\in
C^{k,\alpha}_{\phi,\varphi_1\varphi_2}(g)$.
\end{Lemma}
Lemma~\ref{Lproduit} can be used to show an equivalent of
Lemma~\ref{L4.3} in weighted H\"older spaces.

\item  The last condition will be the contents of  Definition~\ref{def:regul} that
follows. We emphasise that all the conditions spelled out here
will be satisfied in all the applications we have in mind.

\begin{defi}\label{def:regul}
We will say that an operator $L$ from $\zH^{3}_{\phi,\psi}\times
\zH^{4}_{\phi,\psi}$ to $\zH^{1}_{\phi,\psi}\times
\zH^{0}_{\phi,\psi}$ satisfies the {\em weighted elliptic
regularity condition} if there exists a constant $C$ such that for
all $(Y,N)$ in $\zH^{3}_{\phi,\psi}\times \zH^{4}_{\phi,\psi}$
satisfying $L(Y,N)\in C^{k+1,\alpha}_{\phi,\varphi}\times
C^{k,\alpha}_{\phi,\varphi}$ we have $(Y,N)\in
C^{k+3,\alpha}_{\phi,\varphi}\times C^{k+4,\alpha}_{\phi,\varphi}$
with
\bel{erc}
\|(Y,N)\|_{C^{k+3,\alpha}_{\phi,\varphi}\times
C^{k+4,\alpha}_{\phi,\varphi}}\leq C \left(\|L(Y,N)\|_{
C^{k+1,\alpha}_{\phi,\varphi}\times C^{k,\alpha}_{\phi,\varphi}}
+\|(Y,N)\|_{H^{3}_{\phi,\psi}\times H^{4}_{\phi,\psi}}\right)\;.
\ee\end{defi}
\end{enumerate}

Armed with those conditions we can pass to an existence theorem in
weighted H\"older spaces:

\begin{prop}[Existence of solutions in weighted H\"older spaces, I]\label{regularite}
Let $k\in \N$, $0<\alpha<1$, assume that \eq{threecB} with $\ell
\ge k+4$ holds, and that \eq{nscalprop1}-\eq{nscalprop3} and
\eq{threeaB} hold, together with
\bel{cocobis}\psi^2\varphi^{-1}\le C\;.\ee In addition to the
hypotheses of Theorem~\ref{theor:proj}, suppose that $g_0\in
C^{k+4,\alpha}$, and that
$$ \Ricc(g_0)\in\phi^{-2}C^{k+2,\alpha}_{\phi,1}(g_0)\;,\quad
K_0\in C^{k+3,\alpha}_{\phi,1}(g_0)\cap
\phi^{-2}C^{k+2,\alpha}_{\phi,1}(g_0)\;.
$$
We further assume that the weights $\phi$, $\varphi$ and $\psi$
have the \emph{scaling property}, \emph{cf.\/} the end of
\ApSwSw{} and Appendix~\ref{Sscaling}. Suppose, next, that we have
the continuous inclusion
\bel{poidscond}\psi^2 C^{i,\alpha}_{\phi,\varphi^2}(g)\subset
\Lpsikg{i}{g} \ee
for $i=k,k+1$, with the inclusion norms uniformly bounded for $g$
close to $g_0$ in $C^{k+4,\alpha}_{\phi,1}(g_0)$. Assume finally
that ${ \cal L}_{\phi,\psi}(K,g)$ satisfies the weighted elliptic
regularity condition, with a uniform constant $C$ in \eq{erc} for
$(K,g)$ close to $(K_0,g_0)$ in $
\left(C^{k+3,\alpha}_{\phi,1}(g_0)\cap\phi^{-2}C^{k+2,\alpha}_{\phi,1}(g_0)\right)\times
C^{k+4,\alpha}_{\phi,1}(g_0)
$. If the source $(\delta J, \delta\rho)$ is in
$\psi^2\Phi^{-1}(\zH^{k+1}_{\phi,\psi}(g)\times
\zH^{k}_{\phi,\psi}(g))\cap
\psi^2\Phi^{-1}(C^{k+1,\alpha}_{\phi,\varphi}(g)\times
C^{k,\alpha}_{\phi,\varphi}(g))$, with sufficiently small norm,
then the solution obtained  in  Theorem \ref{theor:proj} is in
$$\psi^2(\zH^{k+2}_{\phi,\psi}(g)\times
\zH^{k+2}_{\phi,\psi}(g))\cap\psi^2(C^{k+2,\alpha}_{\phi,\varphi}(g)\times
C^{k+2,\alpha}_{\phi,\varphi}(g))\;.$$
\end{prop}

\proof We start with a lemma, which we leave as an exercise to the
reader (here Lemma~\ref{Lproduit} together with
Equations~\eq{threeaB} and \eq{cocobis} are useful):

\begin{Lemma}
\label{Ldiff} Under the conditions of
Proposition~\ref{regularite},
 the map \be\label{et1projreg}
\begin{array}{c}
C^{k+3,\alpha}_{\phi,\varphi}(g)\times
C^{k+4,\alpha}_{\phi,\varphi}(g) \longrightarrow
C^{k+1,\alpha}_{\phi,\varphi}(g)\times C^{k,\alpha}_{\phi,\varphi}(g)\\
(Y,N) \longmapsto \psi^{-2}\Phi\left\{\left(
\begin{array}{c}
J\\
\rho
\end{array}
\right)[(K,g)+\psi^2 P^*\Phi(Y,N)] -\left(
\begin{array}{c} J\\
\rho
\end{array}
\right)(K,g)\right\}
\end{array}
\ee is smooth in a neighborhood $\mcU_k$ of zero.
\end{Lemma}

Returning to the proof of Proposition~\ref{regularite}, we use the
notations of the  proof of Theorem~\ref{theor:proj} and we apply
Proposition~\ref{prop:regul} with $E_x=
\psi^2(C^{k+2,\alpha}_{\phi,\varphi}(g)\times
C^{k+2,\alpha}_{\phi,\varphi}(g))$, $F_x=
C^{k+1,\alpha}_{\phi,\varphi}(g)\times
C^{k,\alpha}_{\phi,\varphi}(g)$, $$G_x={\maclKzo
}\cap\left(\psi^2(C^{k+1,\alpha}_{\phi,\varphi^2}(g)\times
C^{k,\alpha}_{\phi,\varphi^2}(g))\right)\;,$$ and with $A$ ---  a
neighborhood of $(K_0,g_0)$ in $[W^{k+3,\infty}_\phi(g_0)\cap
\phi^{-2} W^{k+2,\infty}_\phi(g_0)\times
W^{k+4,\infty}_\phi(g_0)]\cap[C^{k+3,\alpha}_{\phi,1}(g_0)\cap
\phi^{-2} C^{k+2,\alpha}_{\phi,1}(g_0)\times
C^{k+4,\alpha}_{\phi,1}(g_0)]$. We have  continuous inclusions
$G_x \subset F_x$ and $G_x \subset W_x$ by \eq{poidscond}. The
condition 1.\ of Proposition~\ref{prop:regul} holds by the
hypothesis that ${\cal L}_{\phi,\psi}$ satisfies the weighted
elliptic regularity condition, and the form of the right inverse
used here. Condition 2.\ and 3.\ there hold because $J$ and $\rho$
are twice-differentiable (actually smooth) functions of their
arguments by Lemma~\ref{Ldiff}. \qed

\begin{remark} There is an intriguing mismatch
between the order of differentiability of the initial data set
$(K,g)$  at which the inverse function theorem is being applied,
and the order of differentiability of the final data $( K+\delta K
,g+\delta g)$. This seems unavoidable in our setup, and leads to
several unpleasant features such as dependence of the
neighborhoods on which we can solve the equations upon the degree
of differentiability, or failure to produce a Banach manifold
structure for the set of solutions, \emph{etc.} In a forthcoming
publication we will give a partial cure to this
problem~\cite{ChDelayHilbert}.
\end{remark}

We continue with H\"older continuous solutions in the setup of
Theorem~\ref{theor:projbis}:

\begin{prop}[Existence of solutions in weighted H\"older spaces, II]\label{regularitebis}
Let $k\in \N$, $0<\alpha<1$, assume that \eq{threecB} with $\ell
\ge k+4$ holds, and that \eq{nscalprop1}-\eq{nscalprop3} and
\eq{threeaB} hold. In addition to the hypotheses of
Theorem~\ref{theor:projbis}, suppose that $g_0\in C^{k+4,\alpha}$,
and that
$$ \Ricc(g_0)\in\phi^{-2}C^{k+2,\alpha}_{\phi,1}(g_0)\;,\quad
K_0\in \phi^{-1}C^{k+3,\alpha}_{\phi,1}(g_0)\;.
$$
We further assume that the weights $\phi$, $\varphi$ and $\psi$
have the \emph{scaling property}, \emph{cf.\/} the end of
\ApSwSw{} and Appendix~\ref{Sscaling}.  Suppose, next, that we
have the continuous inclusions
\bel{poidscond2}\psi^2 \phi^2 C^{i,\alpha}_{\phi,\varphi^2}(g)\subset
\Lpsikg{i}{g} \ee for $i=k,k+1$, with the inclusion norms
uniformly bounded for $g$ close to $g_0$ in
$C^{k+4,\alpha}_{\phi,1}(g_0)$. Assume finally that ${
L}_{\phi,\psi}(K,g)$ satisfies the weighted elliptic regularity
condition, with a uniform constant $C$ in \eq{erc} for $(K,g)$
close to $(K_0,g_0)$ in $
\phi^{-1}C^{k+3,\alpha}_{\phi,1}(g_0)\times
C^{k+4,\alpha}_{\phi,1}(g_0)
$. If the source $(\delta J, \delta\rho)$ is in
$\psi^2(\zH^{k+1}_{\phi,\psi}(g)\times \zH^{k}_{\phi,\psi}(g))\cap
\psi^2(C^{k+1,\alpha}_{\phi,\varphi}(g)\times
C^{k,\alpha}_{\phi,\varphi}(g))$, with sufficiently small norm,
then the solution obtained  in  Theorem \ref{theor:projbis} is in
$$\psi^2(\phi\zH^{k+2}_{\phi,\psi}(g)\times
\phi^2\zH^{k+2}_{\phi,\psi}(g))\cap\psi^2(\phi
C^{k+2,\alpha}_{\phi,\varphi}(g)\times
\phi^2C^{k+2,\alpha}_{\phi,\varphi}(g))\;.$$

\qed
\end{prop}

Propositions~\ref{regularite} and \ref{regularitebis} give
existence of H\"older continuous solutions. We can apply the usual
bootstrap arguments to those solutions to obtain smoothness, when
all the objects at hand are smooth (however, as already pointed
out, the bootstrap does not appear to work for solutions in
Sobolev spaces):

\begin{prop}[Higher regularity]\label{regularite3}
Let $k\in \N$, $\alpha\in (0,1)$, assume that \eq{threecB} with
$\ell \ge k+4$ holds, and that \eq{nscalprop1}-\eq{nscalprop3} and
\eq{threeaB} hold. Suppose moreover  that the scaling property, as
spelled out at the end of \ApSwSw, holds. Assume that $(K,g)\in
C^{k+3,\alpha}\times C^{k+4,\alpha} $ and $(Y,N)\in
C^{3,\alpha}_{\phi,\varphi}(g)\times
C^{4,\alpha}_{\phi,\varphi}(g) $.
\begin{enumerate}\item If  \eq{cocobis} holds and if \be\label{et1projregbis3}
\hspace{-0.5cm}\left(
\begin{array}{c}
J\\
\rho
\end{array}
\right)[(K,g)+\psi^2 P^*\Phi(Y,N)]-\left(
\begin{array}{c}
J\\
\rho
\end{array}
\right)[(K,g)] \in
\psi^2\Phi^{-1}(C^{k+1,\alpha}_{\phi,\varphi}(g)\times
C^{k,\alpha}_{\phi,\varphi}(g))\;, \ee
 then
$(Y,N)\in  C^{k+3,\alpha}_{\phi,\varphi}(g)\times
C^{k+4,\alpha}_{\phi,\varphi}(g)$, and thus
$$(\delta K,\delta
g):=\psi^2 P^*\Phi(Y,N) \in \psi^2(
C^{k+2,\alpha}_{\phi,\varphi}(g)\times
C^{k+2,\alpha}_{\phi,\varphi}(g))\;.$$ \item
 Similarly, if
\be\label{et1projregbis4}\hspace{-0.3cm} \left(
\begin{array}{c}
J\\
\rho
\end{array}
\right)[(K,g)+\psi^2 \Phi^2P^*(Y,N)]-\left(
\begin{array}{c}
J\\
\rho
\end{array}
\right)[(K,g)] \in \psi^2(C^{k+1,\alpha}_{\phi,\varphi}(g)\times
C^{k,\alpha}_{\phi,\varphi}(g))\;, \ee
 then
$(Y,N)\in C^{k+3,\alpha}_{\phi,\varphi}(g)\times
C^{k+4,\alpha}_{\phi,\varphi}(g))$, thus $$(\delta K,\delta g)\in
\psi^2(\phi C^{k+2,\alpha}_{\phi,\varphi}(g)\times
\phi^2C^{k+2,\alpha}_{\phi,\varphi}(g))\;.$$
\end{enumerate}
\end{prop}

\proof It suffices to rewrite the rescaled non-linear elliptic
equation \eq{et1projregbis4} for $(Y,N)$ as a linear elliptic
equation for $(Y,N)$ and freeze coefficients (depending on
$(K+\delta K, g+\delta g)$ hence on $(Y,N)$). The  interior
H\"older estimates~\cite[Theorem~6.2.5, p.~223]{Morrey} on the
sets $\hat \Omega_\alpha$ appearing in the definition of scaling
property give the local regularity, and the scaling property gives
the global weighted regularity. \qed

 In situations in which $P^*$ has trivial kernel the above theorems produce solutions of the constraint
 equations. As made clear by the analysis of Corvino~\cite{Corvino},
 solutions can be obtained even when a non-trivial kernel is present in the
 following circumstances:
\newcommand{\Kgpl}{(K_{Q,\lambda},g_{Q,\lambda})}
\newcommand{\pql}{{Q,\lambda}}
\newcommand{\pqll}{{Q_\lambda,\lambda}}
Suppose that the kernel ${\maclKz   }_0$ of $P^*$ at $(K_0,g_0)$
is non-trivial, set $k=\textrm{dim}{\maclKz   }_0$. Assume we are
given a family of pairs $\Kgpl $, where $\lambda\in [\lambda_0,
\infty[$ and $Q\in U$, where $U$ is an open in $\R^k$, such that
$x_{\pql } :=\Kgpl $ goes to $(K_0,g_0)$ in $A$ when $\lambda$
goes to infinity, uniformly in $Q\in U$. Assume, in the setup of
Theorem~\ref{theor:proj}, further that
$$
\psi^{-2}\Phi \left(
\begin{array}{c}
\delta J_{\pql }\\
\delta \rho_{\pql }
\end{array}
\right)  := \psi^{-2}\Phi \left\{\left(
\begin{array}{c}
J\\
\rho
\end{array}
\right) \Kgpl  - \left(
\begin{array}{c}
J\\
\rho
\end{array}
\right) (K_0,g_0) \right\}\;,
$$
goes to zero in $\zH^{k+1}_{\phi,\psi}(g_{\pql } )\times
\zH^{k}_{\phi,\psi}(g_{\pql } )$ when $\lambda$ goes to infinity,
uniformly in $Q\in U$. If in the setup of
Theorem~\ref{theor:projbis}, assume instead that the same holds
for the family
$$
\psi^{-2} \left(
\begin{array}{c}
\delta J_{\pql }\\
\delta \rho_{\pql }
\end{array}
\right)  := \psi^{-2} \left\{\left(
\begin{array}{c}
J\\
\rho
\end{array}
\right) \Kgpl  - \left(
\begin{array}{c}
J\\
\rho
\end{array}
\right) (K_0,g_0) \right\}\;,
$$
Then for $\lambda$ large enough, $ \left(
\begin{array}{c}
\delta J_{\pql }\\
\delta \rho_{\pql }
\end{array}
\right) $ is less than $\epsilon$ for all $Q\in U$. So, in the
setup of Theorem~\ref{theor:proj}, we can solve
\bel{coc1}f_{x_{\pql } }(\delta x_{\pql } )=- \pi_{{\cal
K}^{\bot_{g_{\pql } }}}\psi^{-2}\Phi \left(
\begin{array}{c}
\delta J_{\pql }\\
\delta \rho_{\pql }
\end{array}
\right) \;, \ee while in the setup of Theorem~\ref{theor:projbis}
we omit the $\Phi$ factor in \eq{coc1};
recall that $f_x$ has been defined in \eq{et1proja}. Let
$e_{(i)}$, $i=1,\ldots,k$, be
any basis of $\maclKz$, we define the family of maps \bel{coc}
\begin{array}{llll}
F_\lambda:&U&\longrightarrow&{ \R^k}\\
&Q&\mapsto &\left(\langle \psi^{-2}\Phi \left\{\left(
\begin{array}{c}
J\\
\rho
\end{array}
\right) (K_{\pql }+\delta K_{\pql },g_{\pql }+\delta g_{\pql })  -
\left(
\begin{array}{c}
J\\
\rho
\end{array}
\right) (K_0,g_0) \right\},e_{(i)}\rangle_{{\maclKz
}_0}\right)\;,
\end{array}
\ee assuming that we are in the context of
Theorem~\ref{theor:proj}. In the case of
Theorem~\ref{theor:projbis} the $\Phi$ factor should be removed
from \eq{coc}. We note the following result:
\begin{lem}\label{Brower}
Let $U$ and $V$ be open sets in $\R^n$,  suppose that $G$ is a
homeomorphism from $U$ to $V$, and consider
 a family $\{G_{\lambda}\}_{\lambda\in\R}$  of continuous
functions from $U$ to $\R^n$ which converge uniformly to $G$ when
$\lambda$ goes to infinity. Then for all $y$ in $V$, if $\lambda$
is large enough, there exists $x_\lambda\in U$ such that
$$G_\lambda(x_\lambda)=y\;.$$
\end{lem}
\begin{proof}
Consider the family of maps $u_\lambda:=\mbox{Id}-G_\lambda\circ
G^{-1}$ from $V$ to $\R^n$, the $u_\lambda$'s converge uniformly
to $0$ when $\lambda$ goes to infinity.  Let $y$ in $V$ and let
$r>0$ be such that the closed ball $\overline{B(y,r)}$ is included
in $V$. If $\lambda$ is large enough,
$\max_{z\in\overline{B(y,r)}}|u_\lambda(z)|\leq r$, then the map
$z\mapsto y+u_\lambda(z)$ is a continuous map from
$\overline{B(y,r)}$ to $\overline{B(y,r)}$. From the Brouwer fixed
point theorem ({\em cf., e.g.}\/~\cite{GuilleminPollack}) there
exists $y_\lambda\in \overline{B(y,r)}$ such that
$y_\lambda=y+u_\lambda(y_\lambda)$, we then set
$x_\lambda=G^{-1}(y_\lambda)$.
\end{proof}\qed

If there exists a function $h(\lambda)$ such that
$G_\lambda:=h(\lambda) F_\lambda$ satisfies the condition of
Lemma~\ref{Brower}, and assuming further that $0$ is in $V$, for
all $\lambda$ large enough we can choose $Q_\lambda$ such that
$F_\lambda(Q_\lambda)=0$, hence
$$
\left(
\begin{array}{c}
J\\
\rho
\end{array}
\right) (K_{\pqll }+\delta K_{\pqll },g_{\pqll }+\delta g_{\pqll
}) =\left(
\begin{array}{c}
J\\
\rho
\end{array}
\right) (K_0,g_0) \;.
$$
It is important to emphasise that if $\Kgpl -(K_0,g_0)$ is not in
$V_{x_{\pql } }$, then $(x_{\pql }+\delta x_{\pql }) \neq
(K_0,g_0)$, \emph{i.e.}, we have constructed a solution different
from the original one. Summarising, we have shown:
\begin{theor}\label{Tremov}
Under the hypotheses just described, the projection operators $
\pi_{\maclKzo }$ in \eq{et1proj1} and \eq{et1proj2} can be removed
for all $\lambda$ large enough.

\qed
\end{theor}

\section{An asymptotic inequality}\label{Sasympto}
 The isomorphism
theorems of the previous section all rely on the asymptotic
estimate \eq{it1ker}. The object of this section is to reduce the
proof of that estimate to two simpler estimates, one involving
only $Y$ and the other involving only $N$. It turns out that some
decay conditions are needed for that:

\begin{Def}\label{asymptocond} We will say
that $(M,g,K,\phi)$ satisfy the \emph{asymptotic condition a)} if
there exists a sequence $U_i\subset M$ of open  relatively compact
sets such that
 $\overline{U}_i
 \subset U_{i+1}$ (closure in $M$, not in $\bM$) with
\bel{cover} M=\cup_{i=1}^\infty U_i\;, \ee  and {\rm
\begin{deqarr}\arrlabel{ac1}\nonumber &
\lim_{i\rightarrow\infty}\|K\|_{L^\infty(M\setminus U_i)}=
\lim_{i\rightarrow\infty}\|\phi^2 K\|_{L^\infty(M\setminus U_i)}=
\lim_{i\rightarrow\infty}\|\phi\nabla K\|_{L^\infty(M\setminus
U_i)}=0\;,&\\\label{ac1a}&& \\ \label{ac1b}
&\lim_{i\rightarrow\infty}\|\phi^2 \Ricc(g)\|_{L^\infty(M\setminus
U_i)}=0.&
\end{deqarr}}
We will say that $(M,g,K,\phi)$ satisfy the \emph{asymptotic
condition b)} if \eq{cover} and \eq{ac1b} hold and if instead of
\eq{ac1a} we have  {\rm \begin{deqarr}\label{ac2a}
\lim_{i\rightarrow\infty}\|\phi K\|_{L^\infty(M\setminus U_i)}=
\lim_{i\rightarrow\infty}\|\phi^2 \nabla K\|_{L^\infty(M\setminus
U_i)}=0\;,
\end{deqarr}}
\end{Def}
For any  vector field $Y$ set \be \label{cb1} S(Y)_{ij}:=
\nabla_{(i} Y_{j)} = \frac 12 \left(\nabla_i Y_j + \nabla _j
Y_i\right)\;. \ee We can now give a sufficient condition for
\eq{it1ker}:
\begin{lem}\label{lem:equivcond}
Under the hypotheses of Lemma \ref{L5}, assume that $(M,g,K,\phi)$
satisfies the {\it asymptotic condition a)}. Then \eq{it1ker} is
equivalent to the requirement that there exists a compact set and
a constant $C$ such that for all smooth $(Y,N)$ supported outside
this compact set we have {\rm
\begin{deqarr} \label{Yca}\arrlabel{Yc} & C\|S(\phi
Y))\|_{\Lpsi}(g_0) \geq \|Y\|_{\Lpsi}(g_0)\;,& \\ & \label{Nca}
C\|\nabla\nabla (\phi^2 N)\|_{\Lpsi}(g_0) \geq
\|N\|_{\Lpsione}(g_0)\;.\end{deqarr}}
\end{lem}
\proof
Setting $Y=0$ or $N=0$ in \eq{it1ker} one obtains \eq{Yc} by
straightforward manipulations (replacing the compact set $\mcK$ of
Proposition~\ref{P:estproj} by a larger compact set if necessary).
In order to prove the reverse implication let us start by
establishing the inequality
\bel{bamg0}\|Y\|_{\Lpsione(g_0)}+ \|N\|_{\Lpsione(g_0)} \le C
\|P^*(\phi Y,\phi^2 N)\|_{\Lpsi(g_0)}\;, \ee for all $(Y,N)$
supported in $M\setminus \overline{U}_i$, for $i$ large enough.
Let $P^*_a$, $a=1,2$, be defined as
\begin{eqnarray}\nn
P^*(Y,  N) &=:& \left(
\begin{array}{l}
P^*_1( Y, N)\\
P^*_2( Y, N)\\
\end{array}
\right)\;;\label{P12bis}
\end{eqnarray} from \Eq{4} one finds
\begin{eqnarray} \|-2\alpha(S(\phi Y))+P_1^*(\phi Y,\phi^2
N)\|_{\Lpsi(g_0)} \le C \|\phi^2 K
\|_{L^\infty}\|N\|_{\Lpsi(g_0)}\;, \label{bamg1}\eea
 where $\alpha$ is as in \eq{P12}. \Eq{Yca}
together with Equation~\eq{fi.1} (with $b=0$ there)
yield\begin{eqnarray} c \|Y\|_{\Lpsione(g_0)} &\le&
2\|Y\|_{\Lpsi(g_0)}+2\|\alpha(S(\phi Y))\|_{\Lpsi(g_0)}
 \nonumber \\ &\le& C\|S(\phi Y)\|_{\Lpsi(g_0)}+2\|\alpha(S(\phi Y))\|_{\Lpsi(g_0)}\le C'\|\alpha(S(\phi Y))\|_{\Lpsi(g_0)}
 \nonumber \\ &=& C'\|\alpha(S(\phi Y))+P_1^*(\phi Y,\phi^2 N)-P_1^*(\phi Y,\phi^2 N)\|_{\Lpsi(g_0)}
 \nonumber \\ & \le &C'\Big(\|P_1^*(\phi Y,\phi^2 N)\|_{\Lpsi(g_0)} +
C\|\phi^2 K \|_{L^\infty}\|N\|_{\Lpsi(g_0)}\Big)\;.
\label{bamg2}\eea Applying \eq{Nca} it holds that \bel{bamg3}
\|N\|_{\Lpsione(g_0)} \le C \|\nabla \nabla (\phi^2
N)\|_{\Lpsi(g_0)}\le C^2 \|\alpha(\nabla \nabla (\phi^2
N)\|_{\Lpsi(g_0)}\;. \ee {}From \Eq{4} we have\ \beal{bamg4}
\lefteqn {\|\alpha(\nabla \nabla (\phi^2 N))-P_2^*(\phi Y,\phi^2
N)\|_{\Lpsi(g_0)}}&& \nn
\\&&\le C\Bigg( \|K\|_{L^\infty(\chi(M\setminus U_i))}
\|\nabla(\phi Y)\|_{\Lpsi(g_0)}+ \|\phi\nabla
K\|_{L^\infty(\chi(M\setminus U_i))}\|Y\|_{\Lpsi(g_0)} \nn
\\ && \qquad+ \Big(\|\phi^{2}\Ricc\|_{L^\infty(\chi(M\setminus U_i))} +
\|(\phi K)^2\|_{L^\infty(\chi(M\setminus
U_i))}\Big)\|N\|_{\Lpsi(g_0)} \Bigg)\nn \\ && \le \epsilon
\Big(\|Y\|_{\Lpsione(g_0)} + \|N\|_{\Lpsi(g_0)}\Big)\;, \eea where
$\epsilon$ can be made as small as desired by choosing $i$ large
enough. It then follows from \Eq{bamg3} that \bel{bamg5}
\|N\|_{\Lpsione(g_0)} \le C\|P_2^*(\phi Y,\phi^2 N)\|_{\Lpsi(g_0)}
+C\epsilon \Big(\|Y\|_{\Lpsione(g_0)}
 + \|N\|_{\Lpsi(g_0)}\Big) \;.\ee Adding
\eq{bamg2} and \eq{bamg5}, and choosing $i$ large enough --- so
that $\epsilon $ is small enough --- one obtains \Eq{bamg0}.
 We note that from \eq{bamg0} by similar
manipulations one can further obtain
\bel{bamg0a}\|Y\|_{\Lpsione(g_0)}+ \|N\|_{\Lpsitwo(g_0)} \le C
\|P^*(\phi Y,\phi^2 N)\|_{\Lpsi(g_0)}\;, \ee but this is
irrelevant for our purposes. \qed

An identical calculation yields:
\begin{lem}\label{lem:equivcondbis}
Under the hypotheses of Lemma~\ref{L5bis}, assume that
$(M,g,K,\phi)$ satisfies the {\em asymptotic condition b)}. Then
\eq{it1kerbis} is equivalent to the requirement that for all
smooth $(Y,N)$ supported outside a compact set we have
$$
C\|\phi S(Y))\|_{\Lpsi}(g_0) \geq \|Y\|_{\Lpsi}(g_0)\;,
$$
$$
C\|\phi^2 \nabla\nabla N\|_{\Lpsi}(g_0) \geq
\|N\|_{\Lpsione}(g_0)\;.
$$

\qed
\end{lem}

\section{Compact boundaries}
\label{Scb}  Let us justify the inequality~\eq{Yca} in a
neighborhood of a compact boundary $\partial M$. We assume that
the metric is as in Appendix~\ref{SWPin}, in particular \Eq{pi0}
holds. We start with the following:

\begin{Proposition}
\label{Pbc} Let $s\ne -1/2$ and suppose that $$ |\Hess
(x)|=o(x^{-1})\;.$$ Then there exists a neighborhood
$\mathcal{O}_s$ of $\partial M$ such that for every $C^1$ vector
field with compact support in $\mathcal{O}_s \setminus \partial M$
we have
\begin{eqnarray}
\int x^{2s} |Y|^2 &\le&C\int x^{2s+2} |S(Y)|^2 \;, \label{cb15}
\end{eqnarray}
for some constant $C$, where $S$ is defined by \Eq{cb1}.
\end{Proposition}
\begin{remark}\label{r6.2} The restriction $s\ne -1/2$ is sharp, which can be
seen by considering the family of vector fields $\chi_n
(1-\chi_{n_0})Y$, $n\ge n_0$, where $Y$ is a Killing vector which
does not vanish on $\partial M$, and where the  cut-off functions
$\chi_n$ are defined as $\chi_n(x)=1$ for $x\ge 1/n$,
$\chi_n(x)=\ln(2nx)/\ln 2$ for $1/(2n)\le x \le 1/n$,
$\chi_n(x)=0$ otherwise. The resulting $\chi_n$'s are not $C^1$,
but this is enough to invalidate \eq{cb15}; in any case, a small
smooth perturbation of $\chi_n$ will yield the required $C^1$
example.
\end{remark}

\proof The result is a straightforward consequence of
Corollary~\ref{Ce4}.
 \qed

We shall consider metrics which can be quite singular near the
boundary; this is mainly motivated by the applications to
conformally compactifiable metrics, see Section~\ref{Scc} below.
To control the boundary behavior of $g$ we thus introduce the
following definition:

\begin{Definition}\label{Dwab}
Let $k\in \N$ and let ${\mathcal W}$ be a space of symmetric
tensors on $M$. We shall say that $g$ has an $({\mathcal
W},k)$--behavior at $\partial M$ if there exists a metric
${g}_{\bM}$ on $\bM$  of class $C^k(\bM)$ such that
$g-{g}_{\bM}\in {\mathcal W}$.
\end{Definition}

 In the remainder of this section we assume that $\bM$ is
a compact manifold with boundary. We take the weight function
$\phi$ as
$$\phi:=x\;,$$
where $x$ is any defining function for $\partial M$. For $k\in\N$
and $s\in\R$, we define \bel{??}
 \hbord^s_k(g):=\zH^k_{x,x^{-s-n/2}}(g).
\ee The labeling of the spaces here is motivated by the following
decay property (\emph{cf., e.g.}, \cite{AndElli})
\bel{sobinc2h} f\in \hbord_k^\beta(g) \;, \ k>n/2 \quad
\Longrightarrow \quad f=o(x^\beta) \;.\ee We also have $$x^\sigma
\in \hbord^s_k(g) \ \mbox{ iff } \ \sigma > s + (n-1)/2\;.$$ Let
us define in the same way
$$ \cbord^s_{k,\alpha}(g):=C^{k,\alpha}_{x,x^{-s}}(g).$$

 When studying boundary behavior of solutions
of PDE's near boundaries, alternative useful classes of weighted
spaces are obtained as follows: in a collar neighborhood of
$\partial M$ one introduces coordinate systems $(x,v^A)$, with
$\partial M$ being given by the equation $\{x=0\}$. Instead of
adding a weight factor $x$ for each derivative, one adds $x$
factors to the $\partial_x$ derivatives only. Functions in such
weighted spaces have more tangential regularity, as compared with
functions in the $\hbord$ spaces or $\cbord$ spaces. However, some
of the simple scaling arguments which we have been using so far do
not apply, and considerably more work is required (see, {\em
e.g.},~\cite{AndChDiss}) to obtain a-priori estimates in such
spaces. While those alternative spaces could probably be used in
our context here, leading to improved regularity of solutions, we
have not attempted to carry through a systematic study.

We start with the following:
\begin{theor}\label{Til1abprojbis} Let $\bM$ be a compact
manifold with boundary, let $k\ge 0$, and suppose that $g_0$ is a
metric on $M$ which has $(W^{k+4,\infty}_x,k+4)$--behavior at
$\partial M$, with {\rm
\begin{deqarr} \label{??1a} &
x^2|\Ricc(g_0)|_{g_0}\to_{x\to 0}0\;, &
\\ \label{??1b} &
x|\nabla\nabla x|_{g_0}\to_{x\to 0}0\;, &
\\ \label{??1c} &
K_0\in x^{-1}W^{k+3,\infty}_{x}(g_0)\;,  \quad
x|K_0|_{g_0}+x^2|\nabla K_0|_{g_0}\to_{x\to 0}0\;.&
\end{deqarr}}
Then for all $s\neq (n-1)/2,(n-3)/2$ and all $(K,g)$ close to
$(K_0,g_0)$ in $x^{-1}W^{k+3,\infty}_x(g_0)\times
W^{k+4,\infty}_x(g_0)$ norm, the map
$$
\pi_{\maclKzo }{L}_{x,x^{s-n/2}} :{\maclKzo }\cap
(\hbord^{-s}_{k+3}(g)\times \hbord^{-s}_{k+4}(g)) \longrightarrow
{\maclKzo }\cap (\hbord^{-s}_{k+1}(g)\times \hbord^{-s}_{k}(g))
$$
is an  isomorphism such that the norm of its inverse does not
depend on $(K,g)$.
\end{theor}
\begin{remark}
Conditions \eq{??1a}-\eq{??1b} will hold if
  there exists
$\alpha>0$ such that $g$ has $(x^\alpha
W^{k+4,\infty}_x,k+4)$--behavior at $\partial M$.
\end{remark}

 \proof We wish to apply Theorem~\ref{Til1projbis}, in order to
do that we need to establish the inequality \eq{it1kerbis} for
$Y$'s and $N$'s supported outside of a sufficiently a large ball.
For $s \ne (n-1)/2$, Proposition~\ref{Pbc} yields \be
\|Y\|_{\hbord^{-s}_0} \le C\|xS(Y)\|_{\hbord ^{-s}_0};. \ee
Applying Proposition~\ref{Pinb} twice we find that for $s\ne
(n-1)/2, (n-3)/2$ it holds that \be \|N\|_{\hbord ^{-s}_0} \le C
\|x\nabla N\|_{\hbord ^{-s}_0}\le C^2 \|x^2\nabla \nabla
N\|_{\hbord^{-s}_0}\le C^3 \|x^2\alpha(\nabla \nabla N)\|_{\hbord
^{-s}_0}\;, \ee where $\alpha$ is as in \eq{alphadef}. Now
$(M,g,K,x)$ satisfy the \emph{asymptotic condition b)} of
Definition~\ref{asymptocond} with $U_i=\{x>{1}/{i}\}$, and
Lemma~\ref{lem:equivcondbis} shows that we can apply
Theorem~\ref{Til1projbis}. \qed

Our first main application of the abstract results of the previous
sections is surjectivity up to kernel of $P^*$  of the constraint
map. In particular surjectivity is obtained if no kernel is
present; a case with kernel will be analysed in
Section~\ref{SSKio}.

\begin{theor}\label{theor:abprojbis}
Under the hypotheses of Theorem~\ref{Til1abprojbis} with $s\geq
n-2$, $s>1$ if $n=3$ , $k>n/2$, the map \bel{abbij}
\begin{array}{c}
 {\maclKzo }\cap (\hbord^{-s}_{k+3}(g)\times  \hbord^{-s}_{k+4}(g))
\longrightarrow
{\maclKzo }\cap (\hbord^{-s}_{k+1}(g)\times \hbord^{-s}_{k}(g))\\
(Y,N) \longmapsto \pi_{\maclKzo }x^{{-2s}+n}\left\{\left(
\begin{array}{c}
J\\
\rho
\end{array}
\right)[(K,g)+x^{{2s}-n}\Phi^2 P^*(Y,N)] -\left(
\begin{array}{c} J\\
\rho
\end{array}
\right)(K,g)\right\}
\end{array}
\ee is bijective in a neighborhood of zero.
 More precisely,
there exists $\epsilon>0$ such that for all  $(K,g)$ in
$x^{-1}W^{k+3,\infty}_x(g_0)\times W^{k+4,\infty}_x(g_0)$ for
which
$$\|(K-K_0,g-g_0)\|_{x^{-1}W^{k+3,\infty}_x(g_0))\times
W^{k+4,\infty}_x(g_0)} < \epsilon
$$
and for all pairs $(\delta J, \delta\rho)\in
\hbord^{{s}-n}_{k+1}(g)\times \hbord^{{s}-n}_{k}(g)$ satisfying
$$
\|(\delta J, \delta\rho)\|_{\hbord^{{s}-n}_{k+1}(g)\times
\hbord^{{s}-n}_{k}(g)}<\epsilon$$
 there exists a solution $(\delta
K,\delta g)=x^{{2s}-n}\Phi^2 P^*(Y,N) \in
\hbord^{{s-n+1}}_{k+2}(g)\times \hbord^{{s}-n+2}_{k+2}(g))$, close
to zero, of the equation \bea & \nonumber \pi_{\maclKzo
}x^{{-2s}+n} \left\{\left(
\begin{array}{c}
J\\
\rho
\end{array}
\right) (K+\delta K,g+\delta g) - \left(
\begin{array}{c}
J\\
\rho
\end{array}
\right) (K,g) \right\}= \pi_{\maclKzo }x^{{-2s}+n}\left(
\begin{array}{c}
\delta J\\
\delta \rho
\end{array}
\right) \;. &\\&& \label{abbij1}\eea
\end{theor}

\proof  The conditions $s\geq n-1$ and $k>n/2$ ensure that the map
of Equation~\eq{abbij} is well defined and differentiable in a
neighborhood of zero; a relatively straightforward though lengthy
check of that can be done using  weighted Moser inequalities
(see~\cite{ChLengardnwe} for proofs in a slightly different
context; the arguments there adapt to the current setting in a
straightforward way). The result follows then from
Theorem~\ref{theor:projbis}. \qed

We also have solutions with H\"older regularity:

\begin{prop}\label{p6.7} Let $\bM$ be a compact
manifold with boundary, let $k\ge k_0:=\lfloor n/2\rfloor +1$ (the
smallest integer strictly larger than $n/2$), $\alpha\in (0,1)$,
and suppose that $g_0$ is a metric on $M$ which has
$(\cbord^0_{k+4,\alpha},k+5)$--behavior at $\partial M$, with {\rm
\begin{deqarr} \label{??1a2} & x^2|\Ricc(g_0)|_{g_0}\to_{x\to
0}0\;, &
\\ \label{??1b2} &
x|\nabla\nabla x|_{g_0}\to_{x\to 0}0\;, &
\\ {\label{??1c2}} &
K_0\in \cbord^{-1}_{k+3,\alpha}(g_0)\;,  \quad
x|K_0|_{g_0}+x^2|\nabla K_0|_{g_0}\to_{x\to 0}0\;.&
\end{deqarr}
} There exists $\epsilon>0$ such that if $(K,g)$ in
$\cbord^{-1}_{k+3,\alpha}(g_0)\times\cbord^{0}_{k+4,\alpha}(g_0)$,
and if
$$\|(K-K_0,g-g_0)\|_{x^{-1}W^{k_0+3,\infty}_x(g_0)\times
W^{k_0+4,\infty}_x(g_0)} < \epsilon
$$
$$\|(\delta J,\delta \rho)\|_{
\cbord^{t}_{k_0+1,\alpha}(g)\times\cbord^{t}_{k_0,\alpha}(g)} +
\|(\delta J, \delta\rho)\|_{\hbord^{t}_{k_0+1}(g)\times
\hbord^{t}_{k_0}(g)}<\epsilon$$ for some $t\geq -2$, $t>-2$ if
$n=3$, then the solution $(\delta K, \delta g)$ given by
Theorem~\ref{theor:abprojbis} (with $s=t+n$) is in $
\cbord^{t+1}_{{k_0}+2,\alpha}(g)\times\cbord^{t+2}_{{k}+2,\alpha}(g)$.
If moreover $(\delta J,\delta \rho)\in{
\cbord^{t}_{k+1,\alpha}(g)\times\cbord^{t}_{k_0,\alpha}(g)}$ then
the solution given by Theorem~\ref{theor:abprojbis} is in $
\cbord^{t+1}_{{k}+2,\alpha}(g)\times\cbord^{t+2}_{{k}+2,\alpha}(g)$.
\end{prop}

\begin{Remark} All the hypotheses in Proposition~\ref{p6.7} will hold if $g_0$ has
$(\cbord^\beta_{k+4,\alpha},k+5)$ behavior at $\partial M$, for
some $\beta>0$, with $K_0\in \cbord^{\beta-1}_{k+3,\alpha}$; in
particular they will hold if $(K_0,g_0)\in C^{k+4}(\bM)\times
C^{k+5}(\bM)$.
\end{Remark}

\proof Under the current hypotheses all the conditions of
Proposition~\ref{regularitebis} with $k$ there equal to $k_0$ are
met. (The weighted elliptic regularity condition of
Definition~\ref{def:regul} is satisfied by the
calculation~\eq{weighcalc}, Appendix~\ref{Sscaling}.) The higher
H\"older regularity follows from Proposition~\ref{regularite3}.
\qed

 A
useful class of solutions is obtained by taking the weight to
decay exponentially at the boundary: the weighting functions are
then chosen to be $\phi=x^2$ and $\psi=e^{s/x}$.  The main
interest of this class of spaces stems from the inclusion
$$\cap_{k\in\N} C^{k+\alpha}_{x^2,e^{s/x}} \subset
C^\infty(\bM)\;,$$ which holds
 on a compact manifold with boundary $\bM$ for any $s>0$. Here the
space $C^\infty(\bM)$ denotes the space of tensor fields which
extend smoothly to $\partial M$, together with  all their
derivatives; in fact all fields belonging to the left-hand-side of
the inclusion above  can be smoothly extended by a zero tensor
field. It is shown at the end of Appendix~\ref{Sscaling} that the
spaces $ H^k_{x^2,e^{s/x}} $ satisfy the hypotheses of
Lemma~\ref{LC1}; the latter asserts that the scaling property
holds for those spaces. This gives:
\begin{theor}\label{theor:6.9} Let $\bM$ be a compact
manifold with boundary, let $s>0$, $k>n/2$ and suppose that $g_0$
is a metric on $M$ which has
$(W^{k+4,\infty}_{x^2},k+4)$--behavior at $\partial M$, with {\rm
\begin{deqarr}\arrlabel{e6.9} \label{??1aexp} &
x^4|\Ricc(g_0)|_{g_0}\to_{x\to 0}0\;, &
\\ \label{??1bexp} &
x|\nabla\nabla x|_{g_0}\to_{x\to 0}0\;, &
\\ \label{??1cexp} &
K_0\in x^{-2}W^{k+3,\infty}_{x^2}(g_0)\;,  \quad
x^2|K_0|_{g_0}+x^4|\nabla K_0|_{g_0}\to_{x\to 0}0\;.&
\end{deqarr}}There exists $\epsilon>0$ such that for all $(K,g)$ in
$x^{-2}W^{k+3,\infty}_{x^2}(g_0)\times W^{k+4,\infty}_{x^2}(g_0)$
for which
$$\|(K-K_0,g-g_0)\|_{x^{-2}W^{k+3,\infty}_{x^2}(g_0)\times
W^{k+4,\infty}_{x^2}(g_0)} < \epsilon\;,
$$
and for all pairs $(\delta J, \delta\rho)\in
H^{k+1}_{x^2,e^{s/x}}(g)\times H^{k}_{x^2,e^{s/x}}(g)$ satisfying
$$
\|(\delta J, \delta\rho)\|_{H^{k+1}_{x^2,e^{s/x}}(g)\times
H^{k}_{x^2,e^{s/x}}(g)}<\epsilon$$
 there exists a solution $(\delta
K,\delta g)=e^{-2s/x}\Phi^2 P^*(Y,N) \in
x^2H^{k+2}_{x^2,e^{s/x}}(g)\times x^4H^{k+2}_{x^2,e^{s/x}}(g))$,
close to zero, of the equation \bea & \nonumber \pi_{\maclKzo
}e^{{2s}/x} \left\{\left(
\begin{array}{c}
J\\
\rho
\end{array}
\right) (K+\delta K,g+\delta g) - \left(
\begin{array}{c}
J\\
\rho
\end{array}
\right) (K,g) \right\}= \pi_{\maclKzo }e^{{2s}/x}\left(
\begin{array}{c}
\delta J\\
\delta \rho
\end{array}
\right) \;. &\\&& \label{abbij1exp}\eea

\end{theor}

 \proof As before, we  apply Theorems~\ref{Til1projbis} and
 \ref{theor:projbis}. We
 first show
the inequality \eq{it1kerbis} for $Y$'s and $N$'s supported in a
sufficiently small neighborhood of $\partial M$. For $s \neq 0$,
consider the equality in Corollary~\ref{CE5}. Taking absolute
values of both sides, and applying Cauchy-Schwarz to the left term
in  the resulting equality one obtains
\be \|Y\|_{H^{0}_{x^2,e^{s/x}}} \le
C\|x^2S(Y)\|_{H^{0}_{x^2,e^{s/x}}}\;. \ee Applying
Proposition~\ref{Pinbexp}  with $u:=x^2\nabla N$ and with $u=N$,
we find that for $s\neq 0$ it holds that
\begin{eqnarray*}
\|x^4\nabla\nabla N\|_{H^{0}_{x^2,e^{s/x}}}
&\geq& C_1\|x^2\nabla N\|_{H^{0}_{x^2,e^{s/x}}}\\
&=&\frac{C_1}{2}\|x^2\nabla
N\|_{H^{0}_{x^2,e^{s/x}}}+\frac{C_1}{2}\|x^2\nabla
N\|_{H^{0}_{x^2,e^{s/x}}}\\
&\geq&\frac{C_1}{2}\|x^2\nabla
N\|_{H^{0}_{x^2,e^{s/x}}}+C_2\|N\|_{H^{0}_{x^2,e^{s/x}}}\\
&\geq&{C_3}\|N\|_{H^{1}_{x^2,e^{s/x}}}\;.
\end{eqnarray*}
The inequality \eq{it1kerbis} is then satisfied. Now $(M,g,K,x)$
satisfy the \emph{asymptotic condition b)} of
Definition~\ref{asymptocond} with $U_i=\{x>{1}/{i}\}$, and
Lemma~\ref{lem:equivcondbis} shows that we can apply
Theorem~\ref{Til1projbis}. The conditions $s>0$ and $k>n/2$ ensure
that the map of Equation~\eq{et1projbis} is well defined and
differentiable in a neighborhood of zero. (Here one should use
weighted Moser inequalities, which can be established by the
methods of~\cite{ChLengardnwe} together with the scaling arguments
of Appendix~\ref{Sscaling}.) The result follows then from
Theorem~\ref{theor:projbis}. \qed

 It is
easy to check that the spaces $H^k_{x^2,e^{s/x}}$ in
Theorem~\ref{theor:6.9} can be replaced by the spaces
$H^k_{x^2,x^{2a}e^{s/x}}$, for any $a\in\R$, we leave the details
to the reader. The need for such a generalisation arises when
wishing to pass from weighted  Sobolev spaces to exponentially
weighted H\"older spaces: indeed, \Eq{weighcalc},
Appendix~\ref{Sscaling}, gives \bel{weighc2}
\|u\|_{C^{k+m,\alpha}_{x^2,x^{2a} e^{s/x}}(M)}\leq C
(\|Pu\|_{C^{k,\alpha}_{x^2,x^{2a} e^{s/x}
}(M)}+\|u\|_{L^{2}_{x^{2(a-n/2)} e^{s/x}}(M)})\;. \ee
 and leads to the following
proposition, the details are left to the reader:

\begin{prop}\label{p6.10} Let $\bM$ be a compact
manifold with boundary, let $k\ge k_0:=\lfloor n/2\rfloor +1$ (the
smallest integer strictly larger than $n/2$), and suppose that
$g_0$ is a metric on $M$ which has
$(C^{k+4,\alpha}_{x^2,1},k+5)$--behavior at $\partial M$. Assume
that \eq{e6.9} holds.  Then there exists $\epsilon>0$ such that if
$(K,g)$ in $x^{-2}C^{k+3,\alpha}_{x^2,1}(g_0)\times
C^{k+4,\alpha}_{x^2,1}(g_0)$ with
$$\|(K-K_0,g-g_0)\|_{x^{-2}W^{k_0+3,\infty}_{x^2}(g_0)\times
W^{k_0+4,\infty}_{x^2}(g_0)} < \epsilon\;,
$$
and if
$$\|(\delta J,\delta
\rho)\|_{C^{k_0+1,\alpha}_{x^2,e^{t/x}}(g)\times
C^{k_0,\alpha}_{x^2,e^{t/x}}(g)} + \|(\delta J,
\delta\rho)\|_{H^{k_0+1}_{x^2,x^{-n}e^{t/x}}(g)\times
H^{k_0}_{x^2,x^{-n}e^{t/x}}(g)}<\epsilon$$ with $t>0$, then the
solution $(\delta K, \delta g)$ given by Theorem~\ref{theor:6.9}
is in $ x^2C^{{k_0}+2,\alpha}_{x^2,e^{t/x}}(g)\times
x^4C^{{k_0}+2,\alpha}_{x^2,e^{t/x}}(g)$. If moreover $(\delta
J,\delta \rho)\in C^{k+1,\alpha}_{x^2,e^{t/x}}(g)\times
C^{k,\alpha}_{x^2,e^{t/x}}(g)$ then the solution given by
Theorem~\ref{theor:6.9} is in $
x^4C^{{k}+2,\alpha}_{x^2,e^{t/x}}(g)\times
x^8C^{{k}+2,\alpha}_{x^2,e^{t/x}}(g) $.
\end{prop}

\qed

Choose some $\alpha>0$ and  define the Fr\'echet spaces
$C^{\infty}_{x^2,e^{s/x}}(g)$ as the collection of all functions
or tensor fields which are in $C^{k,\alpha}_{x^2,e^{s/x}}(g)$
whatever $k\in\N$, equipped with the family of semi-norms
$\{\|\cdot\|_{C^{k,\alpha}_{x^2,e^{s/x}}(g)},k\in\N\}$. We then
have:
\begin{cor}\label{C6.11}
Under the hypotheses of the preceding proposition, if $(\delta
J,\delta \rho)\in C^{\infty}_{x^2,e^{s/x}}(g)\times
C^{\infty}_{x^2,e^{s/x}}(g)$, and if $(K,g)\in C^\infty(\bM)\times
C^\infty(\bM)$, then the solution given by Theorem~\ref{theor:6.9}
is in $$ x^2C^{\infty}_{x^2,e^{s/x}}(g)\times
x^4C^{\infty}_{x^2,e^{s/x}}(g)\subset C^\infty(\bM)\times
C^\infty(\bM)\;. $$ In fact $(\delta K,\delta g)$  can be smoothly
extended by zero across $\partial M$.

\qed
\end{cor}

\newcommand{\chyp}{{\mathcal C}}%
\section{Conformally compactifiable initial data}
\label{Scc}

 A Riemannian manifold $(M,g)$ will be
said to be \emph{conformally compactifiable} if
\begin{enumerate}
\item $\bM=M\cup \partial M$ is a compact manifold with non-empty
boundary;
\item let $x$ be any defining function for $\partial M$, then the
tensor field  $x^2g$ extends by continuity to a continuous
Riemannian metric
 $\bg$ on $\bM$.
\end{enumerate}
This definition encompasses Riemannian manifolds such as
hyperbolic space. The associated initial data are occur in the
context of space-times which are asymptotically flat in lightlike
directions~\cite{AndChDiss,ACF,Friedrich:Pune}, or in that of
asymptotically anti-de Sitter
space-times~\cite{Kannar:adS,friedrich:AdS}.

The topological setup here is thus identical to that of the
previous section, but the metrics $g$ differ from the ones used
there by a rescaling factor $x^2$. It turns out that there is a
simple correspondence of the functional spaces $\hbord^\alpha_k$
from the previous section with a class of natural weighted spaces
associated to  conformally compactifiable metrics: since
$g=x^{-2}\bg$ on $M$ we obviously have
$$L^2_\psi(g)=L^2_{x^{-n/2}\psi}(\bg)\;.$$
Further, assuming that \eq{lcond} holds for $0\le i \le k$ with
$\phi=x$ and $g$ there replaced by $\bg$ (recall that this will
hold if $\bg$ has $(W^{k,\infty}_x,k)$--behavior at $\partial M$
in the sense of Definition~\ref{Dwab}, in particular that will be
the case if $\bg$ is $C^k(\bM)$) it is simple to check that
$$H^k_{1,\psi}(g)=H^k_{x,x^{-n/2}\psi}(\bg)$$ for
 tensors; $0\le i \le k-1$ in \eq{lcond} would suffice for
functions.
 It is therefore natural (see \eq{??}-\eq{sobinc2h})
to define \bel{hypHdef}
\hhyp^s_k(g):=\zH^k_{1,x^{-s}}(g)=\hbord^{s}_k(\bg) \;. \ee

 Let us define in the same way
 $$
 \chyp^s_{k,\alpha}(g):=C^{k,\alpha}_{x,x^{-s}}(\overline{g})=\cbord^s_{k,\alpha}(\overline{g}).$$
We note that $$x^\sigma \in \hhyp^s_k \ \mbox{ iff } \ \sigma > s
+ (n-1)/2\;.$$ Similarly to \eq{sobinc2h} we have
\bel{sobinc2hh} f\in \hhyp_k^\beta(g) \;, \ k>n/2 \quad
\Longrightarrow \quad f=o(x^\beta) \;.\ee We will be mainly
interested in conformally compactifiable metrics such that $x^2g$
has $(\hhyp^\alpha_k,k)$-- or $\chyp^s_{k,\alpha}$--behavior at
the conformal boundary, $\alpha>0$; such metrics arise naturally
when solving the constraint equations via the conformal
method~\cite{AndChDiss,ACF}.

 We will need some estimates on
$P^*(Y,N)$ extending those of Section~\ref{Sasympto}, when
$$
K=\lambda g+L,
$$
where $\lambda$ is a uniformly bounded function on $M$. We will
further assume that $$|L|_{\bg}=o(x^{-2})\ \mbox{ and } \ |\nabla
L|_{\bg}=o(x^{-3})\;.$$ If we use this particular choice of $K$ in
equation \eq{4}, we find
$$ P^*(Y,N)= \left(
\begin{array}{l}
P_1^*(Y,N) \\
P_2^*(Y,N)
\end{array}
\right)$$ \be \label{4h}= \left(
\begin{array}{l}
2(\nabla_{(i}Y_{j)}-\nabla^lY_l g_{ij}-g_{ij}\lambda N+n\lambda\; N g_{ij})+o(x^{-2})N\\
 \\
2\lambda(\nabla^lY_l g_{ij}-\nabla_{(i}Y_{j)})+
-\Delta N g_{ij}+\nabla_i\nabla_j N-N \Ricc(g)_{ij} \\
\; +2N\lambda^2g_{ij}-2Nn\lambda^2 g_{ij}+o(x^{-2})(\nabla
Y)+o(x^{-3})(Y)+o(x^{-2})N
\end{array}
\right) \;,\ee where $o(x^{\alpha})$ denotes a tensor the
$\bg$--norm of which is $o(x^{\alpha})$. We then have
\begin{eqnarray}\nn \lambda P_1^*(Y,N)+P_2^*(Y,N)=-\Delta N
g_{ij}+\nabla_i\nabla_j N-N \Ricc(g)_{ij}\;\;\;\;\;\;\;\\
+o(x^{-2})(\nabla
Y)+o(x^{-3})(Y)+o(x^{-2})N.\label{enh}\end{eqnarray} If we prove
that for all $(Y,N)$ supported in a neighborhood ${\cal O}$ of
$\partial M$,
\begin{deqarr}
\label{Ycha}\arrlabel{Ych} & \|S(Y)\|_{\hhyp^s_0 }\geq
C\|Y\|_{\hhyp^s_0 }\;,& \\ & \label{Ncha} \|-\Delta N
g+\nabla\nabla N-N \Ricc(g)\|_{\hhyp^s_0 }\geq C\|N\|_{\hhyp^s_1
}\;.\end{deqarr} then we will have from \eq{Ycha} and \eq{fi.1}:
\be\label{a} \|Y\|_{\hhyp^s_1 }\leq C(\|Y\|_{\hhyp^s_0
}+\|S(Y)\|_{\hhyp^s_0 })\leq C'\|S(Y)\|_{\hhyp^s_0 }\leq C''(
\|P_1^*(Y,N)\|_{\hhyp^s_0 }+\|N\|_{\hhyp^s_0 })\;,\ee and from
\eq{Ncha} and \eq{enh}, \be\label{b}
\|\lambda\|_{L^\infty}\|P_1^*(Y,N)\|_{\hhyp^s_0
}+\|P_2^*(Y,N)\|_{\hhyp^s_0 }\geq C\|N\|_{\hhyp^s_1 }-\epsilon
\|Y\|_{\hhyp^s_1 }\;,\ee where $\epsilon$ is arbitrary  close to
zero, reducing ${\cal O}$ if necessary. Finally
$\epsilon$\eq{a}+\eq{b} gives
$$
(\|\lambda\|_{L^\infty}+1)\|P_1^*(Y,N)\|_{\hhyp^s_0
}+\|P_2^*(Y,N)\|_{\hhyp^s_0 }\geq C\|N\|_{\hhyp^s_1}\;,
$$
then for $\epsilon$ small, we obtain the asymptotic inequality
\eq{it1kerbis} (with $\Phi=\mathrm{id}$) for $P^*$:
$$
(\|\lambda\|_{L^\infty}+1)\|P_1^*(Y,N)\|_{\hhyp^s_0
}+\|P_2^*(Y,N)\|_{\hhyp^s_0 }\geq C\|Y\|_{\hhyp^s_0 }+ C
\|N\|_{\hhyp^s_1}\;.
$$

Let us justify the inequality~\eq{Ycha} in a neighborhood of a
compact boundary $\partial M$.  We have the following:

\begin{Proposition}
\label{Pbch} Let $s\ne (n-1)/2,(n+1)/2$. Then there exists a
neighborhood $\mathcal{O}_s$ of $\partial M$ such that for every
$C^1$ vector field with compact support in $\mathcal{O}_s
\setminus \partial M$ we have
\begin{eqnarray}
\int x^{2s} |Y|^2 \;d\mu_g &\le&C\int x^{2s} |S(Y)|^2 \;d\mu_g\;,
\label{cb15h}
\end{eqnarray}
for some constant $C$, where $S$ is defined by \Eq{cb1}.
\end{Proposition}
\begin{remark} The argument given in Remark~\ref{r6.2} shows that the restriction $s\ne (n+1)/2$ is sharp.
We suspect that the restriction $s\ne (n-1)/2$ can be removed.
\end{remark}

\proof This is just a rewriting  of
Corollary~\ref{prop:estiPS2ahg}.
 \qed

\begin{theor}\label{Til1abprojbish} Let $\bM$ be a compact
manifold with boundary, let $k\ge 0$, and suppose that $g_0$ is a
conformally compact metric on $M$ such that $g_0=x^{-2}\bg_0$,
with $\bg_0$ having $(W_x^{k+4,\infty},k+4)$--behavior at
$\partial M$. Assume that  \be
K_0=\lambda_0g_0+L_0,\;\;L_0,\lambda_0 \in
W^{k+3,\infty}_{1}(g_0)\;, \; |L_0|_{g_0}\to_{x\to 0}0,\;\;
|\nabla L_0|_{g_0}\to_{x\to 0}0\;. \ee Then for all $s\neq
(n-3)/2, (n-1)/2, (n+1)/2$ and all $(K,g)$ close to $(K_0,g_0)$ in
the $W^{k+3,\infty}_1(g_0)\times W^{k+4,\infty}_1(g_0)$ norm, the
map
$$
\pi_{\maclKzo }{L}_{1,x^{s}} :{\maclKzo }\cap
(\hbord^{-s}_{k+3}(g)\times \hbord^{-s}_{k+4}(g)) \longrightarrow
{\maclKzo }\cap (\hbord^{-s}_{k+1}(g)\times \hbord^{-s}_{k}(g))
$$
is an  isomorphism such that the norm of its inverse does not
depend upon $(K,g)$.
\end{theor}

 \proof We wish to apply Theorem~\ref{Til1projbis}. It follows from the discussion above
  that we only need to establish the inequality \eq{Ych} for $Y$'s and
$N$'s supported outside of a sufficiently a large compact set. For
$s \ne (n-1)/2, (n+1)/2$, Proposition~\ref{Pbch} yields \be
\|Y\|_{\hbord^{-s}_0} \le C\|S(Y)\|_{\hbord ^{-s}_0}\;. \ee
Applying Proposition~\ref{prop:estiN} we find that for $s\neq
(n-3)/2, (n-1)/2, (n+1)/2$ it holds that \be \|N\|_{\hbord
^{-s}_1} \le C \|-\Delta N g+\nabla\nabla N-N \Ricc(g)\|_{\hbord
^{-s}_0}\;, \ee which is what had to be established.  \qed

A proof identical to that of Theorem~\ref{theor:abprojbis} yields:

\begin{theor}\label{theor:abprojbish}
Under the hypotheses of Theorem~\ref{Til1abprojbish} with $s\geq 0
$, $s\neq (n-3)/2, (n-1)/2,  (n+1)/2$, $k>n/2$, the map
 \bel{abbijh}
\begin{array}{c}
 {{\maclKz   }^{\bot_g}}\cap (\hbord^{-s}_{k+3}(g)\times  \hbord^{-s}_{k+4}(g))
\longrightarrow
{{\maclKz   }^{\bot_g}}\cap (\hbord^{-s}_{k+1}(g)\times \hbord^{-s}_{k}(g))\\
(Y,N) \longmapsto \pi_{{\maclKz }^{\bot_g}}x^{{-2s}}\left\{\left(
\begin{array}{c}
J\\
\rho
\end{array}
\right)[(K,g)+x^{{2s}} P^*(Y,N)] -\left(
\begin{array}{c} J\\
\rho
\end{array}
\right)(K,g)\right\}
\end{array}
\ee is bijective in a neighborhood of zero. Thus, there exists
$\epsilon>0$ such that for all $(K,g)$ close to $(K_0,g_0)$ in
$W^{k+3,\infty}_1(g_0)\times W^{k+4,\infty}_1(g_0)$, and for all
pairs $(\delta J, \delta\rho)\in \hbord^{{s}}_{k+1}(g)\times
\hbord^{{s}}_{k}(g)$ with norm less than $\epsilon$,
 there exists a solution $(\delta
K,\delta g)=x^{{2s}} P^*(Y,N) \in \hbord^{{s}}_{k+2}(g)\times
\hbord^{{s}}_{k+2}(g))$, close to zero, of the equation
\begin{eqnarray}\nn \pi_{{\maclKz
}^{\bot_g}}x^{{-2s}} \left\{\left(
\begin{array}{c}
J\\
\rho
\end{array}
\right) (K+\delta K,g+\delta g) - \left(
\begin{array}{c}
J\\
\rho
\end{array}
\right) (K,g) \right\} \\= \pi_{{\maclKz
}^{\bot_g}}x^{{-2s}}\left(
\begin{array}{c}
\delta J\\
\delta \rho
\end{array}
\right) \;. \label{abbij1h}
\end{eqnarray}

\end{theor}

\begin{remark}
For metrics which are sufficiently regular at the conformal
boundary it can be shown that any non-trivial solution of the
equation $P^*(Y,N)=0$ satisfies $|N|+|Y|_g\sim 1/x$ near $x=0$.
This shows that if $0\le s < (n+1)/2$, then $\maclKz =\{0\}$, so
that no projection operator is necessary in \eq{abbij1h}.
\end{remark}

 \proof  The conditions $s\geq 0$ and $k>n/2$ ensure
that the map of Equation~\eq{abbijh} is well defined and
differentiable in a neighborhood of zero; a relatively
straightforward though lengthy check of that can be done using
weighted Moser inequalities (see~\cite{ChLengardnwe} for proofs in
a slightly different context; the arguments there adapt to the
current setting in a straightforward way). The solvability of the
equation \ref{abbij1h} follows then from
Theorem~\ref{theor:projbis}. \qed

A proof identical to that of Proposition~\ref{p6.7} gives
solutions with H\"older regularity:

\begin{prop}\label{reguh}Let $\bM$ be a compact
manifold with boundary, let $k\ge k_0:=\lfloor n/2\rfloor +1$ (the
smallest integer strictly larger than $n/2$), $\alpha\in (0,1)$,
and suppose that $g_0$ is a conformally compact metric on $M$ such
that $g_0=x^{-2}\bg_0$, with $\bg_0$ having
$(\chyp^0_{k+4,\alpha},k+5)$--behavior at $\partial M$. Assume
that  \be K_0=\lambda_0g_0+L_0,\;\;L_0,\lambda_0 \in
\chyp^{0}_{k+3,\alpha}(g_0)\;, \; |L_0|_{g_0}\to_{x\to 0}0,\;\;
|\nabla L_0|_{g_0}\to_{x\to 0}0\;. \ee Then for all $t\geq 0$,
$t\neq (n-3)/2, (n-1)/2, (n+1)/2$
 there exists $\epsilon>0$ such that if $(K,g)$ in
$\chyp^{0}_{k+3,\alpha}(g_0)\times\chyp^{0}_{k+4,\alpha}(g_0)$,
and if
$$\|(K-K_0,g-g_0)\|_{W^{k_0+3,\infty}_1(g_0)\times
W^{k_0+4,\infty}_1(g_0)} < \epsilon
$$
$$\|(\delta J,\delta \rho)\|_{
\chyp^{t}_{k_0+1,\alpha}(g)\times\chyp^{t}_{k_0,\alpha}(g)} +
\|(\delta J, \delta\rho)\|_{\hbord^{t}_{k_0+1}(g)\times
\hbord^{t}_{k_0}(g)}<\epsilon\;,$$  then the solution $(\delta K,
\delta g)$ given by Theorem~\ref{theor:abprojbish} (with $s=t$) is
in $
\chyp^{t}_{{k_0}+2,\alpha}(g)\times\chyp^{t}_{{k_0}+2,\alpha}(g)$.
If moreover $(\delta J,\delta \rho)\in{
\chyp^{t}_{k+1,\alpha}(g)\times\chyp^{t}_{k,\alpha}(g)}$ then the
solution given by Theorem~\ref{theor:abprojbish} is in $
\chyp^{t}_{{k}+2,\alpha}(g)\times\chyp^{t}_{{k}+2,\alpha}(g)$.

\qed
\end{prop}

One has very similar results in exponentially weighted H\"older
and Sobolev spaces as at the end of Section~\ref{Scb}, the details
are left to the reader.

\section{Asymptotically flat initial data}\label{Safm}
\begin{Definition}\label{Dwaf}
Let $W$ be a space of symmetric tensors on $\R^n\setminus
\overline{B(R_0)}$, $R_0\ge 1$, where $B(R_0)$ is an open
coordinate ball of radius $R_0$ in $\R^n$. We shall say that
$(M,g)$ is \emph{$W$--asymptotically flat}if there exists a set
$\mcK\subset M$ and a diffeomorphism $\chi^{-1}:M\setminus \mcK
\to \R^n\setminus \overline{B(R_0)}$ such that
$$(\chi^*g)_{ij}-\delta_{ij} \in W\;.$$
The region $M_{\ext}:= M\setminus \mcK$ will be called an end of
$M$.  $M$ will be said to have  \emph{compact interior} if $\mcK$
is compact. \label{defaf}
\end{Definition}
In the above definition we have assumed that $M$ has only one end,
there is an obvious natural generalisation of the above notion to
any finite number of ends; the results below generalise without
any difficulties to such cases. We emphasise that in
Definition~\ref{defaf} the manifold $M$ is allowed to have a
compact boundary.
We will often use the symbol $r$ to denote a function $f$ on $M$
such that $f\circ \chi$ coincides with  the radius $r$ on
$\R^n\setminus B(R)$ for $R$ large enough. The requirement $R_0\ge
1$ has been made for notational convenience, to guarantee that the
function $r$, which will be used as a weight on $M_{\ext}$, is
strictly positive there.

 The simplest choice for the $W$ spaces above are the
$C_{k}^{\alpha}$ H\"older spaces, defined as the spaces of
functions such that \be
\|f\|_{C^\alpha_k}:=\|(1+r^2)^{-\alpha/2}f\|_{C^0_k} <
\infty\;,\label{Cadef}\ee where $\|\cdot\|_{C^0_k}$ is the  sum of
sup norms of $f$ and its derivatives up to order $k$, with each
derivative entering with a supplementary factor of $r$. If
$$f=f^{i_1\ldots i_k}{}_{j_1\ldots j_\ell}\partial_{i_1}\otimes \ldots
\otimes\partial_{i_k}\otimes dx^{j_1}\otimes \ldots dx^{j_\ell}$$
is a tensor field, then \eq{Cadef} should be used for each entry
$f^{i_1\ldots i_k}{}_{j_1\ldots j_\ell}$ of $f$, with respect to
the natural coordinates $x^i$ on $\R^n$, and a sum over the
$i_p$'s and $j_q$'s of the norms $ \|f^{i_1\ldots
i_k}{}_{j_1\ldots j_\ell}\|_{C^\alpha_k}$ should be made.

Somewhat sharper results can be obtained when working with
manifolds for which $W$ is a weighted Sobolev space. We shall say
that $f\in r^\alpha W^{k,\infty}_r$ if $(1+r^2)^{-\alpha/2}f\in
W^{k,\infty}_r$. This is equivalent to the requirement that $f\in
C^{\alpha}_{k-1}$, and that the distributional $k$'th derivatives
of $f$ satisfy a weighted Lipschitz condition. The metrics,
solutions of the constraint equations which are obtained by our
methods, are $(r^{-\alpha}W^{k+2,\infty}_r+\zHk
_{r,r^{-\sigma}})$--asymptotically flat with some
$\alpha,\sigma>0$, $k\ge 2$.

It is convenient to relabel the $\zHk _{r,r^{\alpha}}$ spaces as
follows: for $k\in \N$ and $\beta\in \R$ we set \be\label{cHdef}
\zmcH _k^\beta := \zHk _{r,r^{-n/2-\beta}}\;,\ee so that the
$\zmcH _k^\beta$ spaces are the weighted Sobolev spaces of
\eq{defHn} with $\phi=r\circ \chi^{-1}$ and $\psi$ a power of
$r\circ \chi^{-1}$; the labeling here follows~\cite{Bartnik:mass},
and is motivated by the simple inclusions~\cite{Bartnik:mass}
\bel{sobinc} C^{\beta'}_k\subset \zmcH _k^\beta\;, \ \beta'<
\beta\;; \qquad\zmcH _k^\beta \subset C^\beta_{\lfloor k-n/2
\rfloor}\;, \ k>n/2\;.\ee In fact~\cite{Bartnik:mass}
\bel{sobinc2} f\in \zmcH _k^\beta \;, \ k>n/2 \quad
\Longrightarrow \quad f=o(r^\beta) \;.\ee

In order to apply Theorem~\ref{Til1projbis} we need to establish
the inequality \eq{it1kerbis} for tensor fields with compact
support in the asymptotic region, and we will use
Lemma~\ref{lem:equivcondbis} for that. In addition to \eq{metfo},
we assume that the
Hessian $\Hess r$ satisfies, in the preferred coordinates on the
asymptotic region,  \be\label{Hessr} (\Hess r)_{ij}=\frac
1r(\delta_i^j-n_i n_j) +o(r^{-1})\;. \ee (\Eq{Hessr} will clearly
hold without the error term for a flat metric; similarly
\eq{Hessr} will hold
for metrics which  are $C^\alpha_1$-asymptotically flat, for some
$\alpha>0$; we note that \eq{Hessr} implies \eq{metfo1}.) For the
convenience of the reader we restate
Proposition~\ref{prop:estiPS2afg}:

\begin{prop}\label{prop:af1}  Let $S$ be defined
by \Eq{cb1}. Assume that $g\in W^{1,\infty}_\loc$, that
 \be\label{metfo1a}
|g_{ij}-\delta^i_j| \le \epsilon \ \  \mbox{on} \ \ \{r\ge
R_\epsilon\}\;, \ee and that \eq{Hessr} holds. Then for
$s\in\R\setminus\{ 0,1\}$ there exist constants $R=R(s)$ and
$C_{s}$ such that for all differentiable vector fields $Y$
compactly supported outside of a
 ball of radius $R$ we have
\begin{eqnarray}
\int r^{-2s-n} |Y|^2 &\le&C_{s}\int r^{-2s-n+2} |S(Y)|^2 \;.
\label{cb11af}
\end{eqnarray}

\qed
\end{prop}
\begin{remark} The result is sharp, compare the argument in
Remark~\ref{r6.2}.
\end{remark}

 Proposition~\ref{prop:af1} gives, in essence, the inequality
needed in Theorem~\ref{Til1projbis} for $s\not\in\{0,1\}$. This
leads to the following rewording of Theorem~\ref{Til1projbis} in
the asymptotically flat context (there is also an  equivalent of
Theorem~\ref{Til1proj} here, we leave the transcription to the
reader):

\begin{theor}\label{Til1afbis}
Let $g_0$ be $r^{-\alpha}W^{k+4,\infty}_r$--asymptotically flat
for some $\alpha>0$ and $k\in\N$, suppose that
\be\label{Kcondafbis} K_0 \in r^{-1} W^{k+3,\infty}_r\;,\qquad
K_0=o(r^{-1})\;.\ee Then for all $\sigma\in \R\setminus\{0,1\}$
and for all $(K,g)$ close to $(K_0,g_0)$ in
$r^{-1}W^{k+3,\infty}_r(g_0)\times W^{k+4,\infty}_r(g_0)$ norm,
the map
$$
\pi_{\maclKzo }{L}_{r,r^{-n/2-\sigma}} :{\maclKzo }\cap (\zmcH
^{\sigma}_{k+3}(g)\times \zmcH ^{\sigma}_{k+4}(g)) \longrightarrow
{\maclKzo }\cap (\zmcH ^{\sigma}_{k+1}(g)\times \zmcH
^{\sigma}_{k}(g))
$$
is an  isomorphism such that the norm of its inverse does not
depend upon $(K,g)$.
\end{theor}

Before passing to its proof, we note that Theorem~\ref{Til1afbis}
immediately implies:

\begin{cor}
\label{CTil1afbis} Under the hypotheses of
Theorem~\ref{Til1afbis}, the space of linearised fields $(\delta
K, \delta g)\in \zmcH ^{-n-\sigma+1}_{k+1}\times\zmcH
^{-n-\sigma+2}_{k+2}$ splits as a direct sum $\textrm{Ker} P\oplus
B$, with the restriction of the linearisation $P$ of the
constraint map to $B$ being an isomorphism of $B$ and of $${\maclK
}^{\bot_{g}}\cap\left(\zmcH ^{-n-\sigma}_{k+1}\times \zmcH
^{-n-\sigma}_{k}\right) \;.$$ In particular if there are no
solutions $(Y,N)\in \zmcH ^{\sigma}_1\times \zmcH ^{\sigma}_2$ of
the equation $P^*(Y,N)=0$, then the map
$$\zmcH ^{-n-\sigma+1}_{k+2}\times\zmcH ^{-n-\sigma+2}_{k+2} \ni (\delta
K, \delta g) \to (\delta J,\delta \rho):=P(\delta K, \delta g) \in
\zmcH ^{-n-\sigma}_{k+1}\times \zmcH ^{-n-\sigma}_{k}$$ is
surjective.

\qed
\end{cor}

\medskip

\noindent{\sc Proof of Theorem~\ref{Til1afbis}:} We wish to apply
Theorem~\ref{Til1projbis}, in order to do that we need to
establish the inequality \eq{it1kerbis} for $Y$'s and $N$'s
supported outside of a sufficiently a large ball. For $\sigma \ne
0,1$, Proposition~\ref{prop:af1} yields \be \|Y\|_{\zmcH
^{\sigma}_0} \le C\|S(Y)\|_{\zmcH ^{\sigma-1}_0};. \ee Applying
Proposition~\ref{PwPiaf} twice we find that for $\sigma\ne 0,1$ it
holds that \bel{bam3} \|N\|_{\zmcH ^{\sigma}_1} \le C \|\nabla
N\|_{\zmcH ^{\sigma-1}_0}\le C^2 \|\nabla \nabla N\|_{\zmcH
^{\sigma-2}_0}\le C^3 \|\alpha(\nabla \nabla N)\|_{\zmcH
^{\sigma-2}_0}\;, \ee where $\alpha$ is as in \eq{P12}. Now
$(M,g,K,r)$ satisfy the asymptotic condition a) of
Definition~\ref{asymptocond} with
$U_i=(\chi^{-1})^{-1}(B(i)\setminus \overline{B(R_0)})\cup U$,
where $U$ is a relatively compact open neighborhood of $\mcK$
($\chi$, $R_0$ and $\mcK$  as in Definition~\ref{Dwaf}), and
Lemma~\ref{lem:equivcondbis} shows that we can apply
Theorem~\ref{Til1projbis}. \qed

Elements in the kernel of $P^*$ are called \emph{Killing initial
data} (KIDs)~\cite{ChBeig1}. Existence of non-zero KIDs implies
existence of non-zero Killing vectors in the associated vacuum
space-time (see~\cite{ChBeig1,ChBeigKIDs} and references therein).
We have the following corollary of Theorem~\ref{Til1afbis}:

\begin{cor}\label{Cbam1}
Let $g_0$ be $r^{-\alpha}W^{k+4,\infty}_r$--asymptotically flat
for some $\alpha>0$ and $k\in \N$ and suppose that \Eq{Kcondafbis}
holds.
\begin{enumerate}\item
If $\sigma\le 0 $, or if \item $\sigma\not\in\{0,1\}$ and the set
of nontrivial KIDs is empty, \end{enumerate} then the conclusions
of Theorem~\ref{Til1afbis} hold without any projection.

\end{cor}
\proof There are no nontrivial KIDs in  $ \zmcH ^{\sigma}_1\times
\zmcH ^{\sigma}_2$ for $\sigma\le 0$, then $\maclK_0=\{0\}$.
$\phantom{ccccccc}$ \qed

\begin{theor}\label{afprojbis}
Let $g_0$ be $r^{-\alpha}W^{k+4,\infty}_r$--asymptotically flat
for some $\alpha>0$ and $k> n/2$, suppose that
\be\label{Kcondafbisbis} K_0 \in r^{-1} W^{k+3,\infty}_r\;,\qquad
K_0=o(r^{-1})\;.\ee Then for all $\sigma\ge 2-n$,
$\sigma\not\in\{0,1\}$
 the nonlinear map
\be\label{afbij}
\begin{array}{c}
{\maclKzo }\cap (\zmcH ^{\sigma}_{k+3}(g)\times \zmcH
^{\sigma}_{k+4}(g)) \longrightarrow {\maclKzo }\cap
(\zmcH ^{\sigma}_{k+1}(g)\times \zmcH ^{\sigma}_{k}(g))\\
(Y,N) \longmapsto \pi_{\maclKzo }r^{n+2\sigma}\left\{\left(
\begin{array}{c}
J\\
\rho
\end{array}
\right)[(K,g)+r^{-n-2\sigma}\Phi^2 P^*(Y,N)] -\left(
\begin{array}{c} J\\
\rho
\end{array}
\right)(K,g)\right\}
\end{array}
\ee is bijective in a neighborhood of zero. Thus,  there exists
$\epsilon>0$ such that for all $(K,g)$ close to $(K_0,g_0)$ in
$r^{-1}W^{k+3,\infty}_r(g_0)\times W^{k+4,\infty}_r(g_0)$, and for
all pairs $(\delta J, \delta\rho)\in
\zmcH^{-n-\sigma}_{k+1}(g)\times \zmcH^{-n-\sigma}_{k}(g)$
 with norm
less than $\epsilon$,
 there exists a solution $(\delta
K,\delta g)=r^{-n-2\sigma}\Phi^2 P^*(Y,N)
\in\zmcH^{-n-\sigma+1}_{k+2}(g)\times
\zmcH^{-n-\sigma+2}_{k+2}(g))$, close to zero, of the equation
\bel{8.13} \pi_{\maclKzo }r^{n+2\sigma} \left\{\left(
\begin{array}{c}
J\\
\rho
\end{array}
\right) (K+\delta K,g+\delta g) - \left(
\begin{array}{c}
J\\
\rho
\end{array}
\right) (K,g) \right\}= \pi_{\maclKzo }r^{n+2\sigma}\left(
\begin{array}{c}
\delta J\\
\delta \rho
\end{array}
\right) \ee
\end{theor}
\proof  The conditions $\sigma\geq 2-n$ and $k>n/2$ ensure that
the map of Equation~\eq{afbij} is well defined and differentiable
in a neighborhood of zero. The result follows then from
Theorem~\ref{theor:projbis}.\qed

Clearly, the projection operator  in \eq{8.13} is not needed when
the hypotheses of Corollary~\ref{Cbam1} are satisfied.

 We also have solutions with H\"older
regularity:

\begin{prop}\label{reguaf}
Let $g_0$ be $C^{-\alpha}_{k+4,\beta}$--asymptotically flat for
some $\alpha>0$, $\beta \in (0,1)$, and $k\ge k_0:=\lfloor
n/2\rfloor +1$ (the smallest integer strictly larger than $n/2$),
suppose that
\be K_0 \in C^{-1}_{k+3,\beta}\;,\qquad K_0=o(r^{-1})\;.\ee Then for all
$t\ge 2$, $t\not\in\{n,n+1\}$ there exists $\epsilon>0$ such that
if $(K,g)$ in $C^{-1}_{k+3,\beta}(g_0)\times
C^0_{k+4,\beta}(g_0)$, and if
$$\|(K-K_0,g-g_0)\|_{r^{-1}W^{k_0+3,\infty}_r(g_0)\times
W^{k_0+4,\infty}_r(g_0)} < \epsilon
$$
$$\|(\delta J,\delta \rho)\|_{C^{-t}_{k_0+1,\beta}(g)\times C^{-t}_{k_0,\beta}(g)}
+ \|(\delta J, \delta\rho)\|_{\zmcH^{-t}_{k_0+1}(g)\times
\zmcH^{-t}_{k_0}(g)}<\epsilon$$ then the solution $(\delta K,
\delta g)$ given by Theorem~\ref{afprojbis} (with $\sigma=t-n$) is
in $C^{-t+1}_{k_0+2,\beta}(g)\times C^{-t+2}_{k_0+2,\beta}(g) $.
If moreover $(\delta J,\delta \rho) \in
C^{-t}_{k+1,\beta}(g)\times C^{-t}_{k,\beta}(g)$ then the solution
given by Theorem~\ref{afprojbis} is in $
C^{-t+1}_{k+2,\beta}(g)\times C^{-t+2}_{k+2,\beta}(g)$.
\end{prop}

\proof Under the current hypotheses all the conditions of
Proposition~\ref{regularitebis} with $k$ there equal to $k_0$ are
met. (The weighted elliptic regularity condition of
Definition~\ref{def:regul} is satisfied by the
calculation~\eq{weighcalc}, Appendix~\ref{Sscaling}.) The higher
H\"older regularity follows from Proposition~\ref{regularite3}.
\qed

\begin{Remark}
One has very similar results in exponentially weighted H\"older
and Sobolev spaces as at the end of Section~\ref{Scb}, the details
are left to the reader. However, the exponentially weighted
conditions seem difficult to verify in the current case, unless
one is in a setting where the results of Section~\ref{Scb} can be
applied. In such a case sharper results are obtained by using the
theorems of that section.\end{Remark}

\section{Applications}

In this section we will give several applications of the general
results proved so far. It should be clear that a key role in this
approach is played by the kernel of $P^*$. As already mentioned in
the previous section, elements of this kernel will be called
Killing Initial Data (KIDs). Thus, a KID is a pair $(Y,N)$ such
that
$$P^*(Y,N)=0.$$

Our first application of the techniques developed so far concerns
the construction of initial data which are \emph{exactly Kerrian}
outside of a compact set:

\subsection{Space-times that are Kerrian near $i^0$}\label{SSKio}

(A version of) the following result has been announced in
\cite{CorvinoOberwolfach}; we assume that the initial data
manifold is three-dimensional:

\begin{theor}\label{Tkerrio}
Let $g$ be $r^{-\alpha}W^{k+4,\infty}_r$--asymptotically flat for
some $\alpha>1/2$ and $2\le k\in \N$, with $ K \in r^{-\alpha-1}
W^{k+3,\infty}_r$, and suppose that $(K,g)$ satisfies the vacuum
constraint equations. We further assume that $(K,g)$ satisfy the
$3+1$ equivalent of the parity conditions \eq{af4}, \bel{af43+1}
|g_{ij}^-|+r|\partial_k (g_{ij}^-)|+r|K_{ij}^-|\le C
(1+r)^{-\alpha_-}\;, \quad \alpha_->\alpha\;,\ \alpha+\alpha_->2
\;, \ee so that all Poincar\'e charges of $(K,g)$ are finite and
well defined, with the ADM four-momentum being timelike. Then
there exists $R_1<\infty$ such that for all $R\ge R_1$ there
exists an initial data set
$$(\hat K_R,\hat g_R)\in  C^{k+2}\times C^{k+2}
$$ satisfying the vacuum constraint
equations everywhere
such that $(\hat K_R,\hat g_R)$ coincides with $(K,g)$ for $ r\le
R$, and $(\hat K_R,\hat g_R)$ coincides with initial data for some
Kerr metric for $r\ge 4R$. If $K$ and $g$ are smooth, then $(\hat
K_R,\hat g_R)$ is smooth.
\end{theor}
\begin{remark}
A family  of $(n+1)$--dimensional generalisations of the Kerr
metric has been found by Myers and Perry~\cite{MyersPerry}, we
expect that it can be used to prove a corresponding result in
$(n+1)$ dimensions. The argument below carries over to any
dimension, the only element missing is an equivalent of
Proposition~\ref{Ppcm} for the family of translated, rotated, and
boosted Myers-Perry metrics.
\end{remark}
\begin{remark}
The factor $4$ has been chosen for definiteness; an identical
result holds with $(\hat K_R,\hat g_R)$ being Kerr for $r\ge
\lambda R$ for any constant $\lambda
>1$.
\end{remark}
\begin{remark}
The ADM momentum and angular momentum of $(\hat K_R,\hat g_R)$
converges to that of $(K,g)$ as $R$ tends to infinity.
\end{remark}
 \proof The required initial data will be constructed by a gluing procedure,  on an annulus,
 using a method due to  Corvino~\cite{Corvino,CorvinoOberwolfach},
 together with
Theorem~\ref{Tremov}. Let $e_{(i)}$, $i=1,\ldots,10$, be any basis
of the space of KIDs for Minkowski space-time, let $Q_{(i)}$
denote the Hamiltonian charge $Q(Y_{(i)},N_{(i)},K,g)$ of $(K,g)$
as given by \eq{mi3a1} with $(Y_{(i)},N_{(i)})=e_{(i)}$. Let, as
in Appendix~\ref{Skn}, $\cKi$ denote the set of initial data for
boosted, rotated, and translated Kerr metrics, and let $(K_Q,g_Q)$
denote an initial data set in $\cKi$ with Hamiltonian charge
$Q=(Q_{(i)})\in\R^{10}$. For $R\in[R_0,\infty)$ let the scale-down
map $\Phi_R$ be defined as
\be\begin{array}{rcl}\Phi_R:\Gamma(R,4R):=\overline{B(0,4R)\setminus
B(0,R)}&\to& \Gamma(1,4)\;,\\
x & \mapsto & x/R\;.
\end{array}\label{blowdown}
\ee Let $\chi\in C^\infty(\R^3)$ be a spherically symmetric
cut-off function such that $0\le \chi \le 1$, $\chi\equiv 1 $ on
$\Gamma(1,2)$, and $\chi\equiv 0$ on $\Gamma (3,4)$. On
$\Gamma(1,4)$ set
$$g_{Q,R}= R^{-2}\left( \chi \Phi^*_Rg + (1-\chi) \Phi^*_R
g_Q\right)\;,$$
$$K_{Q,R}= R^{-1}\left( \chi \Phi^*_R K+ (1-\chi) \Phi^*_R
K_Q\right)\;.$$ Then the $g_{Q,R}$'s form a family of metrics that
converge,  as $R$ tends to infinity, in weighted Sobolev
topologies with arbitrary decay (at the boundary) index $t>1$, to
the Euclidean metric $g_0$ on $\Gamma(1,4)$, while the $K_{Q,R}$'s
converge to $K_0\equiv 0$: \bel{CS1}
\|g_{Q,R}-g_0\|_{\hbord^t_{k+2}}+ \|K_{Q,R}\|_{\hbord^{t-1}_{k+2}}
\le C_{s,k,Q}R^{-\beta}\;,\qquad
\beta:=\min{(\alpha,1)}
\ee (recall that the weighted Sobolev spaces used here have been
defined in \eq{??}). Further the convergence is uniform in $Q$ on
any compact set of $Q$'s. We shall write $x$ for $(K,g)$, $x_Q$
for $(K_Q,g_Q)$, \emph{etc.} We choose  the index $t$ to be larger
than or equal to $k+5$ --- say $t=k+6$, in particular initial data
in the space $\hbord^{t-1}_{k+2}\times \hbord^t_{k+2}$ vanish on
$\partial \Gamma(1,4)$, as well as their first derivatives.
Further, this ensures  the continuous embedding
$\hbound^{t-1}_{k+2}\subset H^{k+2}$.

It follows from \eq{CS1} that for $R$ sufficiently large we have
\bel{CS2}
\|J(x_{Q,R})\|_{\hbord^{t-2}_{k+1}}+\|\rho(x_{Q,R})\|_{\hbord^{t-2}_{k}}\le
CR^{-\beta}\;,\ee and Theorem~\ref{theor:abprojbis} with $s=t-1$
provides a solution $\delta x_{Q,R}\in \hbord^{t-1}_{k+2}\times
\hbord^{t}_{k+2}$ of \Eq{abbij1} satisfying \bel{CS3}\|\delta
x_{Q,R}\|_{ \hbord^{t-1}_{k+2}\times \hbord^{t}_{k+2}}\le C
R^{-\beta}\;.\ee
 Set $\delta J=J(x_{Q,R}+\delta x_{Q,R})$, $\delta \rho=\rho(x_{Q,R}+\delta x_{Q,R})$,
 let the parameter $\lambda$ in Theorem~\ref{Tremov} be
equal to $R$, and consider the map $F_\lambda$  defined in
\eq{coc}. We have
\begin{eqnarray*}
\langle e_{(i)},\psi^{-2} \left(
\begin{array}{c}
\delta J\\
\delta \rho
\end{array}\right)\rangle_{L^2_\psi\oplus L^2_\psi} & = &\langle (Y_{(i)},N_{(i)}),\psi^{-2} \left(
\begin{array}{c}
\delta J\\
\delta \rho
\end{array}\right)\rangle_{L^2_\psi\oplus L^2_\psi}
\\ & = &\int_{\Gamma(1,4)} \left(Y_{(i)}^j \delta J_j + N_{(i)}
\delta \rho \right)\, d\mu_g\;.
\end{eqnarray*}
Note that the weight factor $\psi$ has vanished from the last
integral. By \eq{C4a} we have
\begin{eqnarray}\nonumber
\langle e_{(i)},\psi^{-2} \left(
\begin{array}{c}
\delta J\\\nonumber \delta \rho
\end{array}\right)\rangle_{L^2_\psi\oplus L^2_\psi} & = &
\int_{\{r=4\}} \ourU^{\alpha\beta}(x_{Q,R}+\delta x_{Q,R})
dS_{\alpha\beta} 
\\ && 
-\int_{\{r=1\}}
 \ourU^{\alpha\beta}(x_{Q,R}+\delta x_{Q,R})dS_{\alpha\beta} +O(R^{-2\beta})
 \;,
\nonumber
\\ &&\label{C4.1.0}
\end{eqnarray}
with the error term $O(R^{-2\beta})$ being uniform in $Q$ whenever
$Q$ ranges over a compact set. Now, on $S(0,1)$ all the initial
data considered coincide --- up to a rescaling --- with $(g,K)$
together with their first derivatives, so that by \eq{C7} we have
$$\int_{\{r=1\}}
 \ourU^{\alpha\beta}dS_{\alpha\beta}= R^{-1}\left\{Q^0_{(i)}
 +O(R^{-\delta})\right\}\;,$$
 where $Q^0_{(i)}$ denotes the Hamiltonian charge of $(K,g)$, for some $\delta >0$.
 Similarly, on $S(0,4)$ all the initial
data considered coincide (up to scale-down) with $(K_Q,g_Q)$
together with their first derivatives, so that
$$\int_{\{r=4\}}
 \ourU^{\alpha\beta}dS_{\alpha\beta}=R^{-1}\left\{ Q_{(i)}
 +O((4R)^{-\delta})\right\}\;,$$
 It follows that
\begin{eqnarray}
\langle e_{(i)},\psi^{-2} \left(
\begin{array}{c}
\delta J\\
\delta \rho
\end{array}\right)\rangle_{L^2_\psi\oplus L^2_\psi} & = &R^{-1}\left\{Q_{(i)}-Q^0_{(i)}
+O(R^{-\min(\delta\;,2\beta-1)})\right\}\;. \nonumber \\ &&
\label{C4.1}
\end{eqnarray}
This implies that, up to an additive constant, the maps $\lambda
F_\lambda$ converge as $\lambda=R$ tends to infinity  to the map
$Q$ of Proposition~\ref{Ppcm}; that last map is a homeomorphism,
as desired. The conclusion is obtained now from
Theorem~\ref{Tremov} by taking $V\subset \R^{10}$ to be a ball
around $(Q^0_{(i)})$ of a radius small enough so that $V$ is
included in the image of the map $Q$ of Proposition~\ref{Ppcm},
with $U=Q^{-1}(V)$.

Finally, if $g$ and $K$ are smooth, then smooth solutions can be
obtained by using the exponentially weighted spaces of
Proposition~\ref{p6.10}, compare Corollary~\ref{C6.11}. In the
construction above one should choose the cut-off function $\chi$
to be constant in a neighborhood of the boundary of the annulus
$\Gamma(1,4)$. \qed

\subsection{Gluing asymptotically flat initial data sets}
\label{Sgafids}

The gluing technique used in  Section~\ref{SSKio} does apply to
much more general situations, as follows: a family of vacuum
initial data $\{(K_\omega,g_\omega)\}_{\omega\in\Omega}$ will be
called a \emph{reference family} if the following holds:
\begin{enumerate}
\item There exists $R>0$ such that all the data sets
$(K_\omega,g_\omega)$ are defined on $\R^3\setminus B(0,R)$. \item
The metrics $g_\omega$ are
$r^{-\alpha}W^{k+4,\infty}_r$--asymptotically flat for some
$\omega$-independent constant $\alpha>1/2$ and $2\le k\in \N$,
with $ K_\omega \in r^{-\alpha-1} W^{k+3,\infty}_r$, and with the
norms in those spaces being bounded independently of
$\omega\in\Omega$. \item The parity conditions \eq{af43+1} hold
with $g$ and $K$ there replaced by $g_\omega$ and $K_\omega$, for
some $\omega$-independent constants $\alpha_-$ and $C$. \item The
map which to $(K_\omega,g_\omega)$ assigns its Poincar\'e charges
$(p^\mu(\omega),J^{\mu\nu}(\omega))$ is a diffeomorphism between
$\Omega$ and an open subset of $\R^{10}$: \bel{mcuom}
\mcU_\Omega:=\{(p^\mu(\omega),J^{\mu\nu}(\omega))\}_{\omega\in\Omega}\subset
\R^4\times \R^6\;.\ee
\end{enumerate}

There is an obvious equivalent of the definition above in the time
symmetric context: in this case one assumes that $K_\omega\equiv
0$, and one requires $\mcU_\Omega$ to be an open subset of $\R^4$,
with one parameter corresponding to mass, and three parameters
corresponding to the centre of mass.

It is proved in Appendix~\ref{Sgc} that  the collection of initial
data obtained from boosted and space-translated Kerr space-times
provides an example of a reference family. More generally,
consider any one parameter family of vacuum initial data sets
$(K_\lambda,g_\lambda)$  on $\R^3\setminus B(0,R)$,
$\lambda\in(-\epsilon,\epsilon)$, which satisfies the decay and
parity conditions of the definition of a reference family. For
definiteness we shall suppose that
 the ADM four-momentum $p^\mu(\lambda)$ of $(K_\lambda,g_\lambda)$ is a $\lambda$-independent
  timelike future pointing vector,
 and that the length squared $J^{ij}(\lambda)J_{ij}(\lambda)$ of the ADM angular
 momentum $J^{ij}(\lambda)$ of $(K_\lambda,g_\lambda)$ varies smoothly in some open interval as $\lambda$ changes.
 Then scaling ($x^i \to a x^i$, $g\to a^{-2}g$ for $a\in\R^+$), translating, and rotating
  the initial data, and boosting the initial
 data hypersurface in the associated maximal globally hyperbolic development, leads to a reference family
 such that the associated Poincar\'e charges form a neighborhood of
 $(p^\mu(0),J^{\mu\nu}(0))$;
 this follows from the boost theorem~\cite{christodoulou:murchadha,christodoulou:boost} together with the
 analysis in
 Appendix~\ref{Sgc}
  (here one needs to apply the boost theorem first to the full metric, and then to its odd part).

Similarly, in the time symmetric context, an example of a
reference family is provided by translated Schwarzschild initial
data. More generally, if $g$ is a scalar flat metric on
$\R^3\setminus B(0,R)$ which satisfies the decay and parity
conditions of the definition of a reference family and which has
non-vanishing ADM mass, then scaling and  translating provides a
time-symmetric reference family.

 A repetition of the proof of Theorem~\ref{Tkerrio}
gives:

\begin{theorem}\label{Tkerr2}
Under the hypotheses of Theorem~\ref{Tkerrio}, consider any
reference family $\{(K_\omega,g_\omega)\}_{\omega\in\Omega}$ such
that the associated set $\mcU_\Omega$ defined by \eq{mcuom} forms
a neighborhood of the Poincar\'e charge $(p^\mu,J^{\mu\nu})$ of
$(K,g)$. Then the conclusion of Theorem~\ref{Tkerrio} holds with
the new initial data set  coinciding in the asymptotic region with
one of the members of the reference family rather than with one of
the members of the Kerr family.

\qed
\end{theorem}

The point of Theorem~\ref{Tkerr2} is that it provides large
families of initial data with well controlled asymptotic behavior.
As an example of application, let the reference family consist of
stationary metrics. Such metrics have well understood asymptotic
behavior ({\em cf.,\/ e.g.,\/}~\cite{SimonBeig}), and large
families of non-trivial solutions (defined and smooth outside of a
compact set) have been constructed
in~\cite{Reula:static,Schaudt,Klenk}. Theorem~\ref{Tkerr2} allows
one to modify an arbitrary initial data set in the asymptotic
region so that it coincides with exactly stationary, but not
necessarily Kerrian, data there. Further, there is a rather large
freedom available. Now, a significant result of Dain, Damour and
Schmidt~\cite{Dain:2001kn,Damour:schmidt} implies that the
resulting vacuum space-time will have a smooth $\scrip$ complete
to the past. Thus, initial data so constructed do have reasonably
well controlled maximal globally hyperbolic developments. We will
see in Section~\ref{Slge} below how to construct initial data that
produce space-times with a complete smooth $\scri$, by using a
variation of the technique above.

Another possibility is to choose as the reference family
appropriate subsets of the set of ``almost stationary" metrics
constructed in Section~\ref{Shao} below, see Theorem~\ref{Tae1}.
The metrics there are stationary (or static, in the time-symmetric
case) to an order as high as desired in an asymptotic expansion,
without being exactly stationary outside of a compact set, which
further increases the freedom available. One expects
(compare~\cite{Friedrich:tuebingen}) that some of those metrics
will also admit complete, or past-complete, conformal completions
with a reasonably high degree of differentiability, but no
rigorous statements of this kind are known so far.

All the constructions described so far can be repeated by
specialising to the time-symmetric case, setting
$$Y\equiv K\equiv 0 $$
throughout. In this context Theorem~\ref{Tkerr2} can be rephrased
as:

\begin{theor}\label{Tkerriots}
Let $g$ be $r^{-\alpha}W^{k+4,\infty}_r$--asymptotically flat for
some $\alpha>1/2$ and $2\le k\in \N$, , and suppose that
$$R(g)=0.$$ We further
assume that $g$ satisfies  the parity conditions
\bel{af43+1ts} |g_{ij}^-|+r|\partial_k (g_{ij}^-)|\le C
(1+r)^{-\alpha_-}\;, \quad \alpha_->\alpha\;,\ \alpha+\alpha_->2
\;, \ee so that the mass $m$ and the centre of mass $\vec c$ of
$g$ are finite and well defined, with $m\ne 0$. Consider any
time-symmetric reference family $\{(K_\omega\equiv
0,g_\omega)\}_{\omega\in\Omega}$ such that the associated set
$\mcU_\Omega$ of masses and centres of mass forms a neighborhood
of $(m,\vec c)$. Then there exists $R_1<\infty$ such that for all
$R\ge R_1$ there exists a scalar flat metric
$$\hat g_R \in  C^{k+2}
$$  such that $\hat g_R$ coincides
with $g$ for $ r\le R$, and $\hat g_R$ coincides with a member of
the reference family for $r\ge 4R$. If  $g$ is smooth, and the
$g_\omega$'s are smooth, then $\hat g_R$ can be chosen to be
smooth.

\qed
\end{theor}

\subsection{Initial data which are stationary to high asymptotic order}
\label{Shao}

The results in Section~\ref{Safm} can be used to construct large
classes of asymptotically flat vacuum initial data sets with
controlled asymptotic behavior. As an illustration, let $(K,g)$ be
a solution of the stationary constraint equations defined on
$\R^3\setminus B(0,R)$ for some $R$. Recall that such solutions
are uniquely determined~\cite{BeigSimon2} by an infinite
collection of \emph{Hansen multipole moments}
$\{\mcP_m(K,g)\}_{m\in\N}$, as defined
in~\cite[Equation~(3.5)]{SimonBeig}, see also~\cite{Hansen}. We
will assume that the reader is familiar with~\cite{SimonBeig} and
we will use notation from there. In that reference it has been
shown how to construct approximate solutions of the stationary
equations to any asymptotic order: given any  set $\mcP_m$ one can
find functions $\Phi_M^{(m)}$, $\Phi_S^{(m)}$, $\Phi_K^{(m)}$,
$\gamma^{(m)}$ which satisfy the reduced Einstein equations
\cite[Equations~(3.1)--(3.4)]{SimonBeig} to order $O(r^{-(m+3)})$.
We wish, first, to show that this implies existence of initial
data $(K(\mcP_m),g(\mcP_m))$ satisfying the stationary --- or
static
--- Einstein equations up to terms $O(r^{-m-3})$, provided that
the NUT charge vanishes; in particular this will imply, in the
notation of~\cite{SimonBeig},
\be \label{ea1} \left(
\begin{array}{c}
J\\
  \\
\rho
\end{array}
\right) (K(\mcP_m),g(\mcP_m))= \left(
\begin{array}{l}
O^\infty(r^{-m-3})\\
  \\
O^\infty(r^{-m-3})
\end{array}
\right) \in \zmcH _k^{-(m-\epsilon) -3}\;,\ \epsilon> 0\;, \ k\in
\N\;. \ee We will say that the moments are static to order $m$ if
the associated twist function $\omega^{(m)}$ vanishes. In this
last case the proof of \eq{ea1} is straightforward: we set
$K(\mcP_m):=0$, and \eq{ea1} follows immediately from the
equations in \cite{SimonBeig} with $\sigma=\omega=0$ there. In the
general case some more work is required, we start with a lemma:

\begin{lem}[Approximate Poincar\'e Lemma]\label{lempoinc}
Let $\nu=Ady\wedge dz+Bdz\wedge dx+Cdx\wedge dy$ be a two form on
$\R^3$, with coefficients of order $o(r^{-2})$,   such that
$$d\nu=(\partial_xA+\partial_yB+\partial_zC)dx\wedge dy\wedge dz=O(r^{-(m+3)})dx\wedge dy\wedge dz\;,$$
with $m> 0$. Then there exists a one form $\sigma$ such that
$$d\sigma=\nu+O(r^{-(m+2)}).$$
\end{lem}
\proof  In the proof that follows we use the notation $X=(x,y,z)$
and $r=\sqrt{x^2+y^2+z^2}$. Let us show that the one form $\sigma$
defined by the usual formula
\begin{eqnarray*}
\sigma=-\left\{[\int_1^\infty A(tX)tdt](ydz-zdy)+[\int_1^\infty
B(tX)tdt](zdx-xdz)\right.\hspace{2cm}\\
\left.+[\int_1^\infty C(tX)tdt](xdy-ydx)\right\}
\end{eqnarray*}
satisfies the desired estimate. Indeed, the coefficient of the
term $dx\wedge dy$ in $d\sigma$ is \begin{eqnarray*}
 -\int_1^\infty
\left\{t^2[x\partial_xC(tX)-z\partial_xA(tX)-z\partial_yB(tX)+y\partial_yC(tX)]+2tC(tX)\right\}dt\hspace{1cm}
\\
= -\int_1^\infty
\left\{t^2[x\partial_xC(tX)+y\partial_yC(tX)+z\partial_zC(tX)+zO((tr)^{-(m+3)})]+2tC(tX)\right\}dt
\\
 =-t^2C(tX)\big|_1^\infty-\int_1^\infty
 zt^2O((tr)^{-(m+3)})dt=C(X)+O(r^{-(m+2)}).\hspace{2cm}
 \end{eqnarray*}
A similar calculation for the remaining terms gives
 the result.\qed
\begin{remark}\label{rempoinc}
If $\widetilde{\nu}$ is another two form satisfying conditions of
Lemma~\ref{lempoinc} such that
$\widetilde{\nu}-\nu=O(r^{-(m+2)})$, then if we define
$\widetilde{\sigma}$ as in the preceding proof we will have
$$\widetilde{\sigma}-\sigma=O(r^{-(m+1)}).$$
\end{remark}
 This shows in particular that in a stationary vacuum metric the
approximate solution $\sigma^{(m)}$ as defined below will differ
from the exact one
 by $O(r^{-(m+1)})$.

\medskip

Returning to our construction, let $\Phi_M^{(m)}$, $\Phi_S^{(m)}$,
$\Phi_K^{(m)}$, and $\gamma^{(m)}$ be as in
\cite[Theorem~2]{SimonBeig}, then Beig \& Simon's equations (2.8)
and (2.9) are satisfied modulo $O(r^{-(m+3)})$. In particular
their equation (2.8) with $\Phi=\Phi^{(m)}_K-\Phi^{(m)}_M$ gives
\bel{premeq}
\Delta_{\gamma^{(m)}}(\lambda^{-1})=2\tau^{(m)}\lambda^{-1}+O(r^{-(m+3)})\;,\ee
where $\lambda\equiv\lambda^{(m)}$ is obtained from
$\Phi_M^{(m)}$, $\Phi_S^{(m)}$, and $\Phi_K^{(m)}$ by inverting
\cite[Equation~(2.6)]{SimonBeig} with $\Phi_M$ there replaced with
$\Phi_M^{(m)}$, \emph{etc.}; we define $\omega\equiv\omega^{(m)}$
in a similar way. Using again Beig and Simon's Equation (2.8) with
$\Phi=\Phi^{(m)}_S$ gives
\bel{seceq}
\Delta_{\gamma^{(m)}}(\lambda^{-1}\omega)=2\tau^{(m)}\lambda^{-1}\omega+O(r^{-(m+3)})\;.\ee
Developing \eq{seceq}, and inserting \eq{premeq} in the result one
obtains
$$
\nabla^i(\lambda^{-2}\partial_i\omega)=O(r^{-(m+3)}).$$ Here
$\nabla\equiv\nabla(\gamma^{(m)})$ is the connection of the metric
$\gamma^{(m)}$. Then, if we define $\nu\equiv\nu^{(m)}:=-
\lambda^{-2}*_{\gamma^{(m)}} d\omega$, we have
$$d\nu=O(r^{-(m+3)}).$$
Recall, now, that the coefficient of the power $r^{-1}$ in the
expansion of $\omega$ is proportional to the NUT charge of the
resulting space-time; usual asymptotic flatness forces the
vanishing thereof. From now on we assume that this is the case;
then $\omega=O(r^{-2})$ thus $\nu=O(r^{-3})$ so from
Lemma~\ref{lempoinc} there exists a one form $\sigma^{(m)}$,
solution of \cite[Equation (2.4)]{SimonBeig} modulo
$O(r^{-(m+3)})$ at the right-hand-side. Set
\bel{geemme}
g^{(m)}:=\lambda^{(m)}(dt+\sigma^{(m)}_idx^i)^2-(\lambda^{(m)})^{-1}\gamma^{(m)}_{ij}dx^idx^j\;.\ee
It then follows \emph{e.g.}\/
from~\cite[Section~16.2]{Exactsolutions} that the Ricci tensor of
the stationary space-time metric $g^{(m)}$ has coordinate
components which are $O(r^{-(m+3)})$. By projecting on the initial
data surface $\{t=0\}$ one obtains \eq{ea1}.

In order to continue, note that the collection $\mcP_m$ of
multipole moments up to order $m$ can be viewed as an element of
$\R^{N(m)}$, for some $N(m)$ (the exact value of which is
irrelevant for our purposes); this leads to an obvious way of
measuring the norm of $\mcP_m$.

Next, it should be clear that ten multipole moments out of the
whole set $\mcP_m$ correspond to the global Poincar\'e charges of
the space-time metric. For example, 
the $1/r$   coefficient in the asymptotic expansion of
$\lambda^{(m)}$ is related to the ADM mass of $g^{(m)}$. We denote
by $\mcP^Q$ the relevant multipole moments, and  by $\mcP^*_m$ the
remaining ones, so that $$\mcP_m=(\mcP^Q,\mcP^*_m)\;.$$

 We have the following:

\begin{theorem}\label{Tae1}
Let $m\in \N$ and let $(K_0,g_0)$ be a stationary solution of the
vacuum Einstein equations defined on $\R^3\setminus B(R_0)$ with
timelike ADM momentum and with multipole moments up to order $m$
equal to $\mcP_m:=\mcP_m(K_0,g_0)$. There exists $\eta>0$ such
that for any
$$|\delta \mcP_m^*|<\eta$$ there exists $\delta\mcP^Q$ and a smooth vacuum initial data set $(K,g)$
(not necessarily
stationary)  defined on $\R^3\setminus B(R_0)$ such that
\bel{res1}(K,g)-(K(\mcP_m+\delta
\mcP_m),g(\mcP_m+\delta \mcP_m))\in C^{-m-2+\epsilon}_\infty
\times C^{-m-1+\epsilon}_\infty
\ee for any $\epsilon
>0$. In particular the first $m$ coefficients in an asymptotic
expansion  of $g$ in terms of inverse powers of $r$, and $m+1$
coefficients in that of $K$, coincide with those of the
Simon--Beig approximate solution $(K(\mcP_m+\delta
\mcP_m),g(\mcP_m+\delta \mcP_m))$. An identical result holds in
the class of time-symmetric initial data sets if one restricts
oneself to moments associated to static space-times, provided that
$K_0\equiv 0$.
\end{theorem}

\begin{Remark} The initial data set $(K,g)$ will coincide with
 $(K_0,g_0)$ in a neighborhood of
$S(0,R_0)$. \end{Remark}

\begin{Remark} We emphasise that one is \emph{not} free to choose the Poincar\'e
charges $Q$ of the final initial data set $(K,g)$, those charges
are determined by the original stationary initial data set and by
the $\delta \mcP^*_m$'s in a highly implicit manner. Further, we
will have $Q-Q_0=O(\eta^2)$, where $Q_0$ are the Poincar\'e
charges of $(K_0,g_0)$.
\end{Remark}

\begin{Remark}\label{Rae1} It follows from the calculations of~\cite[Theorem~3]{SimonBeig}, together with
the properties of the weighted spaces in \eq{res1}, that the
``orbit space manifold" $(\hyp,\gamma)$, with $\gamma$ related to
$g$ as in \eq{geemme} with the ``$^{(m)}$'s" removed, admits a
one-point conformal compactification with a $C^{m,\alpha}$
conformally rescaled metric. In particular in the static case
$(\hyp,g)$ has such a compactification.
\end{Remark}

\begin{Remark} Using the Schwarzschild initial data as
$(K_0=0,g_0)$  one obtains a large family of static initial data
with any arbitrarily prescribed finite set of small static
multipole moments, except for the mass which is implicitly
determined by the seed mass and the remaining multipoles. One has
an obvious analogue of this result using the Kerr initial data as
the reference family. Further, using Weyl metrics as $(K_0=0,g_0)$
 one obtains large classes of
time-symmetric initial data sets where the higher order multipoles
are not necessarily small.
\end{Remark}

\proof The proof is essentially identical to that of
Theorem~\ref{Tkerrio}. For definiteness we choose $\epsilon=1/2$,
the proof applies for any $0<\epsilon<1$. Let ${\cal Z}_k^{m-1/2}$
denote a space of functions on $R^3\setminus B(0,R_0+1/2)$ which
are exponentially weighted near the interior boundary
$S(0,R_0+1/2)$ as in Proposition~\ref{p6.10} with $t$ there equal
to one, and which are
 weighted at infinity
as in Theorem~\ref{afprojbis}, with $\sigma$ there equal to
$m-1/2$. (Thus, functions in ${\cal Z}_k^{m-1/2}$ behave as
functions in $\zmcH ^{m-1/2}_k$ for $r$ large.)
 Let $\chi\in C^\infty(\R^3)$ be a spherically symmetric
cut-off function such that $0\le \chi \le 1$, $\chi\equiv 1 $ on
$\Gamma(R_0,R_0+1)$, and $\chi\equiv 0$ on $\R^3\setminus
B(0,R_0+2)$. Choose any $\mcP^Q$ satisfying $$|\mcP^Q|\le \eta$$
and on $\R^3\setminus B(0,R_0)$ set
$$g_{\delta\mcP}=  \chi g_0 + (1-\chi)
g(\mcP+\delta\mcP)\;,$$
$$K_{\delta\mcP}=\chi  K_0+ (1-\chi)
K(\mcP+\delta\mcP)\;.$$  We shall write $x_{\delta\mcP}$ for
$(K_{\delta\mcP},g_{\delta\mcP})$, $x_0$ for $(K_0,g_0)$,
\emph{etc.} Set $\epsilon =1/2$, choose some $k$ large enough so
that the existence and H\"older regularity results proved in the
previous sections apply, it follows from \eq{ea1} that
we
have\newcommand{\hnewm}{{\cal Z}^{-(m-1/2)}}%
\newcommand{\hnewmt}{{\cal Z}^{-(m-1/2+3)}}%
\newcommand{\hnewmtw}{{\cal Z}^{-(m-1/2+2)}}%
\newcommand{\hnewmo}{{\cal Z}^{-(m-1/2+1)}}%
\bel{CS2b}
\|J(x_{\delta\mcP})\|_{\hnewmt_{k+1}}+\|\rho(x_{\delta\mcP})\|_{\hnewmt_{k}}\le
C\eta\;.\ee The arguments given in the proofs of
Theorems~\ref{theor:6.9} and \ref{afprojbis} show that the
hypotheses of Theorem~\ref{theor:projbis} hold, and for $\eta$
small enough we obtain a solution $\delta x_{\delta\mcP}\in
\hnewmtw_{k+2}\times \hnewmo_{k+2}$ of \Eq{abbij1} satisfying
\bel{CS3b}\|\delta x_{\delta\mcP}\|_{ \hnewmtw_{k+2}\times
\hnewmo_{k+2}}\le C \eta\;.\ee
 Set $\delta J=J(x_{\delta\mcP}+\delta x_{\delta\mcP})$, $\delta \rho=\rho(x_{\delta\mcP}+\delta
 x_{\delta\mcP})$.
 By Corollary~\ref{C6.11} $\delta x_{\delta\mcP}$ extends smoothly
 to $\R^3\setminus B(0,R_0)$ when extended by zero.
As in the proof of Theorem~\ref{Tkerrio} we have
\begin{eqnarray*}
\langle e_{(i)},\psi^{-2} \left(
\begin{array}{c}
\delta J\\
\delta \rho
\end{array}\right)\rangle_{L^2_\psi\oplus L^2_\psi} & = &\langle (Y_{(i)},N_{(i)}),\psi^{-2} \left(
\begin{array}{c}
\delta J\\
\delta \rho
\end{array}\right)\rangle_{L^2_\psi\oplus L^2_\psi}
\\ & = &\int_{\R^3 \setminus B(0,R_0)} \left(Y_{(i)}^j \delta J_j + N_{(i)}
\delta \rho \right)\, d\mu_g\;.
\end{eqnarray*} We use the divergence identity \eq{C3-} with $(K_0,g_0)$ as the background (instead of
$(0,\delta)$, as was the case for Theorem~\ref{Tkerrio}).  The
$3+1$ form of this identity reads
\begin{eqnarray}\nonumber
\langle e_{(i)},\psi^{-2} \left(
\begin{array}{c}
\delta J\\\nonumber \delta \rho
\end{array}\right)\rangle_{L^2_\psi\oplus L^2_\psi} & = &
\lim_{R\to \infty} \int_{\{r=R\}}
\ourU^{\alpha\beta}(x_{\delta\mcP}+\delta x_{\delta\mcP})
dS_{\alpha\beta} 
\\ && 
-\int_{\{r=R_0\}}
 \ourU^{\alpha\beta}(x_{\delta\mcP}+\delta x_{\delta\mcP})dS_{\alpha\beta} +O(\eta^2)
 \;.
\nonumber
\\ &&\label{C4.1.0b}
\end{eqnarray} Now, the initial data coincide with $(g_0,K_0)$ in a
neighborhood of $S(0,R_0)$, so that
$$\int_{\{r=R_0\}}
 \ourU^{\alpha\beta}dS_{\alpha\beta}= 0\;.$$
 On the other hand, the limit as $R$ goes to infinity of the
 integral over $S(0,R)$  gives
$$\int_{\{r=\infty\}}
 \ourU^{\alpha\beta}dS_{\alpha\beta}=\delta Q\;,$$ where  $\delta Q$
 is calculated from $\delta \mcP^Q$; we emphasise that $\delta
 x_{\delta \mcP}$ does not give a contribution to this integral
 because of the fast decay.
 It follows that
\begin{eqnarray}
\langle e_{(i)},\psi^{-2} \left(
\begin{array}{c}
\delta J\\
\delta \rho
\end{array}\right)\rangle_{L^2_\psi\oplus L^2_\psi} & = &\delta Q + O(\eta^2)\;. \nonumber \\ &&
\label{C4.1b}
\end{eqnarray}
Let $F(\delta \mcP^Q)$ denote the left-hand-side of \eq{C4.1b}.
Now, the map $\delta \mcP^Q\to \delta Q$ is a linear isomorphism.
Further, it should be clear  that $F$ is a differentiable function
of $\delta \mcP^Q$. The existence of a $\delta \mcP^Q$ such that
$F(\delta \mcP^Q)$ vanishes can thus be inferred, for $\eta$ small
enough, from the inverse function  theorem. Alternatively, set
$$G_\eta(\delta \mcP^Q) = \eta F\left(\frac{\delta \mcP^Q}{\eta}\right)\;,$$
and the existence of the required solution follows from
Lemma~\ref{Brower}.

 \qed

One can repeat the construction of the proof with $m$ replaced by
$m+1$, varying $\delta\mcP_{m+1}^*$ while keeping $\delta\mcP_m^*$
fixed, obtaining a finite dimensional family of distinct solutions
with the same $\mcP_m^*+\delta\mcP_m^*$; this might require
decreasing $\eta$. By induction, one can obtain a family of
arbitrarily high dimension of distinct solutions with the same
$\mcP_m^*+\delta\mcP_m^*$, for $\delta \mcP_m^*$ sufficiently
small.

 The above
initial data are defined only on $\R^3\setminus B(R_0)$; however,
one can now use Theorem~\ref{Tkerr2} to construct initial data on
$\R^3$, or on other asymptotically flat complete manifolds, which
will coincide with the data constructed in Theorem~\ref{Tae1} in
the asymptotic region.

\subsection{Space-times that are Kerrian near $\scrip$}

Space-times that are Kerrian in a neighborhood of a subset of
$\scrip$ are of course obtained by evolution of data which are
Kerrian in a neighborhood of $i^0$. In some situations it might,
however, be convenient to be able to construct such space-times
starting directly from a hyperboloidal initial data hypersurface.
It is not too difficult to adapt the original Corvino-Schoen
technique to the hyperboloidal initial data setting, using the
analysis above together with the relative mass identities
of~\cite{ChHerzlich,ChNagy}; this will be discussed elsewhere.

\subsection{Bondi-type asymptotic expansions at $\scrip$.}
\label{Shao.orig}

Recall that Bondi {\em et al.}\/~\cite{BBM,Sachs} have proposed a
set of free functions parameterising a certain asymptotic
expansion of the metric at $\scrip$. It is of interest to enquire
whether one can construct hyperboloidal initial data sets which
would lead to space-times with a prescribed set of those
functions. The results in Section~\ref{Scc} can be used to give
perturbational answers to such questions, in the spirit of
Theorem~\ref{Tae1}; this will be discussed elsewhere.

\subsection{Local and global extensions of initial data sets}
\label{Slge}

In this section\footnote{A sketchy presentation of the analysis
given in this section has been given in \cite{ChDelay2}.} we
address the \emph{extension problem}, that is, the following
question: let us be given a vacuum initial data set $(M,K,g)$,
where $\bM=M\cup \pM $ has a compact boundary $\pM$, with the data
$(K,g)$ extending smoothly, or in $C^k(\bM)$, to the boundary.
Does there exist an extension across $\pM$ of $(K,g)$ which
satisfies the constraint equations? In the case where $K$ vanishes
and $\partial M$ is \emph{mean convex} an affirmative answer can
be given by using a method\footnote{Smith and Weinstein actually
assume that $\pM$ is a two-sphere, but this hypothesis is
irrelevant for the discussion here.} due to Smith and
Weinstein~\cite{SmithWeinstein}, which proceeds as follows: In a
neighborhood of $\pM$ we can write the metric in the form
\begin{equation}    \label{eq:metric}
    g = u^2 dr^2 + e^{2v} \gammab_{AB}(\betah^A dr + r d\theta^A)
    (\betah^B dr + r d\theta^B),
\end{equation}
where $(\theta^1,\theta^2)$ are local coordinates on $\partial M$,
$\gammab_{AB}$ is a fixed (independent of $r$) metric on $\partial
M$ , and $\betah=\betah^A\partial_A$ is the ``\emph{shift
vector}". Further $r$ is a coordinate on  a $\bM$-neighborhood of
$\pM$ which is, say, negative and vanishes precisely on $\pM$; to
obtain \eq{eq:metric} one needs further to assume that the mean
extrinsic curvature $H$ of $\pM$ has no zeros.  We can extend the
functions $v$ and $\betah^B$ to positive $r$ in an arbitrary way
preserving their original differentiability. When $H>0$,  the
requirement that the extended metric be Ricci-scalar flat becomes
then a semi-linear parabolic equation for $u$ on
$\pM$~\cite{SmithWeinstein}:
\begin{equation} \label{eq:main}
    r\partial_r u  = \Gamma u^2 \Deltash u + \beta\cdot\nablash u+ A u
    - B u^3,
\end{equation}
where the objects above are defined as follows:  we set $
\gamma_{AB}:=e^{2v} \gammab_{AB}$, we write $\Deltash=r^2 e^{2v}
\Deltash_\gamma$ for the respective Laplacians of $\gammab$ and
$\gamma$,  $\nablash u$ is the tangential component of the
gradient of $u$, $\Gamma=e^{-2v}/\Hb$, $A=\Ab/\Hb$ and
$B=\Bb/\Hb$, while
\begin{gather*}
    \label{eq:chib}
    \chib = r u\chi = 
    (1+r v_r)\gamma -\Pi/2 
    \;,\\
    \label{eq:Hb}
    \Hb = r u H = 2 +
    2 r v_r -e^{-2v}\Div_{\gammab}\beta
    \;, \\
   \Ab=r\partial _{r}\Hb -  \beta\cdot\nablash \Hb - \Hb +\frac12 \abs{\chib}_\gamma^2 + \frac12 \Hb^2\;, \\
   \Bb=
   e^{-2v}(1-\Deltash v)
   \;,
\end{gather*}
with $\Pi=\Lie_\betah \gamma$ --- the deformation tensor of
$\betah$ and $\chi$ --- the second fundamental form of the level
sets of $r$.
It follows from the results in~\cite{LSU} that \Eq{eq:main}, with
the obvious initial value,
can always be solved for a small interval of $r$'s when the
initial metric and its extension are in, say, $C^{3}(\bM)$,
obtaining a scalar-flat extension of $(M,0,g)$.
Similarly  the results in~\cite{LSU} can be used to show that $u$
will be of class $C^{k+1}$ on the extended manifold if the
remaining functions there are in $C^{2k+1}$ (thus smooth if the
initial metric is smooth up to boundary, and if the free functions
above are smooth).

Our aim  here is to prove two alternative extension results under
smallness conditions, \emph{without} the hypothesis that $K$
vanishes. Thus, assume we have a solution $(K,g)\in
(C^{k+3,\alpha}\times C^{k+4,\alpha})(\bM)$, $\alpha\in(0,1)$, of
the vacuum constraints on a manifold ${\bM}$ with compact
boundary. Let $M_0$ be another manifold such that $\partial M_0$
is diffeomorphic to $\pM$, and let $M'$ be the manifold obtained
by gluing $M$ with $M_0$ across $\pM$. Let $x$ be any smooth
function defined in a neighborhood $\mcW$ of $\pM$ on $M'$, with
$\pM=\{x=0\}$, with $dx$ nowhere vanishing on $\pM$, and with
$x>0$ on $M_0$. It is convenient to choose $\mcV:=\mcW\cap M_0$ to
be diffeomorphic to $\pM\times [0,x_0]$, with $x$ being a
coordinate along the $[0,x_0]$ factor

 Suppose, next, that there exists on
$M_0$ a solution $(K_0,g_0)$ of the vacuum constraint equations
which is in $(C^{k+3,\alpha}\times C^{k+4,\alpha})(\bM_0)$; we
emphasise that we do not assume that $(K,g)$ and $(K_0,g_0)$ match
across $\partial M$. We first extend $(K,g)$ to a pair $(K_1,g_1)$
defined  on to $M_0$ with the requirement that $(K_1,g_1)$ remains
in $C^{k+3,\alpha}\times C^{k+4,\alpha}$; we do of course not
assume that the extension is vacuum. For the purposes below it is
convenient to make the extension so that
$\|g_1-g_0\|_{C^{k+4,\alpha}(\mcV)}+\|K_1-K_0\|_{C^{k+3,\alpha}(\mcV)}$
is as small as possible. While we are not aware of an optimal
prescription, a possible procedure which at least controls that
norm is as follows: First, by using a partition of unity
subordinate to a finite cover of a neighborhood of $\partial M_0$
the problem is reduced to that of extending functions. Given that,
Corollary~3.3.2 of \cite{AC} with
$$
f_i=\partial_x^i g|_{\partial M}-\partial_x^i g_0|_{\partial
M},\;\;\; i=0,...,k+4,
$$ shows that
there exists a $C^{k+4,\alpha}$ tensor field $f$ on $M_0$ such
that $\partial_x^i f=f_i$ on $\partial M_0$.  On $M_0$ we define
$g_1$, a $C^{k+4,\alpha}$ extension of $g$, by
$$
g_1-g_0:=f,$$ then $\partial_x^ig_1=\partial_x^ig$ for all
$i=0,...,k+4$. The proof of Lemma~3.3.1 and Corollary~3.3.2
in~\cite{AC} show that
\bel{gder}\|g_1-g_0\|_{C^{k+4,\alpha}(\mcV)}\leq C
\sum_{i=0}^{k+4}\|\partial_x^i g|_{\partial M}-\partial_x^i
g_0|_{\partial M}\|_{C^{k+4-i,\alpha}(\partial M)}. \ee The same
procedure applies to extend $K$ to a $C^{k+3,\alpha}$ tensor field
$K_1$ satisfying $\partial_x^iK_1=\partial_x^iK$ for all
$i=0,...,k+3$ on $\partial M$, and
\bel{Kder}\|K_1-K_0\|_{C^{k+3,\alpha}(\mcV)}\leq C
\sum_{i=0}^{k+3}\|\partial_x^i K|_{\partial M}-\partial_x^i
K_0|_{\partial M}\|_{C^{k+3-i,\alpha}(\partial M)}. \ee Let $\phi$
be any smooth function on $M'$ which equals one on $M$ and on a
small neighborhood $\mcU\subset \mcV\approx \partial M\times
[0,x_0]$ of $\pM$ in $M_0$, and vanishes away of  $\mcV$. On
$\mcV$ we set
$$K'= \phi K_1 + (1-\phi) K_0\;,\ g'= \phi g_1 + (1-\phi) g_0.$$
Since $J(K,g)$, $\rho(K,g)$ vanish on $M$, while $J(K_0,g_0)$,
$\rho(K_0,g_0)$ vanish on $M_0$, we will have
$$ |\rho(K',g')|+|J(K',g')|_{g'} \le C\left
(\|g_1-g_0\|_{C^{k+4,\alpha}(\mcV)}+\|K_1-K_0\|_{C^{k+3,\alpha}(\mcV)}\right)x^{k+2}\;;$$
for points at which $\phi=1$ the inequality is justified by Taylor
expanding  in $x$ at $\partial M$ and using the fact that
$(K_1,g_1)$ satisfies the vacuum constraints on $M$; elsewhere
this is justified by Taylor expanding $\rho$ and $J$   in $(K,g)$
around $(K_0,g_0)$ and using the fact that $(K_0,g_0)$ satisfies
the vacuum constraints. In fact, one has
\begin{eqnarray*}
\lefteqn{
|(\nabla')^{(i)}\rho(K',g')|_{g'}+|(\nabla')^{(i)}J(K',g')|_{g'}} && \\
 && \le C\left
(\|g_1-g_0\|_{C^{k+4,\alpha}(\mcV)}+\|K_1-K_0\|_{C^{k+3,\alpha}(\mcV)}\right)x^{k+2-i+\alpha}\;,
\end{eqnarray*}
for all $0\leq i\leq k+2$, with an analogous inequality holding
for the H\"older quotient. So we have
\begin{eqnarray*} \lefteqn{
\|\rho(K',g')\|_{\cbord^{k+2+\alpha}_{k+2,\alpha}(g',\mcV)}+\|J(K',g')\|_{\cbord^{k+2+\alpha}_{k+2,\alpha}(g',\mcV)}}&&\\
&& \le C\left
(\|g_1-g_0\|_{C^{k+4,\alpha}(\mcV)}+\|K_1-K_0\|_{C^{k+3,\alpha}(\mcV)}\right).
\end{eqnarray*}
Assume, first, that  there are no $(Y,N)$'s such that $P^*(Y,N)=0$
on $\mcV$.  If $(K_1,g_1)$ is sufficiently close to $(K_0,g_0)$ in
$(C^{k+3,\alpha}\times C^{k+4,\alpha})(\overline\mcV)$ norm ---
equivalently, if $(K,g)$ and its derivatives up to appropriate
order, as in \eq{gder}-\eq{Kder}, are sufficiently close to
$(K_0,g_0)$ and its derivatives on $\partial M$, then for any
$\alpha'<\alpha$ the norm
$$
\|\rho(K',g')\|_{\hbord^{k+2+\alpha'-(n-1)/2}_{k+2}(g',\mcV)}
+\|J(K',g')\|_{\hbord^{k+2+\alpha'-(n-1)/2}_{k+2}(g',\mcV)}
$$ will also be
small.
If $k\geq [\frac{n}{2}]+1$ we can use Proposition~\ref{p6.7} on
$\mcV$ with $t=k+2+\alpha'-(n-1)/2$, to conclude that there exists
a solution
$$(\delta K,\delta g) \in \cbord^{k+3+\alpha'-(n-1)/2}_{k+2,\alpha}(g',\mcV)\times
\cbord^{k+4+\alpha'-(n-1)/2}_{k+2,\alpha}(g',\mcV) \subset
(C^{k'+2,\alpha}\times C^{k'+2,\alpha})(\overline{\mcV})\;,$$with
all derivatives up to order $k'+2$ vanishing on $\partial \mcV$,
close to zero, of the vacuum constraint equations. Here $k'$ is
any integer satisfying $$k'\le k\;,\ k' \le
k+1+\alpha'-\alpha-(n-1)/2< k+3/2 - n/2\;.$$

The above construction has a lot of if's attached, but it does
provide new non-trivial extensions in the following, easy to
achieve, situation: \begin{enumerate} \item $(K,g)$ belongs to a
one-parameter family of solutions $(K_\lambda,g_\lambda)$ of the
vacuum constraint equations on $M$, \item the vacuum initial data
set $(K_0,g_0)$, assumed above to be defined on $M_0$, arises from
a vacuum initial data set defined on $M'$, still denoted by
$(K_0,g_0)$, with
\item $(K_\lambda,g_\lambda)$ converging to $(K_0|_{M},g_0|_{M})$
as $\lambda $ tends to zero in $(C^{k+3,\alpha}\times
C^{k+4,\alpha})(\bM)$ .
\end{enumerate}
(Replacing $M$ by a  neighborhood of $\partial M$, it is of course
sufficient for all the above to hold in a small neighborhood of
$\partial M$.) In such a set-up, proceeding as above one obtains
an extension for $\lambda$ small enough when $P^*$ has no kernel
on $\mcV$.

The situation is somewhat more complicated when a kernel is
present, though results can be obtained whenever the set-up of
Theorem~\ref{Tremov} applies. As an illustration, we consider a
situation where $\bM$ is a smooth compact submanifold, with smooth
boundary, of $M'=\R^3$. This involves no generality in the
following sense: any two dimensional manifold can be embedded into
$\R^3$, and so can a tubular neighborhood thereof (this will of
course not be an isometric embedding in general). We allow $M$ to
have more than one connected component. We will only be interested
in a component of $\partial M$ which is two-sided, with one side
thereof corresponding to $M$, and the other corresponding to an
unbounded component of $\R^3\setminus M$ (we assume that such a
component exists). A component of $\partial M$ with this property
will be called an \emph{exterior boundary}, and will be denoted by
$\partial_{\ext} M$. We assume $K_0\equiv 0$, and we let $g_0$ be
the Euclidean metric on $\R^3$. Replacing $\bM$ by a tubular
neighborhood $(-x_0,0]\times
\partial_{\ext} M$ we can thus identify $\bM$ with a
subset of $\R^3$. We note that the closure of $\bM$ in $\R^3$ will
then have a boundary with two components, $\{-x_0\}\times
\partial_{\ext} M$ and $\{0\}\times \partial_{\ext} M$, but we will
ignore $\{-x_0\}\times \partial_{\ext} M$ if occurring, and
consider only $\{0\}\times
\partial_{\ext} M$, which is the exterior boundary of the new $M$. From now
on we write $\partial M$ for $\{0\}\times \partial_{\ext} M$. We
assume that $(K,g)$ are close to $(K_0,g_0)$:
\bel{epsineq}\|g-g_0\|_{C^{k+4,\alpha}(\bM)}+\|K-K_0\|_{C^{k+3,\alpha}(\bM)}<\epsilon\;;\ee
such metrics can be constructed by the conformal method. We now
repeat the construction of the proof of Theorem~\ref{Tkerrio} with
$R=1$ there, so that no rescalings of the metrics are performed.
\Eq{C4.1.0} becomes
\begin{eqnarray}\nonumber
\langle e_{(i)},\psi^{-2} \left(
\begin{array}{c}
\delta J\\ \delta \rho
\end{array}\right)\rangle_{L^2_\psi\oplus L^2_\psi} & = &
\int_{\{x_0\}\times \pM} \ourU^{\alpha\beta}(x_{Q,R}+\delta
x_{Q,R})
dS_{\alpha\beta} 
\\ && \nonumber
-\int_{\{0\}\times \pM}
 \ourU^{\alpha\beta}(x_{Q,R}+\delta x_{Q,R})dS_{\alpha\beta} +O(\epsilon^2)
 \;,
\\ &&\label{C4.1a}
\end{eqnarray}
We set
$$Q_{(i)}^0:= \int_{\{0\}\times \pM}
 \ourU^{\alpha\beta}dS_{\alpha\beta}\;.
 $$
It follows from \eq{epsineq} that there exist a constant $C>0$
such that $|Q^0|\leq C\epsilon$. We restrict ourselves to $Q$'s
such that
$$|Q-Q^0|\leq\epsilon\Longrightarrow|Q|\le (1+C)\epsilon \;.$$
Then
$$\int_{\{x_0\}\times \pM}
 \ourU^{\alpha\beta}dS_{\alpha\beta}= Q_{(i)}
 +O(\epsilon^2)\;,$$
 for $\epsilon$ small enough.
We are thus led to
\begin{eqnarray}
\langle e_{(i)},\psi^{-2} \left(
\begin{array}{c}
\delta J\\
\delta \rho
\end{array}\right)\rangle_{L^2_\psi\oplus L^2_\psi} & = &Q_{(i)}-Q^0_{(i)}
+O(\epsilon^2)\;. 
\label{C4.1.2}
\end{eqnarray}
For $\epsilon$ small enough one would like to conclude as before.
There is, however, a difficulty which arises here because the map
of Proposition~\ref{Pcm} degenerates at $m=0$, as is made clear by
the need of dividing by $m$ in \eq{Pcm4} when one wishes to
determine $a_i$ from $J_{0i}$. This leads to further  conditions
if one wishes the argument to go through: roughly speaking, one
needs to assume that $m$ is of order of $\epsilon$, that the ratio
$|\vec p|/m$ is strictly bounded away from one, and that the ratio
$J_{\mu\nu}/m$ is $o(\epsilon)$; if that is the case, we can use
the Lemma~\ref{Brower} with $U=V=B(0,1)$ , $x=(Q-Q^0)/\epsilon$,
$\lambda=1/\epsilon$, $G_\lambda(x)=\frac{1}{\epsilon}(Q-Q^0
+O(\epsilon^2))=x+O(\epsilon)$ and $y=0$ to conclude. Rather than
making general statements along those lines, with hypotheses which
appear difficult to control, we shall assume that the antipodal
map \bel{antipo} x^i\to -x^i\ee preserves $g$ and maps $K$
to $-K$; clearly $(K_1,g_1)$ can be constructed as to preserve
this property, and we will only consider such extensions. Such
data will be referred to as \emph{parity-covariant}.\footnote{One
of the purposes of the parity conditions here is to ensure
vanishing of the centre of mass. This last property also holds
when both $g$ and $K$ are even. However, for even $K$ and small
$m$ there arise some difficulties with  non-zero angular momentum,
essentially identical to those of non-zero centre of mass; see
Section~\ref{SmanyKerr} for an analysis of one such example.}
Nontrivial parity covariant initial data $(K,g)$, as close to the
Euclidean metric as one wishes, can be easily constructed by the
conformal method
--- we do this, for completeness, in Appendix~\ref{Appsid}.

Assume, first, that $K$ --- and hence $K_1$ --- vanishes. Now, the
construction of Theorem~\ref{theor:abprojbis} preserves all
symmetries of initial data, so that gluing together ``up to
kernel" $g_1$ with (\emph{neither} boosted \emph{nor} translated)
Schwarzschild metrics $g_m$ will lead to sets $(K_1+\delta
K_Q=0,g_1+\delta g_Q)$ still being covariant under the antipodal
map \eq{antipo}. One then obtains, by parity considerations,
\bel{Jvan}J_{\mu\nu}(K_1+\delta K_Q,g_1+\delta g_Q)=0\;;\ee
similarly, the left-hand side of \eq{C4.1.2} vanishes for those
projections which are associated with the $J_{\mu\nu}$'s.  Then,
the only possibly non-zero component of the projection on the
kernel is the one which corresponds to the mass. In that case no
difficulties with the crossing of $m=0$ arise, and we can use on
$\Mext$ the family of Schwarzschild metrics $g_m$ with
$m\in(-\delta\;,\delta)$, with any $\delta \le \min(1, 1/R)$.
Rather than invoking the Brouwer fixed point theorem we note that
if the reference Schwarzschild metric $g_m$ has mass
$m=-\min(C\epsilon,\delta)$ we obtain\footnote{At this stage one
could use harmonic coordinates, and invoke the small data
calculations of Bartnik~\cite{Bartnik:mass} to conclude that the
mass $m_0$ as defined by $Q^0$ must be positive, so that
restricting oneself to the family of Schwarzschild metrics with
$m\ge 0$ suffices. However, this is not necessary, and positivity
of $m_0$ is actually a consequence of the positive energy theorem
and of our argument here, regardless of the coordinate systems
used, for data close enough to Minkowski ones.} a strictly
negative value of the projection \eq{C4.1.2} when $\epsilon $ is
small enough. The value $m=m_0+\epsilon$ leads to a strictly
positive value of the projection in \eq{C4.1.2} (decreasing
$\epsilon$ if necessary); since the left-hand-side of \eq{C4.1.2}
depends continuously upon $m$ there exists $m\in
(-C\epsilon,m_0+\epsilon)$ such that the left-hand side of
\eq{C4.1.2} vanishes.

The case of non-vanishing parity-antisymmetric $K$'s is handled as
follows: let $0\le \lambda< 1$ and consider the set of $(K,g)$
satisfying \bel{lambcond} |\vec{p_0}|_\delta \le \lambda m_0
\;.\ee
 \Eq{Jvan} still holds, so that the only projections on the kernel which are non-zero
are those associated with the mass and the momentum. Since the
charges in \eq{C4.1.2} are smaller than $\epsilon$, while the
error term is one order higher, an argument along lines similar to
those of Lemma~\ref{Brower} gives existence of a solution when
$\epsilon$ is small enough

Summarising, we have proved:

\begin{theor}\label{Textg} Let   $k\geq [\frac{n}{2}]+ 3$, and let
$k'$ be the largest integer strictly smaller than $k+(3-n)/2$.
Consider parity-covariant vacuum initial data sets $(K,g)\in
C^{k+2}\times C^{k+3}$ on a compact smooth submanifold $\bM$ of\,
$\R^3$, and let $\Omega$ be any parity invariant bounded domain
with smooth boundary  containing $\bM$. If \eq{lambcond} with some
$\lambda<0$ holds, then there exists $\epsilon>0$ such that if
\eq{epsineq} holds, then there exists a vacuum $C^{k'}\times
C^{k'}$ extension of $(K,g)$ across the exterior component of
$\partial M$, with the extension being a (perhaps boosted)
Schwarzschild solution outside of $\Omega$.

\qed
\end{theor}

Identical results can be similarly obtained when the source fields
$\rho$ and $J$ are prescribed \emph{a priori}, rather than arising
from some field theoretical model which has its own constraint
equations. It is also clear that the arguments generalise to
Einstein-Maxwell electro-vacuum constraint equations, though we
did not attempt to carry through the details of such a
construction.


\subsection{Localised Isenberg-Mazzeo-Pollack gluings}
\label{Slimp}

In important recent papers, Isenberg, Mazzeo and Pollack have
introduced a conformal gluing method for initial data
sets~\cite{IMP,IMP2}; this generalises previous work of
Joyce~\cite{Joyce} which treats the purely Riemannian case. The
problem addressed is the following: let $(M,K,g)$ be a vacuum
initial data set on a not-necessarily connected manifold $M$; for
simplicity we assume in this section that all the fields are
smooth, though the results below can be stated under finite
differentiability conditions. One also assumes that either $M$ is
compact, or $(M,K,g)$ is asymptotically Euclidean, or $(M,K,g)$ is
asymptotically hyperboloidal; on any compact component a
non-degeneracy condition has moreover to be imposed.  Let $p_i\in
M$, $i=1,2$, and for $t$ small let $\hat M_t$ be a manifold
obtained by cutting from $M$ two geodesic balls $B(p_i,t)$ of
radius $t$ centred at $p_i$, and gluing the left-over manifolds by
adding a neck. It is shown in~\cite{IMP2} that when $\trg K$ is
constant over the $B(p_i,t)$'s,  then one can construct a
one-parameter family of new initial data sets $(\hat M_t,\hat
K_t,\hat g_t)$, $t\in (0,t_0)$ with the property that $(K_t,g_t)$
converges uniformly, in any $C^{k,\alpha}$ norm, on any compact
subset of $M\setminus\{p_1,p_2\}$, to $(K,g)$. In fact, $(\hat
K_t,\hat g_t)$ are conformal deformations of $(K,g)$ on
$M\setminus (B(p_1,t_1)\cup B(p_2,t_1))$ for $t<t_1$. The
technique will be referred to as the \emph{IMP gluing}.

Let us show that in \emph{generic} situations the gluing can be
performed so that the new initial data coincide  with the original
ones away from a small neighborhood of the $p_i$'s:

\begin{theorem}\label{TIMP} Let $t_0$ be small enough so that the geodesic
spheres $S(p_i,t)$ are smooth manifolds for $t\le 2t_0$. Suppose
that there exists $0<t_1\le t_0$ such that the set of
 KIDs on $\Gamma_{p_i}(t_1,2t_1):= B(p_i,2t_1)\setminus \overline {B(p_i,t_1)}$ is trivial. Then there exists
 $t_2\le t_1$ and a family of smooth vacuum initial data sets $(\hat M_t, \tilde
 K_t, \tilde g_t)$, $t\le t_2$ such that
 $$(\tilde K_t,\tilde g_t)= (K,g) \ \mbox{ on } \ M\setminus
(B(p_1,2t_1)\cup B(p_2,2t_1)) \;.$$
 \end{theorem}

\begin{remark}
For generic metrics the set of KID's on $\Gamma_{p_i}(t_1,2t_1)$
will be trivial for all $t_1$.
\end{remark}

\begin{remark} The initial data set $(\tilde K_t, \tilde g_t)$
will coincide with the IMP data set $(\hat K_t, \hat g_t)$ in the
neck region.
\end{remark}

\begin{remark}
Suppose that $M$ has two connected components $M_1$ and $M_2$,
with each of the $p_i$'s lying in a different component, say
$p_1\in M_1$ and $p_2\in M_2$. If the set of KIDs on
$\Gamma_{p_1}(t_1,2t_1)$ is trivial, then the construction below
clearly gives $(\tilde K_t,\tilde g_t)= (K,g)$ on $ M_1\setminus
B(p_1,2t_1)$ for $t\le t_2$, regardless of whether or not there
are KIDs on annuli on the other component.
\end{remark}

\proof Let $\chi$ be a positive smooth  radial cut-off function
equal to $1$ in a neighborhood of $t_1$ and equal to zero in a
neighborhood of $2t_1$. On $\Gamma_{p_1}(t_1,2t_1)$ set
$$\mathring{K}_t = \chi \hat K_t + (1-\chi) K\;,$$
$$\mathring{g}_t = \chi \hat g_t + (1-\chi) g\;.$$
Then $(\mathring{K}_t,\mathring{g}_t)$ coincides with the IMP data
$(\hat K_t, \hat g_t)$ in a neighborhood of $S(p_1,t_1)$, and
coincides with the original data $(K,g)$ in a neighborhood of
$S(p_1,2t_1)$. It follows that $\rho
(\mathring{K}_t,\mathring{g}_t)$ and $
J(\mathring{K}_t,\mathring{g}_t)$ are supported away from the
boundary in $\Gamma_{p_1}(t_1,2t_1)$. Since the IMP data converge
uniformly to the original ones on $\Gamma_{p_1}(t_1,2t_1)$ we will
have
$$\lim_{t\to 0} \rho (\mathring{K}_t,\mathring{g}_t) =0=\lim_{t\to 0}
J(\mathring{K}_t,\mathring{g}_t)\;.$$  Theorem~\ref{theor:6.9} and
Corollary~\ref{C6.11} provides  $0<t_2\le t_1$ such that for all
$0<t\le t_2$ there exists a solution
$(\mathring{K}_t,\mathring{g}_t)$ of the vacuum constraint
equations which is smoothly extended by $(\hat K_t,\hat g_t)$
across $S(p_1,t_1)$ and by $(K,g)$ across $S(p_1,2t_1)$, as
desired. \qed

\subsection{Vacuum space-times with a smooth global $\scri$}\label{Sstg}

The results proved so far can be used to establish existence of a
reasonably large class of small-data, vacuum space-times with a
global smooth $\scri$. While we refer the reader
to~\cite{ChDelay2} for the overall details of this construction,
we note the following here: first, in~\cite{ChDelay2} we did not
claim that the resulting space-times will have a smooth $\scri$,
as we did not realise by then\footnote{We are grateful to
J.~Corvino for pointing out that Corollary to us.} that
Corollary~\ref{C6.11} holds. We note that the argument of
Theorem~\ref{Textg} does not seem to work with $k=\infty$.
However, for the construction of the space-times with a smooth
$\scrip$ one can proceed as follows: in the setting of the proof
of Theorem~\ref{Textg}, choose some $k$ large enough so that the
previous existence and regularity results apply, let $g_1$ be an
extension as in \eq{gder}. A small smoothing will lead to an
extension which is $C^\infty$. One then continues the construction
as in Section~\ref{Slge} using an exponentially weighted Sobolev
space, where the exterior region has been slightly increased, so
that its boundary has been moved from $\partial M$ to the set
$\{x=-\epsilon\}$, for some small positive $\epsilon$. The
remaining arguments remain unchanged. Instead of obtaining a
smooth extension of the initial data on $M$ one will have a smooth
extension of the initial data on $M\setminus \{-\epsilon\le x \le
0\}$, but this difference is irrelevant for the purpose of
constructing \emph{some} examples.

In Theorem~\ref{Textg} we have used the  family of boosted
Schwarzschild metrics in the exterior region. It should be clear
that any parity-covariant reference family of stationary metrics
can be used there. This, together with arguments identical to
those of~\cite{ChDelay2}, establishes existence of asymptotically
simple parity-covariant space-times which are stationary near
$i^0$, with metrics which are not necessarily Schwarzschild near
$i^0$.

A generic metric so constructed will have no KID's. Whenever that
occurs, we can use the conformal method to slightly deform the
initial data on $B(0,R)$ so that the new initial data are not
parity symmetric, and then use Theorem~\ref{theor:6.9} and
Corollary~\ref{C6.11} to obtain perturbed initial data on
$B(0,R+1)$ which will not satisfy any parity conditions, and which
will coincide with the starting ones on $\R^3\setminus B(0,R)$.
Making all perturbations small enough one will obtain a maximal
globally hyperbolic development with a global $\scri$, and with a
metric which does not satisfy any parity conditions. In particular
asymptotically simple space-times  which are Kerrian near $i^0$,
with non-vanishing angular momentum, can be obtained in this way.

\subsection{``Many Kerr" initial data}
\label{SmanyKerr} A noteworthy application of the techniques of
Section~\ref{Slge} is the
 construction of initial data containing black-hole regions with
 exactly Kerrian
  geometry both near the apparent horizons, and in the
 asymptotic region. This generalises a construction of~\cite{ChDelay2},
 which leads to ``many Schwarzschild" black holes. More precisely, let $I\in \N$, we will construct
  initial data for a vacuum
space-time with the following properties:
\begin{enumerate}
\item There exists a compact set $\mcK$ such that $(K,g)$ are initial data for a Kerr metric with some
 mass parameter $m$ and some angular momentum parameter $a$
on each connected component of $M\setminus \mcK$ (in general
different $(m,a)$'s for different components);
\item let $\cal S$ denote the usual marginally trapped sphere within the
Carter extension of the Kerr solution, then $M$ contains $I$ such
surfaces, with the space-time metric being exactly a Kerr metric
in a neighborhood of each $\cal S$.
\end{enumerate}
In fact, $(M,g)$ will be obtained by gluing together $I$ Kerr
initial data with small masses. The resulting space-time $(M,g)$
will contain can be thought as having $I$ black holes: Indeed, the
results in~\cite{ChMazzeo} show that for several configurations
the intersection of the black hole region in the associated
maximal globally hyperbolic development of the initial data will
have at least $I$ connected components.

Let us pass to the construction: Let $N$ be the integer part of
$I/2$, choose two strictly positive radii $0<4R_1<R_2<\infty$, and
for $i=1,\ldots, 2N$ let the points
$$\vec x_i \in \Gamma_0(4R_1,R_2):= B(0,R_2)\setminus
\overline{B(0,4R_1)}\;$$ ($B(\vec a, R)$ --- open coordinate ball
centred at $\vec a$ of radius $R$) and the radii $r_i$ be chosen
so that the balls $B(\vec x_i,4r_i)$ are pairwise disjoint, all
included in $\Gamma_0(4R_1,R_2)$. Set \bel{Om1} \Omega:=
\Gamma_0(R_1,R_2) \setminus \left(\cup_i \overline{B(\vec
x_i,r_i)}\right)\;. \ee We shall further assume that $\Omega$ is
invariant under the parity map $\vec x \to - \vec x$. Let
$$\vec Q=((m,a,\vec \omega),(m_0,a_0,\vec \omega(0)),(m_1,a_1,\vec \omega(1)),\ldots ,(m_{2N},a_{2N},\vec \omega(2N))$$
be a set of  numbers and unit vectors satisfying $2m<R_1$, $ 2m_0<
R_1$, $2m_i < r_i$. If $I=2N$ we require $a_0=m_0=0$. Whenever one
of the $m$'s is zero the associated vector  $\vec \omega$ is
irrelevant, and then we forget it altogether. Let $(K_{\vec
Q},g_{\vec Q})$ be constructed as follows:
\begin{enumerate}
\item If $I=2N+1$ then
on $\Gamma_0(R_1,2R_1)$ the initial data $(K_{\vec Q},g_{\vec Q})$
are the initial data for a Kerr metric with mass $m_0$, with
angular momentum $a_0m_0\vec \omega(0)$, centred at $0$; here
\begin{eqnarray}\nonumber a_0m_0\omega(0)^\ell&:=&
\frac{1}{8\pi}\lim_{R\to\infty}\int_{S(0,R)}
\epsilon^{\ell}{}_{jk}x^j((\trg K)g^{kl}-K^{kl})dS_l\\
& = & \frac{1}{8\pi}\int_{S(0,R_1)}
\epsilon^{\ell}{}_{jk}x^j((\trg K)g^{kl}-K^{kl})dS_l
\nonumber\\
&& + \int_{\R^3\setminus
B(0,R_1)}\nabla_l(\epsilon^{\ell}{}_{jk}x^j)((\trg
K)g^{kl}-K^{kl} )\nonumber 
\\
& = & \frac{1}{8\pi}\int_{S(0,R_1)}
\epsilon^{\ell}{}_{jk}x^j((\trg K)g^{kl}-K^{kl})dS_l\nonumber\\
&& + O(m_0a_0(a_0^2 + m_0+ a_0^2m_0))\;. \label{angmomdef}
\end{eqnarray}
 In the third line of \eq{angmomdef}
the covariant derivative $ \nabla_l(\epsilon^{\ell}{}_{jk}x^j)$ is
understood as that of a vector field with vector index $k$, at
$\ell$ fixed. (To obtain the estimate for the error term we are
using Boyer-Lindquist coordinates as discussed in
Appendix~\ref{Skn}. Recall that $K$ is a linear combination of
space-covariant derivatives of  \eq{kerrshift}, which leads to
$K=O(a_0m_0)$. Next, \eq{kerrm} gives the estimate
$\Gamma^i{}_{jk}= O(a_0^2 + m_0+ a_0^2m_0)$ for the space
Christoffel symbols in asymptotically Euclidean coordinates,
leading to \eq{angmomdef}.)
 If $I=2N$
then we take $(K_{\vec Q},g_{\vec Q})=(0,\delta)$ on
$\Gamma_0(R_1,2R_1)$;
\item
on $\Gamma_0(3R_1,R_2) \setminus \left(\cup_i \overline{B(\vec
x_i,4r_i)}\right)$ the initial data $(K_{\vec Q},g_{\vec Q})$ are
the initial data for a Kerr metric with mass $m$, with angular
momentum $am\vec\omega $, centred at $0$. As in \eq{angmomdef} we
have \bel{angmomdef3}am\omega^\ell= \frac{1}{8\pi}\int_{S(0,R_2)}
\epsilon^{\ell}{}_{jk}x^j((\trg K)g^{kl}-K^{kl})dS_l + O(ma(a^2 +
m+ a^2m))\;;\ee
\item
on $\Gamma_0(2R_1,3R_1)$ the tensor fields $(K_{\vec Q},g_{\vec
Q})$ interpolate between the two Kerr initial data already defined
above using a usual cut-off function;
\item on the annuli $\Gamma_{\vec x_i}(r_i,2r_i):=B(\vec
x_i,2r_i)\setminus \overline{B(\vec x_i, r_i)}$ the initial data
$(K_{\vec Q},g_{\vec Q})$ are the initial data for a Kerr metric
with mass $m$, with angular momentum $a_im_i\vec \omega(i) $,
centred at $\vec x_i$. The vanishing of the total momentum of the
Kerr metric implies
\begin{eqnarray}\nonumber
a_im_i\omega(i)^\ell&:=&
\frac{1}{8\pi}\lim_{R\to\infty}\int_{S(\vec x_i,R)}
\epsilon^{\ell}{}_{jk}(x^j-x_i^j)((\trg K)g^{kl}-K^{kl})dS_l\\
\nonumber &=& \frac{1}{8\pi}\lim_{R\to\infty}\int_{S(\vec x_i,R)}
\epsilon^{\ell}{}_{jk}x^j((\trg K)g^{kl}-K^{kl})dS_l\\
\nonumber &=& \frac{1}{8\pi}\int_{S(\vec x_i,r_i)}
\epsilon^{\ell}{}_{jk}x^j((\trg K)g^{kl}-K^{kl})dS_l\\ && +
O(m_ia_i(a_i^2 + m_i+
a_i^2m_i))\;;\label{angmomdef4}\end{eqnarray}
\item on the annulus $\Gamma_{\vec x_i}(2r_i,3r_i)$ the tensor fields
 $(K_{\vec Q},g_{\vec
Q})$ interpolate between the initial data already defined above
using a usual cut-off function;
\item all the parameters are so chosen, and the gluings are so
performed,
  that the resulting initial data set is symmetric under the parity map
$\vec x \to - \vec x$; note that the Kerr initial data are exactly
parity symmetric in the Boyer-Lindquist coordinates, compare the
discussion at the end of Appendix~\ref{Skn}.
\end{enumerate}
Clearly  $g_{\vec Q=0}$ is the flat Euclidean metric on $\Omega$,
in particular it is vacuum.  For $|\vec Q|\le 1$ this implies
$$|\rho(K_{\vec Q},g_{\vec
Q})|\le C |\vec Q|\;.$$ By construction we also have $$
 |J(K_{\vec Q},g_{\vec Q})|\le C
|\vec Q| \left(|a|+\sum_{i=0}^{2N} |a_i|\right)\;.$$ Similar
inequalities hold for derivatives of $J$ and $\rho$.

 Suppose that
\bel{mcondi}|\vec Q| \le \delta\;;\ee
Theorem~\ref{theor:6.9} and Corollary~\ref{C6.11} show that
there exists $0<\delta\le1$ such that for all $Q$ satisfying
\eq{mcondi}
 there exists a set of $C^\infty$ tensor fields $(\hat
K_{\vec Q},\hat g_{\vec Q})$ defined on $\Omega$ which agrees with
$(K_{\vec Q},g_{\vec Q})$ in a neighborhood of $\partial \Omega$,
and which satisfies the constraint equations except for the
projection on the kernel of $P^*$. (Here one should use
Theorem~\ref{theor:6.9} on a domain strictly included in the
interior of $\Omega$; a similar comment applies whenever we are
referring to that theorem below.) Uniqueness implies that the
solution will be even. Parity shows that both the centre of mass
and the total momentum vanish, so that the obstruction is the
non-vanishing of the four integrals
\begin{deqarr}\frac 1 {8\pi}\int_\Omega \rho (\hat K_{\vec Q},\hat g_{\vec Q})& = &m-\sum_{i=0}^{2N} m_i
+O(\delta^2)\;,\label{Om2a}
\\ \frac 1 {8\pi} \int_\Omega \epsilon^\ell{}_{jk}x^jJ^k(\hat K_{\vec Q},\hat g_{\vec Q})& = &
am\omega^\ell -\sum_{i=0}^{2N} a_im_i\omega(i)^\ell
+O(\delta^2)\;. \label{Om2bb}
\end{deqarr}
Now, one would like to apply a fixed point theorem to conclude the
existence of a solution, but this does not seem to work directly
because the error term in \eq{Om2bb} is too large. Instead, we
proceed as follows: Suppose, first, that
$$a=a_0=a_1=\ldots=a_{2N}=0\;,$$ and write $g_{\vec M}$ and $\hat g_{\vec M}$ for the
resulting $g_{\vec Q}$ and $\hat  g_{\vec Q}$. We then have
$K_{\vec Q} = \hat K_{\vec Q}=0$, so that the left-hand-side of
\eq{Om2bb} vanishes identically. Fix any set of $m_i$'s,
$i=0,\ldots,2N$, satisfying
$$\sum_{i=0}^{2N}|m_i| \le \delta/4\;.$$ If $\delta$ is small enough
the right-hand-side of \eq{Om2a} with $m=\delta/2$ will be
strictly positive; it will be strictly negative with
$m=-\delta/2$, by continuity there exists $m$ such that $\hat
g_{\vec M}$ will be Ricci scalar flat.

To continue, suppose that the $\vec x_i$'s  have $m_i\ge 0 0$,
with at least one $m_i>0$, and they are \emph{not} aligned. Then
the vacuum initial data set $(0,\hat g_{\vec M})$ on $\Omega$ has
no KIDs.\footnote{In the case $K=0$ the KID equations decouple, so
if $(N,Y)$ is a KID, then so are $(0,N)$ and $(Y,0)$. The
existence of
 a KID with $Y=0$ would lead to a vacuum static space-time with a non-connected
black hole with all horizons non-degenerate, which is not possible
by~\cite{bunting:masood}. Thus $N=0$. By \cite{ChBeig2} the only
remaining possibility is a single Killing vector field  $Y$ which
is a non-trivial rotation in the region where the metric is
Schwarzschild, which is clearly only possible if all the $\vec
x_i$'s are aligned.} We can therefore use Theorem~\ref{theor:6.9}
and Corollary~\ref{C6.11} around $(0,\hat g_{\vec M})$ to
construct an initial data set $(g,K)$, which coincides with
$(K_{\vec Q},g_{\vec Q})$ near $\partial \Omega$, for any
collection of $a$'s and $\vec \omega$'s satisfying
\bel{aepsi} |a| + \sum_{i=0}^{2N}|a_i|< \epsilon\;,\ee when
$\epsilon $ is small enough.
For further purposes we impose $\epsilon\le \delta$.

Suppose, finally, that all the $\vec x_i$'s are aligned along an
axis, say the $z$-axis. Then the vacuum initial data set $(0,\hat
g_{\vec M})$ on $\Omega$ has exactly one KID $(Y,0)$, where $Y$ is
the Killing vector associated with the rotations around the
$z$-axis. We repeat now the previous construction, with the
following difference: on $\Gamma_0(R_1,R_2) \setminus \left(\cup_i
\overline{B(\vec x_i,r_i)}\right)$ we use the Ricci scalar flat
metric $\hat g_{\vec M}$, $\hat g_{\vec M}$ and in points 3. and
5. above $\hat g_{\vec Q}$ is taken as a combination with cut-off
functions of the relevant Kerr metric and of $\hat g_{\vec M}$.
Assuming \eq{aepsi}, we will have
$$\|\hat g_{\vec Q} - \hat g_{\vec M}\|_{C_k}\le C(k)\epsilon\;,\quad
\|\hat K_{\vec Q} \|_{C_k}\le C(k)\epsilon\;$$ We need a somewhat
more precise version of the calculation in \eq{angmomdef}. By
hypothesis the vector field $Y_k:= \epsilon^z{}_{kl}x^k =
\partial/\partial \varphi$ is a Killing vector of the metric
$g_{\vec M}$, so that
\bel{newepsi} \nabla _{k} Y_l + \nabla_{l} Y_k = \znabla _{k} Y_l +
\znabla_{l} Y_k + 2C^r_{kl} Y_r = 2C^r_{kl} Y_r\;,\ee where
$\znabla$ is the covariant derivative of $g_{\vec M}$, while
$C^r_{kl}$ is the difference of the Christoffel symbols of
$g_{\vec Q}$ and $g_{\vec M}$. It follows that
$$\nabla _{k} Y_l + \nabla_{l} Y_k = O(\epsilon)\;.$$
Applying the divergence theorem on $\R^3\setminus (\cup_i B(\vec
x_i, r_i)\cup B(0,R_1))$ as in the second and third lines of
\eq{angmomdef}, using the last two lines of \eq{angmomdef}
together with \eq{angmomdef3} and \eq{angmomdef4} we therefore
obtain
\bel{zobst}\frac 1 {8\pi} \int_\Omega
\epsilon^\ell{}_{jk}x^jJ^k(\hat K_{\vec Q},\hat g_{\vec Q}) =
am\omega^\ell -\sum_{i=0}^{2N} a_im_i\omega(i)^\ell +O(\epsilon^2
+ \epsilon \delta^2)\;.\ee Here the $\rho$ integral, as well as
the integrals \eq{zobst} with $\ell=x$ and $\ell=y$ are already
identically zero, so that the only remaining obstruction is the
integral at the left-hand-side of \eq{zobst} with $\ell=z$. We
choose the exterior Kerr solution so that $\omega^z\ne 0$.
At this stage we might need to decrease $\delta$ to conclude, so
we suppose that we are working in a family of mass parameters
$(m,m_i)$ so that $\delta/m$ is uniformly bounded from above
independently of $m$ (in particular $m$ is \emph{not} zero). This
gives
\bel{zobst2}\frac 1 {8\pi m \omega^z} \int_\Omega
\epsilon^z{}_{jk}x^jJ^k(\hat K_{\vec Q},\hat g_{\vec Q}) = a
-\sum_{i=0}^{2N} a_i\frac{m_i}m \frac{\omega(i)^z}{\omega^z}
+O(\frac{\epsilon^2}\delta + \epsilon \delta)\;.\ee We can choose
$0<\epsilon'<\epsilon/2$ so that if $$\sum_{i=0}^{2N}
|a_i|<\epsilon'$$ then $$|\sum_{i=0}^{2N} a_i\frac{m_i}m
\frac{\omega(i)^z}{\omega^z}| <\epsilon/8\;.$$ We then require
$\delta$ to be small enough so that $$|O( \epsilon \delta)|<
\epsilon/16\;,$$ and then for sufficiently small $\epsilon$'s we
will have $$|O(\frac{\epsilon^2}\delta )|< \epsilon/16\;,$$ for
$\epsilon$'s small enough. If $a=-\epsilon/2$ the right-hand-side
of \eq{zobst2} will be negative, it will be positive if
$a=\epsilon/2$, and continuity shows existence of an $a$ that
leads to a solution of the full constraint equations.

\Eq{Om2a} shows that $m$ will be close to $\sum_{i=0}^{2N} m_i$,
which gives the desired control of the ratio $\delta/m$ if all the
$m_i$'s are of the same sign. It follows that the previous
construction applies in this case. Clearly the sign condition is
not necessary, and there exist several other  families of $m$
parameters which will give the desired control.

We can now repeat the whole previous construction by gluing
\emph{boosted} Kerr initial data centred on the $\vec x_i$'s, with
a small boost parameter, to the solution $(K,g)$ just obtained
with the same remaining parameters. If $(K,g)$ has no KID's, then
we will obtain a  new smooth solution from Theorem~\ref{theor:6.9}
and Corollary~\ref{C6.11} provided that the boost parameters are
small enough. We note that the initial data $(K,g)$ will have no
KID's except when all the $\vec x_i$'s are aligned along, say, the
$z$-axis, with all the $\vec \omega(i)$'s pointing in the
$\partial_z$ direction. One expects that a variation of the above
arguments would still give existence of solutions, but we have not
investigated this point any further.

The  mass of the solutions obtained above, as seen from the end
$r\ge R_2$, might  be very small. One can now make a usual
rescaling $m\to \lambda m$, $r \to \lambda r$, $g_{\vec m}\to
\lambda^{-2}g_{\vec m}$, to obtain any value of the final mass
$m$.

We emphasise that the mass parameters $m_i$ and $m_0$ are only
restricted in absolute value, so solutions $(0,g_{\vec M})$ with
some of the $m_i$'s negative or zero, and/or $m_0$ negative or
zero, and $m$ negative, can be constructed. For instance, a zero
value of $m_i$ will correspond to metrics which can be $C^k$
matched to a flat metric on $B(\vec x_i,r_i)$. One can actually
also obtain $a=0$, or  $m=0$, or both: it suffices to repeat the
above argument with prescribed values $am=0$ and $m_i$,
$i=1,\ldots,2N$, adjusting $m_0$ and/or $a_0$ rather than $m$ and
$a$. Arguing as before one can obtain a family of non-trivial
vacuum initial data which are Minkowskian on an exterior region
$\R^3\setminus B(0,R)$. (Clearly $m=0$ implies that at least one
of the $m_i$'s, $i\ge 0$, is negative, unless they all vanish.)

\appendix
\section{Weighted Sobolev and  weighted H\"older
spaces}\label{SwSs} Let $\phi$ and $\psi$ be two smooth strictly
positive functions$^{\mbox{\scriptsize \rm \ref{footmanif}}}$ on
$M$.  For $k\in \Nat$ let $\Hkpp(g) $ be the space of $H^k_\loc$
functions or tensor fields such that the norm\footnote{The reader
is referred to \cite{Aubin,Aubin76,Hebey} for a discussion of
Sobolev spaces on Riemannian manifolds.}
 \be \label{defHn}
 \|u\|_{\Hkpp (g)}:=
(\int_M(\sum_{i=0}^k \phi^{2i}|\nabla^{(i)}
u|^2_g)\psi^2d\mu_g)^{\frac{1}{2}} \ee is finite, where
$\nabla^{(i)}$ stands for the tensor $\underbrace{\nabla ...\nabla
}_{i \mbox{ \scriptsize times}}u$, with $\nabla$ --- the
Levi-Civita covariant derivative of $g$; we assume throughout that
the metric is at least $W^{1,\infty}_\loc$; higher
differentiability will be usually indicated whenever needed. For
$k\in \Nat$ we denote by $\zHkpp $ the closure in $\Hkpp$ of the
space of $H^k$ functions or tensors which are compactly (up to a
negligible set) supported in $M$, with the norm induced from
$\Hkpp$.
The $\zHkpp $'s are Hilbert spaces with the obvious scalar product
associated to the norm \eq{defHn}. We will also use the following
notation
$$
\quad \zHk  :=\zHk  _{1,1}\;,\quad
L^2_{\psi}:=\zH^0_{1,\psi}=H^0_{1,\psi}\;,
$$ so that $L^2\equiv \zH^0:=\zH^0_{1,1}$. We  set
$$
W^{k,\infty}_{\phi}:=\{u\in W^{k,\infty}_{\loc} \mbox{ such that }
\phi^i|\nabla^{(i)}u|_g\in L^{\infty}\}\;,
$$
with the obvious norm, and with $\nabla^{(i)}u$ --- the
distributional derivatives of $u$.

For $\phi$ and $\varphi$  --- smooth strictly positive functions
on M, and for $k\in\N$ and $\alpha\in [0,1]$, we define
$C^{k,\alpha}_{\phi,\varphi}$ the space of $C^{k,\alpha}$
functions or tensor fields  for which the norm
$$
\begin{array}{l}
\|u\|_{C^{k,\alpha}_{\phi,\varphi}(g)}=\sup_{x\in
M}\sum_{i=0}^k\Big(
\|\varphi \phi^i \nabla^{(i)}u(x)\|_g\\
 \hspace{3cm}+\sup_{0\ne d_g(x,y)\le \phi(x)/2}\varphi(x) \phi^{i+\alpha}(x)\frac{\|
\nabla^{(i)}u(x)-\nabla^{(i)}u(y)\|_g}{d^\alpha_g(x,y)}\Big)
\end{array}$$ is finite.

We will only consider weight functions with the property that
there exists $\ell \in \N\cup\{\infty\}$ such
that\footnote{Conditions~\eqref{lcond} will typically impose
$\ell$ restrictions on the behavior of the metric and its
derivatives in the asymptotic regions; it is therefore essential
to allow $\ell<\infty$ if one does not wish to impose an infinite
number of such conditions.} for $0\le i< \ell$ we have
\begin{equation}\label{lcond}
 |\phi^{i-1}\nabla^{(i)}\phi|_g\leq C_{i}\;,\;\;\;
|\phi^{i}\psi^{-1}\nabla^{(i)}\psi|_g\leq C_{i}\;,
\end{equation}
for some constants $C_i$. This implies that for $0\le i < \ell$
 and for all
$k\in \N$ it holds that \be \label{0.1}
|\phi^{i-k}\nabla^{(i)}\phi^k|_g\leq C_{i,k}\;,\;\;\;
|\phi^{i}\psi^{-k}\nabla^{(i)}\psi^k|_g\leq C_{i,k}\;. \ee It
follows that for $m,s\in\N$ and for $0\le i +k< \ell$ the maps
$$
\psi^{-m}\phi^{i-s}\nabla^{(i)}(\phi^s\psi^m\cdot)
:\zH^{k+i}_{\phi,\psi}\longmapsto \zHkpp \;,
$$
$$
\psi^{-m}\phi^{i-s}\nabla^{(i)}(\phi^s\psi^m\cdot)
:W^{k+i,\infty}_{\phi}\longmapsto W^{k,\infty}_{\phi}\;,
$$
$$
\psi^{-m}\phi^{-s}\nabla^{(i)}(\phi^{i+s}\psi^m\cdot)
:\zH^{k+i}_{\phi,\psi}\longmapsto \zHkpp \;,
$$
\be\label{mpp}
\psi^{-m}\phi^{-s}\nabla^{(i)}(\phi^{i+s}\psi^m\cdot)
:W^{k+i,\infty}_{\phi}\longmapsto W^{k,\infty}_{\phi}\;, \ee are
continuous and bounded.  If the function $\varphi$ satisfies the
same condition \eq{0.1} as  $\psi$, then we can replace
$\zH^j_{\phi,\psi}$ by $C^{j,\alpha}_{\phi,\varphi}$ in \eq{mpp}.

The following situations will be of main interest to us:
\begin{itemize}
\item If $\partial M$ is compact, smooth, and non-empty (see
section \ref{Scb}), we will use for $\phi=x$ a function which is a
defining function for the boundary, at least in a neighborhood of
the boundary; that is, any smooth non-negative function on $\bM$
such that $\partial M$ is precisely the zero-level set of $x$,
with $dx$ without zeros on $\partial M$. Then $\psi$ will be a
power of $x$ on a neighborhood of $\partial M$. Condition
\eq{lcond} will hold if $g$ has
$(W^{\ell-1,\infty}_x,l-1)$--behavior at $\partial M$ in the sense
of Definition~\ref{Dwab} below.

\item If $M$ contains an asymptotically flat region (see Section~\ref{Safm}), $\phi$ will behave as $r$
and $\psi$ will behave as a power of $r$ in the asymptotically
flat region; \eq{lcond} will hold if $g$ is
$W^{\ell-1,\infty}_r$--asymptotically flat.

\item If $M$ contains a conformally compactifiable region (see
Section~\ref{Scc}), then in a neighborhood of the conformal
boundary  $\phi$  will be taken to be $1$, while $\psi$ will be a
power of the defining function of the conformal boundary.

\item Exponentially weighted versions of the above will also be
considered.
\end{itemize}
In all those situations one can obtain elliptic estimates in
weighted spaces for the equations considered here by covering and
scaling arguments together with the standard interior elliptic
estimates on compact balls ({\em cf.,
e.g.}\/~\cite{choquet-bruhat:christodoulou:elliptic,Bartnik:mass,AndElli,AndChDiss,GL,Lee:fredholm}).
We will refer to this as \emph{the scaling property}.

More precisely, we shall say that \emph{the scaling property}
holds (with respect to some weighted Sobolev spaces with weight
functions $\psi$ and $\phi$, and/or weighted H\"older spaces with
weight functions $\varphi$ and $\phi$, whichever ones are being
used will always be obvious from the context) if there exists a
covering of $M$ by a family of sets $\Omega_\alpha$, for $\alpha$
in some index set $I$, together with scaling transformations
$\phi_\alpha:\Omega_\alpha\to \hat\Omega_\alpha$ on each of the
sets $\Omega_\alpha$, such that
the transformed fields $(\hat
K_\alpha,\hat g_\alpha)$ on $\Omega_\alpha$ are in\footnote{It is
conceivable that in some situations less \emph{a priori}
regularity on the $(\hat K_\alpha,\hat g_\alpha)$'s can be
assumed, but this is the setup which seems to play the most
important role in our paper; the reader should be able to adapt
the differentiability conditions to his needs if required.} in
$W^{3,\infty}(\hat \Omega_\alpha)\times W^{4,\infty}(\hat
\Omega_\alpha)$, and such that the usual interior elliptic
estimates on the $\hat \Omega_\alpha$'s can be pieced together to
a weighted estimate, such as \eq{erc}, for the original fields.
Some sufficient conditions for the scaling property are discussed
in Appendix~\ref{Sscaling}. We note that the scaling
transformation of the fields on $\hat \Omega_\alpha$,
$(K,g)\to(\hat K_\alpha, \hat g_\alpha)$,  will typically consist
of a pull-back of the fields, accompanied perhaps by a constant
conformal rescaling.
  The ``scaling property" is a
condition {both} on the {metric} $g$, the {extrinsic curvature
tensor} $K$, and on the {weight functions} involved: indeed, both
the metric coefficients, the connection coefficients, as well as
their derivatives, \emph{etc.}, \/which appear in our equations
must have appropriate behavior under the above transformations so
that the required estimates can be established.

\section{Sufficient conditions for  the scaling
property} \label{Sscaling}

 In this section we present some sufficient
conditions to the functions $\phi $ and $\varphi $ which guarantee
that the spaces $C^{k,\alpha}_{\phi ,\varphi }(g)$ satisfies the
scaling property. We give some examples of such spaces.  We assume
that the manifold $M$ is an open subset of $\R^n$, and that the
elliptic operator we work with is an operator on functions. The
result generalises to tensor fields on manifolds by using
coordinate patches, together with covering arguments.

We assume that $\phi $ and $\varphi $ verify \eq{lcond}. For all
$p\in M$, we denote by $B_p $, the open ball of centre $p$ with
radius $\phi(p) /2$. We require that\footnote{It suffices to
assume that there exists $\mu>0$ such that for all $p\in M $,
$B(p,\mu\phi(p) ) \subset M$, as changing $\phi$ to $\mu \phi$ for
a positive constant $\mu$ leads to equivalent norms. This is
actually the condition needed in the asymptotically flat case, as
\eq{scalprop0} will typically not be satisfied there. For
convenience we assume in \eq{scalprop0} that any such rescalings
have already been done.} for all $p\in M $,
\bel{scalprop0}B(p,\phi(p) ) \subset M\;.\ee

 For $p\in M $, we define
$$
\varphi_p  :B(0,1/2)\ni z\mapsto p+\phi(p) z\in B_p .
$$
For all functions $u$ on $M$ and all multi-indices $\gamma$ we
have
$$
\partial_z^\gamma(u\circ
\varphi_p )=\phi(p) ^{|\gamma|}(\partial^\gamma u)\circ \varphi_p
.
$$
Let $P(p,\partial)$ be a strictly elliptic (\emph{e.g.}, in the
sense of Agmon-Douglis-Nirenberg) operator of order $m$ on $M$ and
set
$$
(P_\phi u)(p):=[P(\cdot,\phi \partial)u](p),
$$
note that in our context $P_\phi$ will be elliptic uniformly
degenerate whenever $\phi(p) $ approaches zero in some regions. We
assume that the coefficients of $P$ are in $C^{k,\alpha}_{\phi,1}
(M)$. For all $p\in M $, we define the elliptic operator $Q_p$ on
$B(0,1/2)$ by
$$
Q_p(z,\partial):=P(\varphi_p (z),(\phi(p) )^{-1}\phi
\circ\varphi_p (z)\partial),
$$
we then have
$$
Q_p(u\circ\varphi_p )=(P_\phi u)\circ\varphi_p .
$$
We assume that there exist a constant $C_1>0$ such that for all
$p\in M $ and all $y\in B_p $, we have \bel{scalprop1}
C_1^{-1}\phi(p) \leq \phi (y)\leq C_1\phi(p) .\\
\ee Then the $C^{k,\alpha}(B(0,1/2))$ norm of the coefficients of
$Q_p$ are bounded by the $C^{k,\alpha}_{\phi,1} (M)$ norm of the
coefficients of $P$. On other hand, $Q_p$ is strictly elliptic and
by the usual interior elliptic estimates,there exist $C>0$ which
does not depend on $p$ and $v$  such that for all functions $v\in
L^{2}(B(0,1/2))$, such that $Q_pv$ is in $C^{k,\alpha}(B(0,1/2))$
we have $v\in C^{k+m,\alpha}(B(0,1/4))$ and
$$
\|v\|_{C^{k+m,\alpha}(B(0,1/4))}\leq
C(\|Q_pv\|_{C^{k,\alpha}(B(0,1/2))}+\|v\|_{L^{2}(B(0,1/2))}).
$$
So if $u$ is in $ L^{2}_{\varphi\phi^{-n/2}} (M)$ with $Pu\in
C^{k,\alpha}_{\phi ,\varphi }(M)$, then $u\in
C^{k+m,\alpha}_{loc}$. Now, we assume that there exist a constant
$C_2>0$ such that for all $p\in M $ and all $y\in B_p $, we have
\bel{scalprop2}
C_2^{-1}\varphi(p)\leq \varphi (y)\leq C_2\varphi(p).\\
\ee For $p\in M $, we  define $B'_p $ the ball of centre $p$ and
radius $(1/4)\phi(p) $. It follows from \eq{scalprop1} that there
is a $p$--independent number $N$ such that each $B_p$ is covered
by $N$ balls $B_{p_i(p)}'$, $i=1,\ldots,N$. We then have (the
second and the last inequalities  come from \eq{scalprop2})
\begin{eqnarray}\nonumber
\|u\|_{C^{k+m,\alpha}_{\phi ,\varphi }(M)}&\leq& C \sup_{p\in M
}\|u\|_{C^{k+m,\alpha}_{\phi ,\varphi }(B'_p )}\\\nonumber &\leq&
C \sup_{p\in M }(\varphi(p)\|u\|_{C^{k+m,\alpha}_{\phi(p) ,1}(B'_p
)})\\\nonumber &\leq& C \sup_{p\in M }(\varphi(p)\|u\circ\varphi_p
\|_{C^{k+m,\alpha}(\varphi_p ^{-1}(B'_p ))})
\\\nonumber &=& C
\sup_{p\in M }(\varphi(p)\|u\circ\varphi_p
\|_{C^{k+m,\alpha}(B(0,1/4))})
\\\nonumber &\leq& C \sup_{p\in M
}[\varphi(p)(\|P_\phi u\circ\varphi_p \|_{C^{k,\alpha}(B(0,1/2))}
+\|u\circ\varphi_p \|_{L^{2}(B(0,1/2))})]\\\nonumber &\leq& C
[\sup_{p\in M }(\varphi(p)\|Pu\|_{C^{k,\alpha}_{\phi(p) ,1}(B_p
)})
+\sup_{p\in M }(\|u\|_{L^{2}_{\varphi\phi^{-n/2}}(B_p  )})]\\
&\leq& C (\|Pu\|_{C^{k,\alpha}_{\phi ,\varphi
}(M)}+\|u\|_{L^{2}_{\varphi\phi^{-n/2}}(M)})\;.
\label{weighcalc}\end{eqnarray} In particular $u\in
C^{k+m,\alpha}_{\phi ,\varphi }(M)$. An identical calculation
gives
\begin{eqnarray*}
\|u\|_{C^{k+m,\alpha}_{\phi ,\varphi }(M)} &\leq& C
(\|Pu\|_{C^{k,\alpha}_{\phi ,\varphi
}(M)}+\|u\|_{L^{\infty}_{\varphi}(M)})\;.
\end{eqnarray*}
A similar scaling calculation, together with a summation over a
set of $B_{p_i}$'s  forming an appropriate covering of $M$, gives
the corresponding inequality in weighted Sobolev space. We thus
obtain:
\begin{lem}\label{LC1}
If $\phi $ and $\varphi $ satisfy the condition \eq{lcond} with
$\psi$ replaced by $\varphi$, together with
\eq{scalprop0},\eq{scalprop1} and \eq{scalprop2}, then the spaces
$C^{k,\alpha}_{\phi ,\varphi }$ verify the scaling property.
\end{lem}

 As already mentioned, near a compact boundary a standard example of
 functions satisfying the above requirement  with $x$
--- a defining function for the boundary, $\phi=x$, and $\varphi$ ---
a power of $x$. Another example is $\varphi =e^{-s/x}$,
%
where $s\in\R$, and $\phi = x^2$. In fact in that context,  $x$ is
equivalent to $d(\cdot,\partial M)$. For sufficiently regular
metrics ({\em e.g.}, $g\in C^\ell$) we have \eq{lcond}, while the
choice of $\phi$ guarantees \eq{scalprop0}. For \eq{scalprop1} we
compute for all $q\in B_p$: by the triangle inequality,
$$
d(p,\partial M)-d(p,q)<d(q,\partial M)\leq d(p,\partial
M)+d(p,q)\;.
$$
Then, since $d(p,q)<x(p)^2/2$ for $q\in B_p$,
$$
d(p,\partial M)-x(p)^2/2<d(q,\partial M)\leq d(p,\partial
M)+x(p)^2/2\;.
$$
From \eq{scalprop0} we have that $x(p)^2<d(p,\partial M)$, giving
$$
d(p,\partial M)/2<d(q,\partial M)\leq 3d(p,\partial M)/2\;,
$$
and as $x$ is equivalent to $d(.,\partial M)$ we obtain
\eq{scalprop1}. Now for all $q\in B_p$,
$$
e^{-s/x(p)}e^{s/x(q)}=e^{-s(x(p)-x(q))/x(p)x(q)},
$$
but $|x(p)-x(q)|$  is bounded by some constant times $x(p)^2$ and
$x(p)x(q)$ is equivalent to $x(p)^2$ so we obtain \eq{scalprop2}.

We note that if $\varphi_1$ and $\varphi_2$ satisfy
\eq{scalprop2}, then $\varphi_1 \varphi_2$ also will. It follows
that $\varphi= x^\alpha e^{s/x}$ can also be used as a weighting
function in our context with $\phi=x^2$ for all $\alpha,s\in\R$.

In   asymptotically flat regions the standard choice is $\varphi =
r^\alpha$, for some $\alpha \in \R$, and $\phi = r$. Another one
is    $\phi =1$ and $\varphi =e^{sr}$, where $s\in\R$; in that
case \eq{scalprop0}, \eq{scalprop1} and  \eq{scalprop2} are
evident.

\section{Weighted Poincar\'e inequalities}
\label{SWPin} We start with some general inequalities on an open
manifold $M$, then we will apply them to all the cases  of
interest to us. All the integrals are always calculated with
respect to the natural Riemannian measure $d\mu=d\mu_g=d\mu(g)$
with respect to the metric at hand, in local coordinates $d\mu =
\sqrt{\det g_{ij}}\;d^n x$.

We start with a lemma:
\begin{lem}
Let $u$ be a $C^1$ compactly supported tensor field on $M$, and
let $w$ be a $C^2$ function defined in a neighborhood of the
support of $u$, then
 \be\label{poinc1}
 \int_M |\nabla u|^2\geq  \int_M (-|\nabla w|^2+\Delta w) |u|^2\;.
\ee
\end{lem}
\begin{proof}
$$ \int_M |\nabla u|^2 +|\nabla w|^2|u|^2+2 u\nabla_i u \nabla^iw =\int_M |\nabla u +(\nabla w) u|^2\geq0\;.$$
 (By an abuse of notation, here and below we write $u\nabla_i u
\nabla^iw$ for $h(u,\nabla_{\nabla w}u)$, where $h$ is the metric,
constructed using $g$, on the tensor bundle relevant to the tensor
character of  $u$.) An integration by parts leads to
$$ \int_M |\nabla u|^2 +|\nabla w|^2|u|^2-\Delta w |u|^2\geq 0\;,$$
so that
$$ \int_M |\nabla u|^2 \geq \int_M -|\nabla w|^2|u|^2+\Delta w |u|^2\;.$$
\end{proof}
\qed
\begin{Proposition}\label{prop:poinc}
Let $u$ be a $C^1$ compactly supported tensor field on $M$, and
let $w,v$ be two $C^2$ functions defined on a neighborhood of the
support of $u$, then
$$ \int_M e^{2v}|\nabla u|^2
\geq \int_M e^{2v}\left[\Delta v+\Delta w +|\nabla v|^2-|\nabla
w|^2 \right]|u|^2\;. $$
\end{Proposition}
\proof Returning to the proof of Lemma~\ref{poinc1},  with $u$
replaced by $e^v u$ gives
\begin{eqnarray}\nonumber
\int_M e^{2v}[|\nabla u|^2 +|\nabla v|^2|u|^2+2u \nabla_i u
\nabla^iv]&=& \int_M e^{2v}|\nabla u +u\nabla v |^2
\\ &\geq &\int_M e^{2v}(-|\nabla w|^2+\Delta w)|u|^2\;.
\nonumber \\ && \label{poinc4}
\end{eqnarray}
An integration by parts transforms the left-hand-side of the first
line of \Eq{poinc4} into
$$
\int_M e^{2v}\left\{[|\nabla u|^2 +|\nabla v|^2|u|^2]-\Delta v
|u|^2 -2|\nabla v|^2|u|^2\right\}\;,$$ so that \eq{poinc4} can be
rewritten as
$$ \int_M e^{2v}|\nabla u|^2
\geq \int_M e^{2v}\left[\Delta v+\Delta w +|\nabla v|^2-|\nabla
w|^2 \right]|u|^2\;.
$$
\qed

\subsection{Application: compact boundaries}
\label{SD.1}

Let  $x$ be any twice-differentiable defining function for
$\partial M$. We shall consider metrics $g$ which are in
$W^{1,\infty}_\loc$ on $M$ and continuous on $\bM$. We shall
further suppose that the Hessian $\Hess x:= \nabla \nabla x$ of
$x$ satisfies \be \label{pi0} |\Hess x |= o(x^{-1})\;. \ee
\Eq{pi0} will obviously hold if $g$ is smooth on $\bM$; it is,
however, natural to consider metrics of lower differentiability
class when $\partial M$ corresponds to a conformal boundary at
infinity. (Actually, in this section it would suffice to assume
that $\Delta x =o(x^{-1})$; however, the stronger hypothesis
\eq{pi0} will be required in our further considerations.)

We will work in a neighborhood of $\partial M$ small enough so
that $|dx|$ is bounded away from zero there. The following result
is well known (compare \cite[Lemma~1, Section~3.2.6]{Triebel}), we
give a proof since we need to control the constant in \Eq{pinb}
below; the calculation can be traced back to those in
\cite{Lee:fredholm}:

\begin{Proposition}
\label{Pinb} For any $\epsilon >0$ and $s\ne -1/2$ there exists
$x_{\epsilon,s}>0$  such that for any differentiable tensor field
$u$ with compact support in
 $\{ 0<x<x_{\epsilon,s}\}$
we have \be \label{pinb}\int_M x^{2s +2} |\nabla u|^2 d\mu \ge
\left\{(s+1/2)^2-\epsilon\right\} \int_M x^{2s } | u|^2 |dx|^2d\mu
\;. \ee
\end{Proposition}

\proof We use Proposition \ref{prop:poinc}, choosing
$v=(s+1)\ln(x)$ one has $dv=(s+1)dx/x$ and \be \label{pi2} \Delta
v=-(s+1)|dx|^2/x^2+(s+1)\Delta x/x=-(s+1+o(1))|dx|^2/x^2\;. \ee It
follows that
$$|dv|^2+\Delta v=((s+1)^2-s-1+o(1))|dx|^2/x^2\;.$$
Choosing  $w=-\frac{1}{2}\ln(x)$ we have that
$$-|dw|^2+\Delta w=(1/4+o(1))|dx|^2/x^2\;.$$
\qed
\begin{Proposition}
\label{Pinbexp} For any $\epsilon >0$, $t,s\in\R$ there exists
$x_{\epsilon,s,t}>0$  such that for any differentiable tensor
field $u$ with compact support in
 $\{ 0<x<x_{\epsilon,s,t}\}$
we have \be \label{pinbexp1}\int_M e^{-2s/x}x^{2t} |\nabla u|^2
d\mu \ge \left\{s^2-\epsilon\right\} \int_M e^{-2s/x } | u|^2
x^{2t-4}|dx|^2d\mu \;. \ee
\end{Proposition}
\proof We again use Proposition \ref{prop:poinc} with $v=-s/x+t\ln
x$ and $w=0$, one then has $dv=sdx/x^2+tdx/x$ and
$$\Delta v=-2s|dx|^2/x^3+s\Delta x/x^2-t|dx|^2/x^2+t(\Delta x)/x= o(1)|dx|^2/x^4\;. $$ It
follows that
$$|dv|^2+\Delta v=(s^2+o(1))|dx|^2/x^4\;.$$
then we obtain \eq{pinbexp1}.
 \qed

\subsection{Application: asymptotically flat metrics}

We shall also need a weighted Poincar\'e inequality for
 metrics $g$ defined on $\R^3\setminus\{r\ge R\}$ for some $R$,
satisfying the following requirement: for every $\epsilon>0$ there
exists $R_\epsilon<\infty$ such
 that
 \be\label{metfo}
|g_{ij}-\delta_{ij}| \le \epsilon \ \  \mbox{on} \ \ \{r\ge
R_\epsilon\}\;. \ee We shall also require that \be\label{metfo1}
\Delta r-(n-1)|\nabla r|^2/r=o(1) \ee (recall that the
right-hand-side above is zero for a flat metric). One then has the
following~\cite{Bartnik:mass,ChSob}; we give a proof for
completeness\footnote{Actually the case $n=2$  does not seem to
have appeared in the published literature so far.}:
\begin{Proposition}
\label{PwPiaf} Suppose that \eq{metfo}-\eq{metfo1} hold. Then for
any $s\in\R$ and $\epsilon>0$ there exists $R_{s,\epsilon} <
\infty $ such that for any $C^1$ tensor field $u$ with compact
support included in $\{r> R_{s,\epsilon}\}$ it holds that
\be\label{wPiaf} \int r^{-2s-n+2}|\nabla u|^2 d\mu\ge
(s^2-\epsilon)\int r^{-2s-n}|u|^2d\mu\;. \ee
\end{Proposition}
\proof We use Proposition \ref{prop:poinc} with $v=(-s+1-n/2)\ln
r$ and $w=[(n-2)\ln r]/2 $. We just recall that  when $f=c\ln r$,
we have
$$
\nabla f=c\nabla r/r$$ and $$ \Delta f=c\Delta r/r-c|\nabla
r|^2/r^2=c(n-2)|\nabla r|^2/r^2+o(1/r).$$ So

$$\Delta v+|\nabla v|^2=(s^2-(n-2)^2/4)|\nabla r|^2/r^2+o(1/r),$$
and
$$\Delta w-|\nabla w|^2=((n-2)^2/4)|\nabla r|^2/r^2+o(1/r).$$
\qed
\begin{Proposition}
\label{PwPiafexp} Suppose that \eq{metfo}-\eq{metfo1} hold. Then
for any $\epsilon>0$ there exists $R_{s,\epsilon} < \infty $ such
that for any $C^1$ tensor field $u$ with compact support included
in $\{r> R_{s,\epsilon}\}$ it holds that \be\label{wPiafexp} \int
e^{-2sr}|\nabla u|^2 d\mu\ge (s^2-\epsilon)\int e^{-2sr}|\nabla
r|^2|u|^2d\mu\;. \ee
\end{Proposition}
\proof We use Proposition \ref{prop:poinc} with $v=-sr$ and $w=0$.
Then $ \nabla v=-s\nabla r$ and $ \Delta v=-s\Delta
r=-s(n-1)|\nabla r|^2/r+o(1)=o(1)|\nabla r|^2.$ So
$$\Delta v+|\nabla v|^2=(s^2+o(1))|\nabla r|^2.$$
\qed

\subsection{Application: conformally compact manifolds}
Here, as elsewhere, $n$ denotes the dimension of $M$. We recall
that we have $g=x^{-2}\overline{g}$ then
\bel{gammaah}
(\Gamma-\overline{\Gamma})^k_{ij}=-x^{-1}(2\delta^k_{(i}\overline{\nabla}_{j)}x-\overline{g}_{ij}\overline{\nabla}^kx).
\ee In particular, we have \bel{hessah} \nabla_i\nabla_j
x=\overline{\nabla}_i\overline{\nabla}_jx
+x^{-1}(2\overline{\nabla}_ix\overline{\nabla}_jx-\overline{g}_{ij}|dx|^2_{\overline
g}). \ee Throughout this section we  use the symbol $|\cdot |$ for
$|\cdot |_g$, but we write explicitly $|\cdot |_{\overline g}$
when the ${\overline g}$ metric is involved.
\begin{Proposition}
\label{prop:estiDNah} For any $\epsilon >0$ and $s\in\R$, there
exists $x_{\epsilon,s}>0$  such that for any differentiable tensor
field $u$ with compact support in
 $\{ 0<x<x_{\epsilon,s}\}$
we have \be \label{pinbah}\int_M x^{2s} |x^{-2}\nabla (xu)|^2 d\mu
\ge \left\{[s-(n+3)/2]^2-\epsilon\right\} \int_{{\mathcal
O}_{\epsilon}} x^{2s-2} | u|^2 |dx|^2_{\overline{g}}d\mu \;. \ee
\end{Proposition}

\proof We use Proposition \ref{prop:poinc}, choosing $v=(s-2)\ln
x$ one has $dv=(s-2)dx/x$ and $$ \Delta v=(s-2)[-|dx|^2/x^2+\Delta
x/x]=(s-2)(1-n)|dx|^2/x^2+o(1)\;. $$ It follows that
$$|dv|^2+\Delta v=[(s-2)^2+(s-2)(1-n)+o(1)]|dx|^2/x^2\;.$$
Choosing  $w=[(1-n)\ln x]/2$ we have that
$$-|dw|^2+\Delta w=[-(1-n)^2/4+(1-n)^2/2+o(1)]|dx|^2/x^2\;.$$
\qed

\section{Weighted estimates for vector fields}
\label{ApL2} In this section we give some estimates for the
operator $S$ which associates to a vector field $Y$ one half of
the Lie derivative of the metric along $Y$ :
$$
S(Y)_{ij}=({\mathcal L}_{Y}g)_{ij}:=\frac{1}{2}(\nabla_i
Y_j+\nabla_jY_i).
$$
As it will be often used, we recall that
$$
\tr(S(Y))=\divr Y=\nabla^iY_i.
$$

\begin{lem}\label{lem:estiPS}
For all vector fields $V$ and all vector fields $Y$ with compact
support we have the equality
$$
\int_M [S(Y)+\frac{1}{2}\tr(S(Y))g](Y,V)= -\frac{1}{2}\int_M
\nabla V(Y,Y)+\frac{1}{2}\divr(V)|Y|^2.
$$
\end{lem}
\begin{proof}
We integrate by parts the two terms on the right-hand side of the
equality
$$
S(Y)_{ij}Y^iV^j=\frac{1}{2}(\nabla_iY_j Y^iV^j+\nabla_jY_i
Y^iV^j).
$$
\qed
\end{proof}
\begin{prop}\label{prop:estiPS}
For all functions $u$, all vector fields $V$ and all vector fields
$Y$ with compact support we have the equality
\begin{eqnarray*}
\lefteqn{ \int_M e^{2u}[S(Y)+\frac{1}{2}\tr(S(Y))g](Y,V)}&&
\\ &&=
-\frac{1}{2}\int_M e^{2u}\left\{\right.\nabla
V(Y,Y)+\frac{1}{2}\divr(V)|Y|^2+\langle  du,V\rangle  |Y|^2
+2\langle  du,Y\rangle  \langle V,Y\rangle  \left.\right\}.
\end{eqnarray*}
\end{prop}
\begin{proof}
We use Lemma \ref{lem:estiPS} with $Y$ replaced by $e^u Y$, so
that
\begin{eqnarray*}
\lefteqn{ \int_M [S(e^uY)+\frac{1}{2}\tr(S(e^u Y))g](e^uY,V)}&&
\\ &&=\int_M e^{2u}\left\{\right.[S(Y)+\frac{1}{2}\tr(S(Y))g](Y,V)
\\ &&\qquad+\frac{1}{2}(\nabla_iuY_j+\nabla_juY_i
+\langle  du,Y\rangle   g_{ij})Y^iV^j\left.\right\}\\
&&= \int_M
e^{2u}\left\{\right.[S(Y)+\frac{1}{2}\tr(S(Y))g](Y,V)+\frac{1}{2}\langle
du,V\rangle  |Y|^2+ \langle du,Y\rangle \langle  V,Y\rangle
\left.\right\}.
\end{eqnarray*}
\end{proof}
\vspace{2mm}


\begin{prop}\label{prop:estiPS2}
For all  vector fields $Y$ with compact support and functions $u$
and $v$ supported in a neighborhood of the support of $Y$, we have
\begin{eqnarray*}
\lefteqn{
 -2\int_M ve^{2u}S(Y)(\nabla v,\nabla v)\langle dv,Y\rangle
 \hspace{5cm}}&&
\\&&=\int_M e^{2u}\langle dv,Y\rangle  \left[\langle dv,Y\rangle  (|dv|^2+v\Delta
v+2v\langle dv,du\rangle  )\right. \left.+2v\nabla\nabla
v(Y,\nabla v)\right].
\end{eqnarray*}
\end{prop}
\begin{proof}
Integrating $\nabla_j[(\langle dv,Y\rangle)^2ve^{2u}\nabla^jv]$
over $M$ one has
\begin{eqnarray*}
-\int_M \langle dv,Y\rangle  ^2|dv|^2e^{2u}= \int_M\left\{
2\langle dv,Y\rangle  \left[\nabla\nabla v(Y,\nabla
v)+\nabla Y(\nabla v,\nabla v)\right]v e^{2u}\right.\\
\hspace{5cm}+\langle dv,Y\rangle  ^2\left.\left[v\Delta v
e^{2u}+2v\langle dv,du\rangle e^{2u}\right]\right\},
\end{eqnarray*}
and $\nabla Y(\nabla v,\nabla v)=S(Y)(\nabla v,\nabla v).$
\end{proof}\qed

\subsection{Application: compact boundaries}\label{SE.1}
We use here the notations of Section \ref{Scb}. Similarly to
Section~\ref{SD.1} we assume that \eq{pi0} holds.

\begin{cor}\label{Ce4}
For  all $s\in\Reel$ and all $\epsilon>0$ there exists
$x_{s,\epsilon}>0$ such that for
 all vector fields $Y$ with compact support in $\{0<x<x_{s,\epsilon}\}$ we have
\begin{eqnarray*}
\int_M x^{2s}
[S(Y)+\frac{1}{2}\tr(S(Y))g](Y,\nabla x/x)=\hspace{4cm}\mbox{ }\\
\;\;\; \frac{1}{2}\int_M
x^{2s-2}\Big[\left.(\frac{1}{2}-s)\Big(\right.|dx|^2|Y|^2+2\langle
Y,\nabla x\rangle  ^2\Big)+ o(1)|Y|^2\left.\right.\Big]
\end{eqnarray*}
\end{cor}
\begin{proof}
We apply Proposition \ref{prop:estiPS} with the vector field
$V=\nabla x/x$ and the function $u=s\ln(x)$, so that $du=sdx/x$
and $\nabla\nabla u=-s\nabla x\nabla x/x^2+s\nabla\nabla x/x
=-s\nabla x\nabla x/x^2+o(x^{-2})$.
\end{proof}
\begin{cor}\label{CE5}
For  all $s,t\in\Reel$ and all $\epsilon>0$ there exists
$x_{s,t,\epsilon}>0$ such that for
 all vector fields $Y$ with compact support in $\{0<x<x_{s,t,\epsilon}\}$ we
 have
\begin{eqnarray*}
\int_M x^{2t} e^{-2s/x}
[S(Y)+\frac{1}{2}\tr(S(Y))g](Y,\nabla x/x)=\hspace{4cm}\mbox{ }\\
\;\;\; -\int_M x^{2t-4}e^{-2s/x}\Big[\left.\frac s2
\Big(\right.|dx|^2|Y|^2+2\langle Y,\nabla x\rangle  ^2\Big)+
o(1)|Y|^2\left.\right)\Big]\;.
\end{eqnarray*}
\end{cor}
\begin{proof}
We apply Proposition \ref{prop:estiPS} with the vector field
$V=\nabla x/x^2$ and the function $u=-s/x+t\ln x$, so that we have
$du=-sdx/x^2+t dx/x$ and  $\nabla V=o(x^{-4})$.
\end{proof}
\qed
\subsection{Application: asymptotically flat metrics}
In this section we assume that \eq{metfo} holds, while \eq{metfo1}
will be strengthened to \bel{metfo2} r\nabla\nabla r=\delta-\nabla
r\nabla r +o(1/r)\;.\ee
\begin{cor}\label{cor:PSaf}
For all $s\in\Reel$ and
 all vector fields $Y$ with compact support  near infinity we have
\begin{eqnarray}\nonumber
\int_M r^{-2s-n+2}
[S(Y)+\frac{1}{2}\tr(S(Y))g](Y,\nabla r/r)=\hspace{4cm}\mbox{ }\\
\;\;\; \frac{1}{2}\int_M r ^{-2s-n}\left[(s-1)|Y|^2+(2s+n)\langle
Y,\nabla r\rangle  ^2+ o(1)|Y|^2\right]. \label{PSaf}
\end{eqnarray}
\end{cor}
\begin{proof}
We apply  Proposition \ref{prop:estiPS} with the vector field
$V=\nabla r/r=\nabla(\ln(r))$ and the function
$u=(-s-n/2+1)\ln(r)$, then $du=(-s-n/2+1)\nabla r/r$ and
$\nabla\nabla u=(-s-n/2+1)(-\nabla r\nabla r/r^2+\nabla\nabla r/r)
=(-s-n/2+1)(-2\nabla r\nabla r/r^2+\delta/r^2) +o(r^{-2})$, recall
that  $\lim_{r\rightarrow\infty}|\nabla r|^2=1$.
\end{proof}\qed

\begin{cor}\label{cor:PSafexp}
For all $s\in\Reel$ and
 all vector fields $Y$ with compact support  near infinity we
 have
\begin{eqnarray}\nonumber
\int_M e^{-2sr}
[S(Y)+\frac{1}{2}\tr(S(Y))g](Y,\nabla r)=\hspace{4cm}\mbox{ }\\
\;\;\; -\int_M e ^{-2sr}\left[s(|Y|^2+2\langle Y,\nabla r\rangle
^2)+ o(1)|Y|^2\right]. \label{PSafexp}
\end{eqnarray}
\end{cor}
\begin{proof}
We apply the Proposition \ref{prop:estiPS} with the vector field
$V=\nabla r$ and the function $u=-sr$.
\end{proof}

\begin{cor}\label{prop:estiPS2af}
For all  vector fields $Y$ with compact support , we have
$$
 \int_M r^{-2s-n+1}S(Y)(\nabla r,\nabla r)\langle dr,Y\rangle  =\int_M r^{-2s-n}(s+o(1))\langle dr,Y\rangle  ^2.
$$
\end{cor}
\begin{proof}
We use Proposition~\ref{prop:estiPS2} with $v=r$ and
$u=(-s-n/2)\ln r$, together with the fact that
$$
r\nabla\nabla r=-\nabla r\nabla r+\delta+o(1),
$$
and $|dr|^2=1+o(1)$.\qed
\end{proof}

We obtain finally the desired inequalities:

\begin{prop}\label{prop:estiPS2afg}
For all $s\neq 0,1$, there exist $C_s>0$ and $R(s)$ such that  for
all vector fields $Y$ with compact support in $\{r>R(s)\}$, we
have
$$
\int_M r^{-2s-n+2}|S(Y)|^2\geq C_s\int_M r^{-2s-n}|Y|^2.
$$
\end{prop}
\begin{proof}
From Corollary \ref{prop:estiPS2af},  for all $b>0$, we have
$$
\frac{b}{2}\int_M
r^{-2s-n+2}|S(Y)|^2+\frac{1}{2b}\int_Mr^{-2s-n}|\nabla r|^2\langle
dr,Y\rangle  ^2\geq\int_M r^{-2s-n}(|s|+o(1))\langle dr,Y\rangle
^2.
$$
We conclude by using Corollary \ref{cor:PSaf} and the inequality
$$
|[S(Y)+\frac{1}{2}\tr(S(Y))g](Y,\nabla r/r)|\leq
\frac{a}{2}|S(Y)+\frac{1}{2}\tr(S(Y))g|^2+\frac{1}{2a}r^{-2}|Y|^2|dr|^2,
$$
for all $a>0$, together with the inequality
$$
|S(Y)|^2\geq \frac{1}{n}|\tr S(Y)|^2.
$$
\end{proof}\qed

\subsection{Application: conformally compact manifolds}

We recall that we have $g=x^{-2}\overline{g}$.
\Eq{hessah} gives
\begin{eqnarray}\nonumber
\nabla_i\nabla_j (x^{-1})&=&2x^{-3}{\nabla}_ix{\nabla}_jx
-x^{-2}{\nabla}_i\nabla_jx=x^{-3}\overline{g}_{ij}|dx|^2_{\overline{g}}
-x^{-2}\overline{\nabla}_i\overline{\nabla}_jx
\\&=&x^{-1}|dx|^2_{\overline{g}}g_{ij}+\mbox{l.o.}\,\label{Hessah2}\end{eqnarray}
where ``l.o." denotes terms which are small compared to the
remaining ones.
\begin{cor}\label{cor:PSah} For all $s\in\Reel$ and
 all vector fields $Y$ with compact support  near the boundary we
 have
\begin{eqnarray*}
\int_M x^{2s}
[S(Y)+\frac{1}{2}\tr(S(Y))g](Y,\nabla x/x)=\hspace{4cm}\mbox{ }\\
\;\;\; \frac{1}{2}\int_M
x^{2s}\left.\Big(\right.(\frac{1+n}{2}-s)|dx|^2_{\overline{g}})|Y|^2-(2s+1)\langle
dx/x,Y\rangle  ^2|dx|^2_{\overline{g}}+
o(1)|Y|^2\left.\right.\Big)\;.
\end{eqnarray*}
\end{cor}
\begin{proof}
We apply Proposition \ref{prop:estiPS} with the vector field
$V=\nabla (x^{-1})=-x^{-2}\nabla x$ and the function
$u=(s+1/2)\ln(x)$, using \eq{Hessah2} one then has $\nabla
V=\nabla\nabla(x^{-1})=x^{-1}|dx|^2_{\overline{g}}g+o(x^{-3})$.\qed
\end{proof}


\begin{cor}\label{prop:estiPS2ah}
For all  vector fields $Y$ with compact support , we have
$$
 2\int_M x^{2s}S(Y)(\nabla x/x,\nabla x/x)\langle dx/x,Y\rangle
 =\int_M x^{2s}(n-2s-1+o(1))\langle dx/x,Y\rangle  ^2
|dx|^2_{\overline{g}}\;.$$
\end{cor}
\begin{proof}
We apply Proposition \ref{prop:estiPS2} with $v=x^{-1}$,
$u=(s+2)\ln x$, making use of \eq{Hessah2}. \qed
\end{proof}

\begin{prop}\label{prop:estiPS2ahg}
For all $s\neq (n+1)/2, (n-1)/2$ there exist constants $C_s>0$,
$x(s)>0$ such that for all differentiable vector fields $Y$ with
compact support in $\{x<x(s)\}$ we have
$$
\int_M x^{2s}|S(Y)|^2\; d\mu_g\geq C_s\int_M
x^{2s}|Y|^2\;d\mu_g\;.
$$
\end{prop}
\begin{proof}
From Corollary \ref{prop:estiPS2ah},  for all $b>0$, we have
\begin{eqnarray*}
\lefteqn{ \frac{b}{2}\int_M
x^{2s}|S(Y)|^2+\frac{1}{2b}\int_Mx^{2s}|\nabla x/x|^2\langle
dx/x,Y\rangle  ^2}&&
\\ &&\geq\int_M x^{2s}(|n-1-2s|+o(1))\langle
dx/x,Y\rangle ^2|dx|^2_{\overline{g}}.
\end{eqnarray*}
We conclude by using Corollary \ref{cor:PSah} and the inequality
$$
|[S(Y)+\frac{1}{2}\tr(S(Y))g](Y,\nabla x/x)|\leq
\frac{a}{2}|S(Y)+\frac{1}{2}\tr(S(Y))g|^2+\frac{1}{2a}|Y|^2|dx/x|^2,
$$
for all $a>0$, together with
$$
|S(Y)|^2\geq \frac{1}{n}|\tr S(Y)|^2.
$$\qed
\end{proof}

From the last result we also get an inequality governing  the
Hessian operator:

\begin{prop}\label{prop:estiN}
For all $s\neq (n+1)/2,(n-1)/2, (n-3)/2$ there exist constants
$C_s>0$ and $x(s)$ such that for all differentiable functions $N$
with compact support in $\{x<x(s)\}$, we have
$$
\int_M x^{2s}|\nabla\nabla N-\Delta N g-N \Ric g|^2\;d\mu_g\geq
C_s\int_M x^{2s}(|N|^2+|\nabla N|^2)\;d\mu_g\;.
$$
\end{prop}
\begin{proof}
We will use Proposition \ref{prop:estiPS2ahg} with
$$ Y=x^{-1}\nabla N-N\nabla(x^{-1})=x^{-2}\nabla(xN).$$
By \eq{Hessah2} we have
$$
\nabla_{(i}Y_{j)}=x^{-1}\nabla_i\nabla_j
N-N\nabla_i\nabla_j(x^{-1})=x^{-1}(\nabla_i\nabla_j
N-N|dx|^2_{\overline{g}}g)+Nx^{-2}\overline{\nabla}_i\overline{\nabla}_jx\;,
$$
then
$$S(Y)-\mathrm{div}Y g=
x^{-1}\Big[\nabla\nabla N-\Delta N
g+(n-1)N|dx|^2_{\overline{g}}g+N(x^{-1}\overline{\nabla}\overline{\nabla}x
-x^{-1}\Delta_{\overline{g}}x \overline{g}) \Big].
$$
On the other hand we have
\begin{eqnarray*}
\Ric g&=&\Ric
\overline{g}+x^{-1}[(n-2)\overline{\nabla}\overline{\nabla}x+(\overline{\Delta}x)\overline{g}]
-(n-1)|dx|^2_{\overline{g}}x^{-2}\overline{g}
\\ &=&-(n-1)|dx|^2_{\overline{g}}{g}+\mathrm{l.o.}
\end{eqnarray*}
Finally, we obtain
$$S(Y)-\mathrm{div}Y g=
x^{-1}[\nabla\nabla N-\Delta N g-N(\Ric g+\mathrm{l.o.})]\;.
$$
Now, we use the inequality
$$
|S(Y)-\tr S(Y) g|^2=|S(Y)|^2+(n-2)(\tr S(Y))^2\geq|S(Y)|^2,
$$
and Proposition \ref{prop:estiPS2ahg} with $s$ there replaced by
$s+1$ yields
$$
\int x^{2s}\Big(|\nabla\nabla N-\Delta N g-N\Ric g|^2
+o(1)N^2\Big) \ge C \int x^{2s-2 } |\nabla (xN)|^2\;.
$$ The result follows now from the following calculation, where
Proposition~\ref{prop:estiDNah} with $s$ there equal to $s+1$ is
used when going from the second to the third line:
$$
\begin{array}{lll}
\|x^{s-1}\nabla(xN)\|_{L^2}&=&\epsilon\|x^{s-1}\nabla(xN)\|_{L^2}+(1-\epsilon)\|x^{s-1}\nabla(xN)\|_{L^2}\\
&\geq&\epsilon\|x^{s}\nabla N\|_{L^2}-\epsilon\|x^{s-1}N\nabla x\|_{L^2}+(1-\epsilon)\|x^{s-1}\nabla(xN)\|_{L^2}\\
&\geq&\epsilon\|x^{s}\nabla N\|_{L^2}-\epsilon\|x^{s-1}N\nabla x\|_{L^2}+(1-\epsilon)c\|x^sN\|_{L^2}\\
&\geq&C(\|x^{s}\nabla N \|_{L^2}+\|x^sN\|_{L^2})\\
\end{array}
$$
 \qed
\end{proof}

\medskip

\section{Poincar\'e  charges}\label{Sgc}
Let $\hyp$ be an
$n$-dimensional spacelike hypersurface in a $n+1$-dimensional
Lorentzian space-time $(\mcM,g)$, $n\ge 2$. Suppose that $\mcM$
contains an open set $\mcU$ with a global time coordinate $t$
(with range not necessarily equal to $\R$), as well as a global
``radial'' coordinate $r\in[R,\infty)$, leading to local
coordinate systems $(t,r,v^A)$, with $(v^A)$ --- local coordinates
on some compact $n-1$ dimensional manifold $\mn$. We further
require that $\hyp\cap \mcU=\{t=0\}$. Assume that the metric $g$
approaches (as $r$ tends to infinity, in a sense which is made
precise below) a background metric $b$. The Hamiltonian analysis
of vacuum general relativity in~\cite{ChAIHP} (see also
\cite[Section 5]{ChruscielSimon} or \cite[Appendix~A]{ChNagy})
leads to the following expression for the Hamiltonian associated
to the flow of a vector field $X$, assumed to be a Killing vector
field of the background $b$:\footnote{The integral over $\partial
\hyp$ should be understood by a limiting process, as the limit as
$R$ tends to infinity of integrals over the sets $t=0$, $r=R$. $d
S_{\alpha\beta}$ is defined as
  $\frac{\partial}{\partial x^\alpha}\lrcorner
  \frac{\partial}{\partial x^\beta}\lrcorner \rd x^0 \wedge\cdots
  \wedge\rd x^{n} $, with $\lrcorner$ denoting contraction; $g$ stands
  for the space-time metric  unless explicitly indicated
  otherwise. Further, a semicolon denotes covariant differentiation
  \emph{with respect to the background metric $b$}.}
\begin{eqnarray}
  H(\hyp,g,b, X)&= &\frac 12 \int_{\partial\hyp}
 \ourU^{\alpha\beta}dS_{\alpha\beta}\;,
\label{toto}
\end{eqnarray}\begin{eqnarray}
  \ourU^{\nu\lambda}&= &
{\ourU^{\nu\lambda}}_{\beta}X^\beta + \frac 1{8\pi}
\left(\sqrt{|\det g_{\rho\sigma}|}~g^{\alpha[\nu}-\sqrt{|\det
b_{\rho\sigma}|}~ b^{\alpha[\nu}\right) {X^{\lambda]}}_{;\alpha} \
,\label{Fsup2new}
\\ {\ourU^{\nu\lambda}}_\beta &= & \displaystyle{\frac{2|\det
  \bmetric_{\mu\nu}|}{ 16\pi\sqrt{|\det g_{\rho\sigma}|}}}
g_{\beta\gamma}(e^2 g^{\gamma[\nu}g^{\lambda]\kappa})_{;\kappa}
\;,\label{Freud2.0} 
\\
  \label{mas2}
e&=& 
{\sqrt{|\det g_{\rho\sigma}|}}/{\sqrt{|\det\bmetric_{\mu\nu}|}}\;
.
\end{eqnarray}
(The question of convergence of the right-hand-side of \eq{toto}
will be  considered shortly. The last term in \eq{Fsup2new} is
actually identically zero for asymptotically Euclidean
hypersurfaces, but does not vanish for hyperboloidal hypersurfaces
and is necessary there to ensure convergence of the integral.) The
form \eq{toto} is most convenient when trying to establish
formulae such as \eq{Pcm2a} below, expressing the
Poincar\'e--covariance of the Hamiltonians.

\subsection{Initial data asymptotically flat in spacelike directions}
\label{Ssafio}
 Consider, to start with, Lorentzian metrics which
are asymptotically flat in the following sense: there exists a
coordinate system $x^\mu$ covering a set which contains $$\hyp_0:=
\{x^0=0\;, r(x):=\sqrt{\sum (x^i)^2}> R\}\;,$$ and assume that the
tensors $g_{\mu\nu}:=g(\partial_\mu,\partial_\nu)$ and
$b_{\mu\nu}:=b(\partial_\mu,\partial_\nu)$ satisfy along $\hyp_0$
{\rm \beadl{af1a}
&b_{\mu\nu}=\eta_{\mu\nu}:=\textrm{diag}(-1,+1,\ldots,+1)\;,& \\ &
|g_{\mu\nu}- b_{\mu\nu}|\le Cr^{-\alpha}\;, \ |\partial_\sigma
g_{\mu\nu}|\le Cr^{-\alpha-1}\;,  \quad n/2-1<\alpha\le n- 2
\;.&\label{af1b}\eeadl{af1}}If one further assumes that the
energy-momentum tensor $T_{\mu\nu}$ of $g$ is in $L^1(\hyp_0)$,
then the {\em ADM energy-momentum vector} defined as
\bel{mi1}p_\mu(\hyp_0):=H(\hyp_0,g,b,\partial_\mu)\ee is finite
and well defined~\cite{Bartnik:mass,ChErice,Chmass}. The
finiteness of the Lorentz charges,
\bel{mi2}J_{\mu\nu}(\hyp_0):=H(\hyp_0,g,b,x_\mu\partial_\nu-x_\nu\partial_\mu)\;,\ee
where $x_\mu:=\eta_{\mu\nu}x^\nu$, requires further restrictions
-- there are various ways to
proceed~\cite{Changmom,BOM:poincare,Solovyev,RT}, the following is
convenient for our purposes: let $\Omega\subset \R^{1,n}$ be
invariant under the transformation \bel{af3}x^\mu\to -x^\mu\;,\ee
for any $f:\Omega\to \R$ we set
$$f^{+}(x)= \frac 12 \left(f(x)+f(-x)\right)\;,\quad
f^{-}(x)= \frac 12 \left(f(x)-f(-x)\right)\;. $$ We shall
henceforth only consider metrics defined on  domains of coordinate
systems which are invariant under \eq{af3}, and we will assume
that in addition to \eq{af1} we have \bel{af4} |g_{\mu\nu}^-|\le C
(1+r)^{-\alpha_-}\;, \ |\partial_\sigma (g_{\mu\nu}^-)|\le C
(1+r)^{-1-\alpha_-}\;, \quad \alpha_->\alpha\;,\
\alpha+\alpha_->n-1\;.\ee We note that in dimension $n+1=3+1$ ,
Equations \eq{af1} and \eq{af3} hold for the Schwarzschild metric
in the usual static coordinates, with $\alpha=1$ and $\alpha_-$
--- as large as desired. Similarly \eq{af1}, \eq{af3} hold for the
Kerr metric in the Boyer-Lindquist coordinates, discussed in
Section~\ref{Skn} below, with $\alpha=1$ and $\alpha_-=2$.

Recall that a {\emph{boost-type domain}}
$\Omega_{R,T,\theta}\subset \R^{1,n}$ is defined as\bel{btd}
\Omega_{R,T,\theta}:=\{r > R\;,\ |t| <\theta r+ T\}\;,\ee with
$\theta\in (0,\infty]$. We have the following:
\begin{Proposition}\label{Pbtd} Let $g_{\mu\nu}$ be a Lorentzian metric
 satisfying \eq{af1} and \eq{af4} on a boost-type domain $\Omega_{R,T,\theta}$,
 and suppose that the coordinate components  $\mcT_{\mu\nu}:=\mcT(\partial_\mu,\partial_\nu)$
 of the energy-momentum
tensor density, \bel{P10.0} \mcT_{\mu\nu}:= \frac {\sqrt{|\det
g_{\alpha\beta}|}} {8 \pi} \left(\Ricc_{\mu\nu}-\frac 12
\tr_g\Ricc g_{\mu\nu}\right)\;, \ee satisfy \bel{P10}
|\mcT_{\mu\nu}|\le C (1+r)^{-n-\epsilon}\;,\qquad
|\mcT_{\mu\nu}^-|\le C (1+r)^{-n-1-\epsilon}\;,\qquad \epsilon
>0.\ee Let $\hyp\subset\Omega_{R,T,\theta} $ be the hypersurface $\{y^0=0\}\cap \Omega_{R,T,\theta} $,
where the coordinates $y^\mu$ are obtained from the $x^\mu$'s by a
Poincar\'e transformation, \bel{P1} x^\mu\to y^\mu:=
\Lambda^\mu{}_\nu x^\nu +a^\mu\;,\ee so that $\Lambda^\mu{}_\nu $
is a constant-coefficients Lorentz matrix, and $a^\mu$ is a set of
constants, set $\hyp_0:= \{x^0=0\}$. Then:
\begin{enumerate}
\item The integrals defining  the ``Poincar\'e charges" \eq{mi1}-\eq{mi2} of $\hyp$
and $\hyp_0$ converge.
\item We have
\begin{eqnarray}\nonumber (p_\mu(\hyp),J_{\mu\nu}(\hyp))&=&(\Lambda_\mu{}^\alpha
p_\alpha(\hyp_0), \Lambda_\mu{}^\alpha\Lambda_\nu{}^\beta
J_{\alpha\beta}(\hyp_0) \\ &&  + a_\mu \Lambda_\nu{}^\alpha
p_\alpha(\hyp_0) - a_\nu \Lambda_\mu{}^\alpha
p_\alpha(\hyp_0))\;.\label{Pcm2a}\end{eqnarray}
\end{enumerate}
Here $\Lambda_\alpha{}^\beta:=\eta_{\alpha\mu}\Lambda^\mu{}_\nu
\eta^{\nu\beta} $ and $p_\mu(\hyp_0)=
H(\hyp_0,g,b,\partial/\partial x^\mu)$, while $p_\mu(\hyp)=
H(\hyp,g,b,\partial/\partial y^\mu)$, similarly for $J_{\mu\nu}$.
\end{Proposition}

\proof We have~\cite{ChAIHP}
\begin{equation}
   \int_{\{x^0=0,r=R\}}
 \ourU^{\alpha\beta}dS_{\alpha\beta}=  2\int_{\{x^0=0,R_0\le r\le R\}}
 \znabla_\beta \ourU^{\alpha\beta}dS_{\alpha}  + \int_{\{x^0=0,r=R_0\}}
 \ourU^{\alpha\beta}dS_{\alpha\beta}\;,
\label{C2}
\end{equation}
with
\begin{eqnarray}
  16\pi\znabla_\beta \ourU^{\alpha\beta} &= &   \mcT^\alpha{}_\beta X^\beta + \sqrt{|\det    b|}
  \;\left( Q^\alpha{}_{\beta}
X^\beta + Q^{\alpha\beta}_{}{ \gamma}\;
    \znabla_\beta X^\gamma\right)\;, \label{C3-}\end{eqnarray} where
    $Q^\alpha{}_{\beta}$ is a quadratic form in $\znabla_\sigma g_{\mu\nu}$, and
    $Q^{\alpha\beta}{}_{ \gamma}$ is bilinear in $\znabla_\sigma g_{\mu\nu}$ and
    $g_{\mu\nu}-b_{\mu\nu}$, both with bounded coefficients which are constants plus terms $O(r^{-\alpha})$.
    For $p_\mu$ and for $R\ge R_0$ one immediately
    obtains
\begin{eqnarray}
   \int_{\{x^0=0,r=R\}}
 \ourU^{\alpha\beta}dS_{\alpha\beta}&=& \int_{\{x^0=0,r=R_0\}}
 \ourU^{\alpha\beta}dS_{\alpha\beta} + O(R_0^{n-2-2\alpha})
  \nonumber\\ && +\frac 1 {8\pi} \int_{\{x^0=0,R_0\le r\le R\}}
  \mcT^\alpha{}_\beta X^\beta dS_{\alpha}\label{C3}\\&=& \int_{\{x^0=0,r=R_0\}}
 \ourU^{\alpha\beta}dS_{\alpha\beta} + O(R_0^{n-2-2\alpha})+O(R_0^{-\epsilon})\;.
  \nonumber \\ &&
\label{C4}\end{eqnarray} For $J_{\mu\nu}$ simple parity
considerations lead instead to
\begin{equation}
   \int_{\{x^0=0,r=R\}}
 \ourU^{\alpha\beta}dS_{\alpha\beta}= \int_{\{x^0=0,r=R_0\}}
 \ourU^{\alpha\beta}dS_{\alpha\beta} + O(R_0^{n-1-\alpha-\alpha_-})+O(R_0^{-\epsilon})\;.
\label{C5}\end{equation} Passing to the limit $R\to \infty$ one
obtains convergence of $p_\mu(\hyp_0)$ and of
$J_{\mu\nu}(\hyp_0)$. For further reference we note the formulae
\begin{deqarr}
  p_\mu(\hyp_0)&= &\int_{\{x^0=0,r=R_0\}}
 \ourU^{\alpha\beta}dS_{\alpha\beta} + \frac 1 {16 \pi} \int_{r\ge R_0} \mcT^\mu{}_\nu X^\nu dS_\mu +
 O(R_0^{n-2-2\alpha})\;,
 \nonumber \\ &&
\label{C7a}\\
   J_{\mu\nu}(\hyp_0)&= &\int_{\{x^0=0,r=R_0\}}
 \ourU^{\alpha\beta}dS_{\alpha\beta}
  + \frac 1 {16 \pi} \int_{r\ge R_0} \mcT^\mu{}_\nu X^\nu dS_\mu
   + O(R_0^{n-1-\alpha-\alpha_-})\;.
   \nonumber \\ &&
\label{C7b}\arrlabel{C7}\end{deqarr} Because Lorentz
transformations commute with the antipodal map \eq{af3} the
boundary conditions \eq{af1} and \eq{af4} are preserved under
them, and convergence of the Poincar\'e charges of $\hyp$ for
transformations of the form \eq{P1} with $a^\mu=0$ follows. In
order to establish point 2., still for  $a^\mu=0$, we use Stokes'
theorem on a set $\mcT_R$ defined as \bel{C9} \mcT_R=\{r=R, 0 \le
t \le -(\Lambda^0{}_0)^{-1}\Lambda^0{}_i x^i\}\cup \{r=R, 0\ge t
\ge -(\Lambda^0{}_0)^{-1}\Lambda^0{}_i x^i\}\;,\ee so that the
boundary $\partial\mcT_R$ has two connected components, the set
$\hyp_0\cap \{r=R\}$ and the set $\hyp\cap\{r=R\}$. This leads to
\begin{eqnarray}
   \int_{\hyp\cap\{r=R\}}
 \ourU^{\alpha\beta}dS_{\alpha\beta}=  2\int_{\mcT_R}
 \znabla_\beta \ourU^{\alpha\beta}dS_{\alpha}  + \int_{\hyp_0\cap\{r=R\}}
 \ourU^{\alpha\beta}dS_{\alpha\beta}\;,
\label{C10}
\end{eqnarray}
The boundary conditions ensure that the integral over $\mcT_R$
vanishes in the limit $R\to\infty$ (for $p_\mu$ this is again
straightforward, while for $J_{\mu\nu}$ this follows again by
parity considerations), so that
\begin{eqnarray}
  H(\hyp,g,b, X)&= &H(\hyp_0,g,b, X)\;.
\label{toto2}
\end{eqnarray}
We consider finally a translation; Stokes' theorem on the
$n$--dimensional region
$$\{y^\mu=x^\mu+sa^\mu\;,s\in[0,1]\;,x^\mu\in \hyp\;, r(x^\mu)=R\}$$
leads again --- in the limit $R\to\infty$ --- to \eq{toto2}, in
particular $H(\hyp,g,b, X)$ converges. The transformation law
\eq{Pcm2a} follows now from \eq{toto2} by the following
calculation:
\begin{eqnarray*}
J_{\mu\nu}(\hyp)&:=& H(\hyp,g,b,
y_\mu\textstyle{{\frac{\partial}{\partial y^\nu}}}-
y_\nu\textstyle{{\frac{\partial}{\partial y^\mu}}}) \\
&=& H(\hyp_0,g,b, y_\mu\textstyle{{\frac{\partial}{\partial
y^\nu}}}- y_\nu\textstyle{{\frac{\partial}{\partial y^\mu}}}) \\&
=&H(\hyp_0,g,b,
(\Lambda_{\mu}{}^{\alpha}x_\alpha+a_\mu)\Lambda_\nu{}^\beta\textstyle{{\frac{\partial}{\partial
x^\beta}}}-
(\Lambda_{\nu}{}^{\alpha}x_\alpha+a_\nu)\Lambda_\mu{}^\beta\textstyle{{\frac{\partial}{\partial
x^\beta}}})\;.\end{eqnarray*} \qed

\newcommand{\ngm}{{}^{n+1}g}%
It is convenient to have a $n+1$ version of \eq{toto}, in the
asymptotically flat vacuum case this is easily implemented as
follows: let $(\hyp,K,g)$ be an asymptotically flat vacuum initial
data set, if the data are sufficiently differentiable there exists
a vacuum development $(M,\ngm)$ of the data so that $\hyp$ can be
isometrically identified with a hypersurface $t=0$ in $M$, with
$K$ corresponding to the second fundamental form of $\hyp$ in
$(M,\ngm)$. We can introduce Gauss coordinates around $\hyp$ to
bring $\ngm$ to the form
$$\ngm=-dt^2 +g_t\;$$ where $g_t$ is a family of Riemannian
metrics on $\hyp$ with $g_0=g$. We then set
$$b=-dt^2+e\;,$$ where $e$ is the Euclidean flat metric equal to diag$(+1,\ldots,+1)$ in
asymptotically flat coordinates on $\hyp$. Let $n_b$ be the future
directed $b$-unit normal to $\hyp$ and let $(Y,N)$ be the KID
determined on $\hyp$ by the $b$-Killing vector $X$; by definition,
\bel{kiddef} X=Nn_b+Y\;, \ b(n_b,Y)=0 \ \textrm{ along $\hyp$}
\;.\ee Since the future pointing  $g$-unit normal to $\hyp$, say
$n_g$, coincides with $n_b$, we also have \bel{kiddef2}X=N n_g+Y
\;, \ g(n_g,Y)=0\;.\ee We define the \emph{Poincar\'e charges} $Q$
by the formula \bel{mi3a1} Q((Y,N),(K,g)):= H(\hyp,\ngm,b,
VN+Y)\;. \ee It is well known that the integrand of \eq{mi3a1} can
be expressed in terms of $K$, $g$, as well as the first
derivatives of $g$. The $n+1$ form of \eq{C3} reads
\begin{eqnarray}
   \int_{\{x^0=0,r=R\}}
 \ourU^{\alpha\beta}dS_{\alpha\beta}&=& \int_{\{x^0=0,r=R_0\}}
 \ourU^{\alpha\beta}dS_{\alpha\beta} 
 \nonumber \\ && +\frac 1 {8\pi} \int_{\{x^0=0,R_0\le r\le R\}}
  \left(Y^i J_i + N \rho+q\right) d\mu_g\;,
  \nonumber \\ &&
\label{C4a}\end{eqnarray}
where $q$ is a quadratic form in $g_{ij}-\delta_{ij}$, $\partial_k
g_{ij}$, and $K_{ij}$, with uniformly bounded coefficients
whenever $g_{ij}$ and $g^{ij}$ are uniformly bounded. This follows
immediately from \eq{C2}-\eq{C3-}, together with the $n+1$
decomposition of the energy-momentum tensor density \eq{P10.0},
and of the error term in \eq{C3-}. One can also work directly with
the $n+1$ equivalents of the boundary integrals in \eq{C4a} ---
{\em cf.,\/ e.g.},\/~\cite{BOM:poincare} --- but those are
somewhat cumbersome when studying behavior of the charges under
Lorentz transformations.

\section{The reference family of Kerr metrics}\label{Skn}
 Let us denote by $\cKi$ the family of Cauchy data $(g,K)$
 obtained
 as follows: let $\fg$ be a Kerr metric with $m\ne 0$,
 $a\in\R$;
 in Boyer-Lindquist coordinates
$(t,r,\theta,\varphi)$ we have \cite[p.~100]{LPPT} (see also
\url{http://grdb.org}) \bea \nonumber & \fg_{tt} = -1 +{2mr \over
\rho ^2} \;, \;\;\;\; \fg_{t\varphi } = -{2mra\sin ^2 \theta \over
\rho ^2} \;,\;\;\;\; \fg_{rr} = {\rho ^2 \over \triangle} \;,
\;\;\;\; \fg_{\theta \theta} = \rho^2 \;, &\\& \fg_{\varphi
\varphi } =
\sin^2\theta\left(r^2+a^2+{2mra^2\sin^2\theta\over\rho^2}\right)\;,&
\label{kerrm}\end{eqnarray} where
\[
\;\;\;\;
 \rho ^2 = r^2 + a^2 \cos ^2 \theta \;,\;\;\;\; \triangle = r^2 -2mr +a^2 \;.\]
Introduce a ``quasi-Minkowskian" coordinate system
$(x^\mu)=(t,x^i)$ by setting $$x^1=r\sin\theta \cos
\varphi\;,\quad x^2=r\sin\theta \sin \varphi\;,\quad
x^3=r\cos\theta\;,$$ which brings $\fg_{\mu\nu}$ to the form
$\eta_{\mu\nu}+O(r^{-1})$ for $x^\mu$'s in a set $r\ge R_0$ for
some $R_0$, and apply to it a Poincar\'e transformation \eq{P1}.
We further assume that $(\Lambda^\mu{}_\nu,a^\mu) $ belongs to the
connected component $G_0$ of the identity of the Poincar\'e group.
Then $(g,K)$ are defined on the set \bel{P2}
\left\{\sqrt{\sum_i(y^i)^2} > R\right\}\ee for some
$R=R(m,a,\Lambda^\mu{}_\nu, a^\mu)$ by extracting the
gravitational initial data out of the metric $\fg$ on the
hypersurface $y^0=0$. The function $R(m,a,\Lambda^\mu{}_\nu,
a^\mu)$ can be chosen to be continuous, in particular for any set
$(m_0,a_0,\Lambda_0{}^\mu{}_\nu, a_0^\mu)$ there exists a
neighborhood $\mcO_0$ thereof such that $R(m,a,\Lambda^\mu{}_\nu,
a^\mu)$ can be chosen independently of $(m,a,\Lambda^\mu{}_\nu,
a^\mu)\in \mcO_0$. We equip $\cKi$ with the topology of uniform
convergence on relatively compact open sets; any weighted Sobolev
topology on the set of initial data will lead, by restriction, to
this topology on $\cKi$.

We wish to show that the set $\cKi$ can be uniquely
parameterized\footnote{The construction of the set $\cKi$ involves
twelve free parameters, however two of them are redundant because
of the existence of the two-parameter group of isometries of the
Kerr metric.}  by the Poincar\'e charges $(p_\mu,J_{\mu\nu})$
defined in \eq{mi1}-\eq{mi2} , with $p_\mu$ ranging over the set
of timelike vectors $I(0)$ in the Minkowski space-time $\R^{1,3}$,
and $J_{\mu\nu}$ ranging over all anti-symmetric two-covariant
tensors. In other words:
\begin{Proposition}\label{Ppcm}
 The map \bel{Pcm} Q:\cKi\ni (g,K)\to(p_\mu,J_{\mu\nu})
\in I(0)\times\R^6\subset \R^4\times\R^6 \ee is a continuous
bijection.
\end{Proposition}
\proof Let $(g,K)$ be the Cauchy data on $\{x^0=0, r\ge R_0 \}$
for a Kerr metric as above with some parameters $m\in \R^*$ and
$a\in\R$, we then have \bel{Pcm1}p_\mu=(m,0,0,0)\;, \ J_{\mu\nu}=
2ma \delta^1_{[\mu}\delta^2_{\nu]}\;.\ee
The transformation law \eq{Pcm2a} shows that for any vector
$n^i\in\R^3$ satisfying $\sum_i(n^i)^2=1$ we can obtain a pair
$(p_\mu,J_{\mu\nu})$ of the form \bel{Pcm3} p_\mu = m
\delta^0_\mu\;,\quad J_{0i}=0\;, \ J_{ij}=\pm
ma\epsilon_{ijk}n^k\;,\ee by
\begin{itemize} \item either performing a rotation by an angle
less than or equal to $\pi/2$ in the plane
$\textrm{Span}(\partial_z, n^i\partial_i)$ which brings
$n^i\partial_i$ to $\partial_z$, then we choose the sign $+$, or
\item we perform a rotation by an angle less than or equal to $\pi/2$ in
the plane $\textrm{Span}(\partial_z, n^i\partial_i)$ which brings
$n^i\partial_i$ to $- \partial_z$, then we choose the sign $-$.
\end{itemize} In the overlapping case $n^i\partial_i\perp \partial_z$ the choice does not matter because the
resulting metrics (and thus initial data) are identical ``modulo
gauge"
--- the corresponding transformation $a\to -a$, $(t,r,\varphi,\theta)\to (t,r, -\varphi,\pi-\theta)$
 is an isometry of the Kerr metric. Next, a
space-translation $a^i\in\R^3$ produces out of \eq{Pcm3} a pair
$(p_\mu,J_{\mu\nu})$ \bel{Pcm4} p_\mu = m \delta^0_\mu\;,\quad
J_{0i}=-ma_i\;, \ J_{ij}=\pm ma\epsilon_{ijk}n^k\;.\ee It follows
that any set $(p_\mu=m\delta^0_\mu,J_{\mu\nu})$ can be obtained in
a unique way by calculating the charges  \eq{mi1}-\eq{mi2} using
initial data in $\cKi$ by the operations just described. Now, for
any timelike $p_\mu$ there exists precisely one boost
transformation $\Lambda^\mu{}_\nu$ in the plane
$\textrm{Span}(\delta_\mu^0,p_\mu)$ which maps $m\delta_\mu^0$ to
$p_\mu$, provided $m$ is suitably chosen, and we conclude by
noting that, at fixed $\Lambda^\mu{}_\nu$, the map
$$\R^6\ni J_{\mu\nu}\to \Lambda_\mu{}^\alpha{}\Lambda_\nu{}^\beta{}
J_{\alpha\beta}\in\R^6$$ is a linear isomorphism.\qed

We end this section by verifying that the initial data for the
Kerr metric in Boyer-Lindquist coordinates are parity symmetric.
First, we note that  $g_{ij}$ is obviously even. Next, we have
\bel{kerrshift} ^4g_{ti} dx^i = \fg_{t\varphi }d\varphi  = -{2mra\sin ^2 \theta
\over \rho ^2}d\varphi =-{2ma\over r\rho ^2}(xdy-ydx)\;,\ee so
that the coordinate components of the shift vector are odd. Now,
the lapse function is symmetric under parity. Further, the
derivatives of an even function are odd and vice-versa; in
particular the Christoffel symbols are odd while the partial
derivatives of the coordinate components of the shift vector are
even. The usual formula for $K_{ij}$ in terms of the derivatives
of the shift vector yields the result.

\section {Uniform local invertibility}

\begin{prop}\label{prop:invloc}
Let $(V_x,\|\cdot\|_{V_x})_{x\in A}$ and
$(W_x,\|\cdot\|_{W_x})_{x\in A}$ be two families of  Banach
spaces. Let $r>0$ and let $\{f_x:B_{V_x}(0,r)\rightarrow
W_x\}_{x\in A}$ be a family of differentiable functions such that:
\begin{enumerate}\item $Df_x(0):V_x\longrightarrow W_x$ has a
right inverse for all $x\in A$ which is bounded independently of
$x\in A$.\item $\|f_x(v+h)-f_x(v)-Df_x(v)h\|_{W_x}/\|h\|_{V_x}^2$
is bounded independently of $x\in A$, $v\in B_{V_x}(0,r)$ and
$h\in V$ such that $v+h\in B_{V_x}(0,r)$.\item
$\|Df_{x}(v+h)-Df_x(v)\|_{L(V_x,W_x)}/\|h\|_{V_x}$ is bounded
independently of $x\in A$, $v\in B_{V_x}(0,r)$ and $h\in V$ such
that $v+h\in B_{V_x}(0,r)$.\end{enumerate} Then there exists
$\epsilon>0$ and $C>0$ such that for all $x\in A$ and all $\delta
f\in W_x$, $\|\delta f\|_{W_x}<\epsilon$, there exists a solution
$\delta x\in V_x$ of the equation
$$
f_x(\delta x)-f_x(0)=\delta f,
$$
which satisfies $\|\delta x\|_{V_x}\leq C \|\delta f\|_{W_x}$.
\end{prop}
\begin{proof}
{}From 1),  there exist a constant $C_1$ such that  for all $x\in
A$ and all $w\in W_x$, the equation
$$Df_x(0)h=w\;,$$
has a solution $h\in V_x$ such that
$$\|h\|_{V_x}\leq C_1\|w\|_{W_x}.$$
{}From 2) and 3), there exist constants $C_2$ and $C_3$ such that
for all $x\in A$, all $v\in B_{V_x}(0,r)$ and all $h\in V$,
$$
\|f_x(v+h)-f_x(v)-Df_x(v)h\|_{W_x}\leq C_2 \|h\|_{V_x}^2,
$$
$$
\|Df_{x}(v+h)-Df_x(v)\|_{L(V_x,W_x)}\leq C_3 \|h\|_{V_x}.
$$
Let $x\in A$ and $\delta f\in W_x$. We will construct a Picard
sequence $\{ h_n\}$ such that $ \sum h_n$  converges to a solution
when $\delta f$ is small enough. {}From 1), we have  a solution
$h_0\in V_x$  of
$$
Df_x(0)h_0=\delta f,
$$
which satisfies $\|h_0\|_{V_x}\leq C_1 \|\delta f\|_{W_x}.$ Let
$\delta x_1:=h_0$ which is in $B_{V_x}(0,r)$ if $\|\delta
f\|_{W_x}$ is small enough. Let us now define the sequence
$h_{i+1}$, solution of
\begin{eqnarray*}
Df_x(0)h_{i+1}=f_x(0)-f_x(\delta x_{i+1})+\delta f\;,
\end{eqnarray*}
where $\delta x_{i+1}=\delta x_i+h_i$ (we assume that $\delta
x_{i+1}\in B_{V_x}(0,r)$, it will be justified at the end of the
proof). We have that
\begin{eqnarray}\nonumber
Df_x(0)h_{i+1}&=&f_x(0)-f_x(\delta x_{i})+f_x(\delta
x_{i})-f_x(\delta x_{i+1})+\delta f\\\nonumber
&=&Df_x(0)h_i-[f_x(\delta x_{i+1})-f_x(\delta x_i)]\\\nonumber
&=&\sum_{p=0}^{i-1}[Df_x(\delta x_{p})-Df_x(\delta x_{p+1})]h_i\\
&&\;\;+[Df_x(\delta x_i)h_i+f_x(\delta x_i)-f_x(\delta
x_{i}+h_i)]\;,\label{grosseequation}
\end{eqnarray}
with $\delta x_0=0$, so by hypothesis 1), 2) and 3), we have
$$
\|h_{i+1}\|_{V_x}\leq
C_1\left(\sum_{p=0}^{i-1}C_3\|h_p\|_{V_x}\|h_i\|_{V_x}+C_2\|h_i\|_{V_x}^2\right)\;.
$$
Let $K:=\max(C_1C_2,C_1C_3)$, then we have
$$
\|h_{i+1}\|_{V_x}\leq K\|h_i\|_{V_x}\sum_{p=0}^{i}\|h_p\|_{V_x}.
$$
Choose any $\delta\in]0,1[$, let $\epsilon$ be small enough so
that
$$
KC_1\epsilon<1\;,\mbox{  }
\frac{(KC_1\epsilon)^{1-\delta}}{1-(KC_1\epsilon)^{\delta}}\leq1\;,
$$
and such that for all $t\in[0,\epsilon[$,
$$
\frac{C_1 t}{1-(KC_1 t)^{\delta}}\leq 2C_1 t<r,
$$
and let $C:=2C_1$. If $\|\delta f\|_{W_x}\leq \epsilon,$ from
Lemma~\ref{lem:suite} with $a_i=\|h_i\|_{V_x}$,
 the sequence $\delta x_{n+1}:=\sum_{i=0}^n h_i$ is convergent
in $V_x$ to a limit $\delta x$ which satisfies
$$
\|\delta x\|_{V_x}\leq
\frac{\|h_0\|_{V_x}}{1-(K\|h_0\|_{V_x})^{\delta}}\leq
\frac{C_1\|\delta f\|_{W_x}}{1-(KC_1\|\delta f\|_{W_x})^{\delta}}
\leq C\|\delta f\|_{W_x}< r.
$$
Note that for all $n\geq0$, $\|\delta x_{n+1}\|_{V_x}<r$. On the
other hand, as $h_{i+1}$ goes to zero in $V_x$ we have that
$f_x(0)-f_x( \delta x_i)-\delta f=Df_x(0)h_{i+1}$ goes to zero in
$W_x$.
\end{proof}
\qed

{The following result is needed to be able to obtain weighted
H\"older regularity of the solutions obtained, to start with, in
weighted Sobolev spaces. In our applications the spaces $E_x$ will
be the weighted H\"older spaces
$C^{k+2,\alpha}_{\phi,\varphi}\times
C^{k+2,\alpha}_{\phi,\varphi}$, the $F_x$'s will be
$C^{k+1,\alpha}_{\phi,\varphi'}\times
C^{k,\alpha}_{\phi,\varphi'}$, the $G_x$'s will correspond to
$C^{k+1,\alpha}_{\phi,\varphi''}\times
C^{k,\alpha}_{\phi,\varphi''}$, for  appropriate weights
$\varphi,\varphi',\varphi''$, see the proof of
Proposition~\ref{regularite}. Finally, $A$ should be thought of as
a neighborhood of $x_0=(K_0,g_0)$ in
$(C^{k+3,\alpha}_{\phi,1}\times C^{k+4,\alpha}_{\phi,1})\cap
(W_\phi^{k+3,\infty}\times W_\phi^{k+4,\infty})$.}

For the following  result we shall denote by
$$Df_x(0)^{-1}_r$$ the
right inverse of $Df_x(0)$, the existence of which has been
assumed in point1. of the preceding proposition.

\begin{prop}\label{prop:regul}
Under the hypotheses of Proposition~\ref{prop:invloc}, consider
three families of Banach spaces $(E_x,\|\cdot\|_{E_x})_{x\in A}$,
$(F_x,\|\cdot\|_{F_x})_{x\in A}$ and $(G_x,\|\cdot\|_{G_x})_{x\in
A}$ such that $G_x$ is continuously embedded both in $F_x$  and in
$W_x$, with the norms of the embeddings uniformly bounded in $x\in
A$. Assume there exist a $r'>0$ such that
$\{f_x:B_{E_x}(0,r')\rightarrow F_x\}_{x\in A}$ is defined,
differentiable and verifies:
\begin{enumerate}\item
 if $h$ is in the image of $Df_x(0)^{-1}_r$ and
$Df_x(0)h\in F_x$ then $h\in E_x$ and
 $$\|h\|_{E_x}\leq C (\|h\|_{V_x}+\|Df_x(0)h\|_{F_x}),$$ where $C$ does not depend
 on $x\in A$.\item $\|f_x(v+h)-f_x(v)-Df_x(v)h\|_{G_x}/\|h\|_{E_x}^2$
is bounded independently of $x\in A$, $v\in B_{E_x}(0,r')$ and
$h\in E_x$ such that $v+h\in B_{E_x}(0,r')$.
\item
$\|Df_{x}(v+h)-Df_x(v)\|_{L(E_x,G_x)}/\|h\|_{E_x}$ is bounded
independently of $x\in A$, $v\in B_{E_x}(0,r')$ and $h\in V$  such
that $v+h\in B_{E_x}(0,r')$.\end{enumerate} Then there exists
$\epsilon>0$ and $C'>0$ such that for all $x\in A$ and all $\delta
f\in W_x\cap F_x$ satisfying
$$\|\delta f\|_{W_x}+\|\delta
f\|_{F_x}<\epsilon$$ there exists a solution $\delta x \in E_x $
satisfying
$$\|\delta x\|_{E_x}\leq C'( \|\delta
f\|_{W_x}+\|\delta f\|_{F_x})\;.$$
\end{prop}
\begin{proof}
The constant $C$ which appears in the proof may change from term
to term and line to line. The solution is constructed by the same
method as in the proof  of Proposition~\ref{prop:invloc}. Let,
thus, $h_i$ be the sequence defined there, by hypothesis 1. for
all $i\geq -1$ we have $h_{i+1}\in E_x$ and
\begin{eqnarray*} \|h_{i+1}\|_{E_x}&\leq&
C(\|h_{i+1}\|_{V_x}+\|Df_x(0)h_{i+1}\|_{F_x}) \\
&\leq& C(\|Df_x(0)h_{i+1}\|_{W_x}+\|Df_x(0)h_{i+1}\|_{F_x})\leq
C\|Df_x(0)h_{i+1}\|_{G_x}\;,\end{eqnarray*} which is clearly true
regardless of whether or not the last term is finite. On the other
hand, from equation \eq{grosseequation} together with the
hypotheses 2. and 3.
 we have that
 $$\|Df_x(0)h_{i+1}\|_{G_x}\leq C
 \|h_i\|_{E_x}\sum_{k=0}^{i}\|h_k\|_{E_x}.$$
 So from Lemma \ref{lem:suite} with $a_i=\|h_i\|_{E_x}$, if $\delta f$
 is sufficiently small in $F_x$ norm, then
 the sequence $\sum_{k=0}^i h_k$ is convergent
 in $E_x$ to some element $\delta x\in E_x $, with
 $$\|\delta x\|_{E_x}\leq C\|h_0\|_{E_x}
\leq C(\|h_{0}\|_{V_x}+\|Df_x(0)h_{0}\|_{F_x}) = C (\|\delta
f\|_{W_x}+\|\delta f\|_{F_x}).$$
\end{proof}
\qed

\subsection{A sequence adapted to the Picard method}

\begin{lem}\label{lem:suite}
Let $K>0$, $\delta\in ]0,1[$,  and let $\{a_n\}_{n\in\Nat}$ be a
sequence with non-negative terms
 which
verifies, for all $n>0$,
$$
a_{n+1}\leq K a_n \sum_{i=0}^{n}a_i\; .
$$
If $a_0$ is small enough to verify
$$
Ka_0<1\mbox{ and }\frac{(Ka_0)^{1-\delta}}{1-(Ka_0)^{\delta}} \leq
1,
$$
then the sequence $S_n(a_0):=\sum_{i=0}^{n}a_i$ is convergent to a
limit denoted $S(a_0)$ which satisfies
$$
0\leq S_n(a_0)\leq S(a_0)\leq \frac{a_0}{1-(Ka_0)^{\delta}},
$$
in particular, $S$ is continuous at $0$.

\end{lem}
\begin{proof}
Let $b_n:=Ka_n$, we have
$$
b_{n+1}\leq b_n \sum_{i=0}^{n}b_i.
$$
We will show by induction that
\begin{eqnarray}\label{ui0}
b_n\leq b_0^{1+n\delta}.
\end{eqnarray}
Equation \ref{ui0} holds for $n=0$,  assume it hold for all
integers less than or equal to $n$, we then have
\begin{eqnarray*}
b_{n+1}&\leq&
b_0^{1+n\delta}\sum_{i=0}^n b_0^{1+i\delta}\\
&\leq&b_0^{2+n\delta}\frac{1-b_0^{(n+1)\delta}}{1-b_0^{\delta}}\\
&\leq&b_0^{2+n\delta}\frac{1}{1-b_0^{\delta}}\\
&\leq&b_0^{1+(n+1)\delta}\frac{b_0^{1-\delta}}{1-b_0^{\delta}}\\
&\leq&b_0^{1+(n+1)\delta}\;, \\
\end{eqnarray*}
the last inequality following from the second hypothesis on $a_0$.
To conclude, it suffices to remark that
$$
0\leq\sum_{i=0}^n b_i\leq \sum_{i=0}^\infty
b_i\leq\sum_{i=0}^\infty b_0^{1+i\delta}
=\frac{b_0}{1-b_0^{\delta}}.
$$
\end{proof}

 \section{Small initial data on a bounded domain in $\R^3$}
\label{Appsid}\newcommand{\tg}{{\widetilde
g}}\newcommand{\tD}{{\widetilde D}}  Let $\Omega$ be a bounded
domain in $\R^3$ with smooth boundary, and let $\tg$ be any smooth
up-to-boundary Riemannian metric on $\bar \Omega$ such that
\bel{LE0}\frac 12 e(X,X)\le \tg(X,X)\le 2 e(X,X)\;,\ee where $e$
is the Euclidean metric. It can be seen that there are no
conformal Killing vectors which vanish on $\partial \Omega$ ({\em
cf., e.g.}\/~\cite[Proposition~6.2.2]{AndChDiss}) which implies
that the operator
$$\mathring{H}_2\ni X \to \tD_i\left(\tD^i X^j + \tD ^j X^i
- \frac 23 \tD_k X^k \tg ^{ij}\right)\in L^2
$$ has no kernel ($\tD$ --- the Levi-Civita connection of $\tg$),
and can thus be used to construct $\tg $-transverse ($\tD_i
L^{ij}=0$) traceless ($\tg^{ij}L_{ij}=0$) tensors $L_{ij}$ on
$\Omega$ in the usual way. When $\tilde g$ is parity-symmetric,
then parity-antisymmetric $L_{ij}$'s can be obtained by replacing
$L_{ij}$ with $(L_{ij}(x)-L_{ij}(-x))/2$. Let, thus, any
parity-antisymmetric, transverse, traceless, $L_{ij}$ be given,
for $\sigma \in [0,1]$ consider the Lichnerowicz
equation:\bel{LE1} 8 \Delta_\tg \phi - R(\tg) \phi + \sigma ^2
|L|_\tg ^2 \phi^{-7}=0\;,\ee which we rewrite as \bel{LE6} Lu:=
(\Delta_\tg + s)u = F(u)\;,\ee where $u:= \phi- 1$, while
$\Delta_\tg +s$ is the linearisation of $\frac 18$\eq{LE1} at
$\phi=1$,
$$ s:=- \frac {R(\tg)}8  -7 \sigma ^2
|L|_\tg ^2 \;. $$One will obtain a solution $$
(K_{ij}:=\sigma\phi^{-2} L_{ij},g_{ij}:=\phi^4 \tg_{ij})$$ of the
vacuum constraint equations using the inverse function  theorem
in, \emph{e.g.}, weighted H\"older spaces, if one can show that
the operator $L$ appearing at the left-hand-side of \eq{LE6} has
no kernel. In order to show that this is indeed the case for $g-e$
small enough in $C_2(\bar \Omega)$, and for $\sigma$ small enough,
let $C_P$ be the constant appearing in the Poincar\'e inequality
for $\Omega$: \bel{SE3} \forall \ u \in \mathring{H}_1(\Omega)
\quad \int_\Omega u^2 d^3x \le C_P \int_\Omega |du|^2_e d^3x
\;,\ee it follows from \eq{LE0} that we also have \bel{SE4}
\forall \ u \in \mathring{H}_1(\Omega) \quad \int_\Omega u^2
\,\sqrt{\det\, \tg }\, d^3x \le 4 \sqrt{2} C_P \int_\Omega
|du|^2_\tg  \sqrt{\det\, \tg }\, d^3x \;,\ee If $Lu=0$, by
integration by parts one obtains
$$
\int_\Omega \left(- |du^2|_\tg  + su^2\right)\sqrt{\det\, \tg }\,
d^3x=0\;,$$ and the Poincar\'e inequality gives
\begin{eqnarray}\nonumber
\int_\Omega u^2 \,\sqrt{\det\, \tg }\, d^3x &\le& 4 \sqrt{2} C_P
\int_\Omega |du|^2_\tg  \sqrt{\det\, \tg }\, d^3x
\\
& \le& 4 \sqrt{2} C_P \sup |s| \int_\Omega u^2 \sqrt{\det\, \tg
}\, d^3x \;,\end{eqnarray} hence $u=0$ if
$\|s\|_{L^\infty(\Omega)}$ is small enough, and the inverse
function theorem applies. Clearly the resulting $(K,g)$ will be
non-trivial as soon as $\tg$ is not conformally flat.

Let $(m,\vec p)$ be the ADM four-momentum of $(K,g)$ obtained by
integrating $\ourU$ given by \eq{Fsup2new} (expressed in terms of
$g$ and $K$) over $\partial \Omega$; here $b$ should be taken as
the Minkowski metric, and space coordinates harmonic for $g$
should be used --- such coordinates can be found globally on
$\Omega$ if $g$ is close enough to $e$. At $\sigma=0$ we have
$\vec p=0$, while it follows from the calculations in
\cite{Bartnik:mass} that $m>0$ (choosing $g$ closer to $e$ if
necessary). Continuity then shows that choosing $\sigma$ small
enough we will obtain
$$|\vec p|_e \le \frac 12 m\;.$$ The initial data set $ (K,g)$ will then fulfill all the requirements set
forth in Theorem~\ref{Textg}.

\medskip

\noindent{\sc Acknowledgements:}  We thank R.~Beig, J.~Corvino,
H.~Friedrich and W.~Simon for useful comments or discussions, as
well as a referee for detailed criticism.

\bibliographystyle{amsplain}
\bibliography{
../../../references/newbiblio,%
../../../references/reffile,%
../../../references/bibl,%
../../../references/Energy,%
../../../references/hip_bib,%
../../../references/netbiblio}
\end{document}